\documentclass[fleqn,usenatbib]{mnras}

\usepackage{newtxtext,newtxmath}
\usepackage[T1]{fontenc}
\usepackage{graphicx}
\usepackage{amsmath}
\usepackage{xcolor}
\usepackage{siunitx}
\usepackage{subfigure}
\usepackage{multirow}
\usepackage{multicol}

\DeclareSIUnit\parsec{pc}
\DeclareSIUnit\ph{ph}
\DeclareSIUnit\year{yr}
\DeclareSIUnit\simsun{M_\odot}
\DeclareSIUnit\ergs{ergs}
\DeclareSIUnit\ev{eV}
\DeclareSIPrefix\comovmega{cM}{6}
\DeclareSIPrefix\comovkilo{ck}{3}
\DeclareSIUnit\h{h}
\DeclareSIUnit\angs{\textup{\AA}}

\newcommand{\fesc}{$f_{\rm esc}$}
\newcommand{\xhi}{$x_{\rm HI}$}
\newcommand{\mstel}{$M_\star$}
\newcommand{\lintr}{$L_{\rm intr}$}

\newcommand{\fescglob}{$f_{\rm esc}^{20\%}$}
\newcommand{\fescevol}{$f_{\rm esc}(z)$}
\newcommand{\feschalo}{$f_{\rm esc}(M_{\rm h})$}
\newcommand{\mean}[1]{\langle #1 \rangle}
\newcommand{\dtb}{$\delta T_{\rm b}$}
\newcommand{\avgdtb}{$\overline{\delta T_{\rm b}}$}

\newcommand{\dtbPS}{$\overline{\delta T_{\rm b}^2}  \Delta_{21}^2$}
\newcommand{\PS}{$\Delta_{21}^2$}

\newcommand{\figref}[1]{Fig. \ref{#1}}
\newcommand{\code}[1]{\textsc{#1}}

\defcitealias{collaboration_planck_2021}{Planck VI (2018)}
\defcitealias{planck_collaboration_planck_2016}{Planck XIII (2016)}
\defcitealias{planck_cosmo_2013}{Planck XVI (2013)}

\title[Galaxy formation models and the 21cm signal]{The signature of galaxy formation models in the power spectrum of the hydrogen 21cm line during reionization}
\author[J. Lewis et al.]{Joseph S. W. Lewis$^{1,2}$\thanks{E-mail: lewis@iap.fr},
Annalisa Pillepich$^{2}$,
Dylan Nelson$^{1}$,
Ralf S. Klessen$^{1,3}$,
Simon C.~O.~Glover$^{1}$
\\\\
$^{1}$ Universit\"{a}t Heidelberg, Zentrum f\"{u}r Astronomie, Institut f\"{u}r Theoretische Astrophysik, Albert-Ueberle-Str. 2, 69120 Heidelberg, Germany\\
$^{2}$ Max-Planck-Institut f\"{u}r Astronomie, K\"{o}nigstuhl 17, 69117 Heidelberg, Germany\\
$^{3}$ Universit\"{a}t Heidelberg, Interdisziplin \"{a}res Zentrum f\"{u}r Wissenschaftliches
Rechnen, Im Neuenheimer Feld 225, D-69120 Heidelberg, Germany
}
\date{}
\pubyear{2023}

\begin{document}
\label{firstpage}
\pagerange{\pageref{firstpage}--\pageref{lastpage}}
\maketitle

\begin{abstract}
Observations of the 21cm line of hydrogen are poised to revolutionize our knowledge of reionization and the first galaxies. However, harnessing such information requires robust and comprehensive theoretical modeling. We study the non-linear effects of hydrodynamics and astrophysical feedback processes, including stellar and AGN feedback, on the 21cm signal by post-processing three existing cosmological hydrodynamical simulations of galaxy formation: Illustris, IllustrisTNG, and Eagle. Despite their different underlying galaxy-formation models, the simulations return similar predictions for the global 21cm brightness temperature and its power spectrum. At fixed redshift, most differences are attributable to alternative reionization histories, in turn driven by differences in the build-up of stellar sources of radiation. However, several astrophysical processes imprint signatures in the 21cm power spectrum at two key scales. First, we find significant small scale ($k \geq 10\, \rm {Mpc}^{-1}$) differences between Illustris and IllustrisTNG, where higher velocity winds generated by supernova feedback soften density peaks, leading to lower 21cm power in TNG. Thus, constraints at these scales could rule out extreme feedback models. Second, we find more 21cm power at intermediate scales ($k \approx 0.8\, \rm {Mpc}^{-1}$) in Eagle, due to ionization differences driven by highly effective stellar feedback, resulting in lower star formation, older and redder stellar populations, and lower ionizing luminosities  for $M_h > 10^9 \rm M_\odot$. Different source models can manifest similarly in the 21cm power spectrum, leading to often ignored degeneracies. These subtle features could allow future observations of the 21cm signal, in conjunction with other observables, to constrain theoretical models for galactic feedback at high redshift.
\end{abstract}



\section{Introduction}
\label{sec:intro}

Hydrogen in the intergalactic medium (IGM) was progressively ionized by light from the first galaxies during the epoch of reionization \citep[][]{barkana_physics_2007}. This process began a few \SI{100}{\mega\year} after the Big Bang (roughly $z\approx 15-30$), and ended when the Universe was over \SI{1}{\giga\year} old, by $z\approx 5.3$, as inferred in \citetalias{collaboration_planck_2021}. Direct observations of galaxies at these high redshifts are challenging, requiring long integration times, techniques such as strong lensing \citep[][]{atek_extreme_2018}, or new telescopes such as JWST. Our knowledge of early galaxy populations and the sources of cosmic reionization is correspondingly limited.

Because reionization is driven by the ionizing light of galaxies, the progress of reionization is coupled to the population of galaxies in the early Universe. Therefore, we can learn about these first galaxies by observing the gas in the IGM. For instance, studies of the Lyman-alpha transmission in the spectra of distant quasars \citep[e.g.][]{fan_constraining_2006, becker_refined_2013, eilers_opacity_2018, yang_measurements_2020, bosman_hydrogen_2021} inform us of the time-frame of reionization, as well as the topology of the ionized regions as reionization completes. However, these techniques will not allow us to probe far beyond $z=6$ due to the increasingly large column densities of neutral hydrogen. This is where the 21cm line of atomic hydrogen plays a crucial role.

The 21cm signal amplitude from a particular volume traces the density of neutral hydrogen gas, its temperature and velocity, as well as the local radiation field \citep[see][for a review of the physics]{furlanetto_cosmology_2006}. Thus, the sky-averaged signal traces the global progress of reionization and constrains first light. Already, there has been a reported detection of the global signal for $z>10$ by EDGES \citep[][]{monsalve_results_2017}, though this is strongly contested by results from SARAS-3 \citep{bevins_saras-3_2022}. At the same time, interferometry and 21cm tomography can probe the heterogeneity of reionization: experiments such as LOFAR \citep[][]{mertens_improved_2020}, MWA \citep[][]{trott_angular_2022} and HERA \citep[][]{abdurashidova_first_2022} have begun to provide upper limits on the power spectrum of the 21cm signal. Over the coming decades, new 21cm observatories will see first light \citep[e.g. the Square Kilometer Array;][]{koopmans_cosmic_2015}.

In order to interpret this new data, we must understand the astrophysical processes and underlying reionization scenarios that shape the 21cm signal. Existing theoretical work can be divided into several categories based on the approach. Analytical models predict the evolution of the global 21cm signal \citep[such as][]{furlanetto_cosmology_2006,mirocha_interpreting_2015} and various statistics thereof. Such models are computationally cheap, and can be used to explore a wide range of cosmological and universal astrophysical parameters. 

Cosmological hydrodynamical simulations, on the other hand, directly model baryonic matter, the formation of stars, and the complex non-linear coupling of physical processes across spatial scales and matter components. In particular, the implementation of stellar and supermassive black hole (SMBH) formation, and the corresponding feedback onto gas, are essential to regulate star formation and reproduce realistic galaxies \citep[e.g.][]{dubois_dancing_2014,schaye_eagle_2015,pillepich_simulating_2018}. Star formation and feedback models can fundamentally alter, for example, the stellar mass to halo mass relationship, in turn modifying the number and distribution of ionizing sources that drive reionization, and shape the 21cm signal at $z\lesssim10$. Feedback from star formation and SMBHs can also vary locally and over time, significantly affecting the distribution of matter in and around galaxies \citep[e.g.][]{ayromlou_feedback_2022}, modulating the matter clustering also on $\gtrsim$ Mpc spatial scales \citep[e.g.][]{springel_first_2018}, heating the gas \citep[e.g.][]{zinger_feedback_2020} and determining the gas velocity structure in and around galaxies and haloes \citep{nelson_first_2019, pillepich_bubbles_2021}. All this can manifest itself differently at different cosmic epochs. However, simulations that do not model radiation transfer on the fly must treat reionization in post-processing, either with excursion set techniques \citep[][]{hutter_accuracy_2018, hutter_21cm_2019} or by full radiative transfer methods \citep[e.g.][]{georgiev_large-scale_2022,semelin_21ssd_2017, bauer_hydrogen_2015}. As a result, running cosmological hydrodynamical simulations as well as variants to explore the relationship between the 21cm signal and different underlying galaxy formation models is more expensive than with analytical or semi-analytical models.

Cosmological hydrodynamical galaxy simulations must also account for the coupled physics between gas and radiation that are relevant for reionization and the 21cm signal. Heating from ionizing (UV and X-ray) photons directly impacts the strength of the 21cm signal, and also suppresses star formation in low mass haloes \citep[][]{dawoodbhoy_suppression_2018, wu_simulating_2019}, again potentially leaving an imprint on the 21cm signal. This effect can be captured with fully coupled radiative transfer simulations, where the gas ionization and heating are tracked on the fly. Second, in order to self-consistently derive the 21cm signal before IGM heating, one must track the production and transfer of $\rm Lyman-\alpha$ photons to account for the Wouthuysen-Field effect \citep[][]{wouthuysen_excitation_1952, field_excitation_1958}. Finally, correctly modeling the formation of the first stars is needed to determine the 21cm signal at early times \citep[e.g. ][]{magg_effect_2022, sartorio_population_2023}, and relies on following the formation of $\rm H_2$-cooled mini-haloes, subsequent photo-dissociation of the molecular gas, and the transition from Pop III to Pop II stars \citep[][]{klessen_first_23}. 

Ideally, simulations must track the radiative transfer of all of the relevant photons in a fully coupled radiation and hydrodynamics (RHD) simulation. To date, several simulation projects have done so, focusing on reionization in a representative volume (for most EoR statistics i.e. $\gtrsim 100^3 \rm Mpc^3$; see \cite{iliev_simulating_2014}), such as: CROC \citep[][]{gnedin_cosmic_2014}, CoDa \citep[][]{ocvirk_cosmic_2016,ocvirk_cosmic_2020,lewis_short_2022}, Aurora \citep[][]{pawlik_aurora_2017}, and THESAN \citep[][]{kannan_introducing_2021}. These efforts are increasingly successful at reproducing constraints on the ionization of the IGM. One should also note the SPHINX simulations \citep[][]{rosdahl_sphinx_2018, katz_introducing_2021}, that focus on more detailed but smaller volumes, and have been leveraged to investigate e.g. $\rm Lyman-\alpha$ emission during the EoR \citep[][]{garel_lyman-alpha_2021}.
Because of the high computational expense of fully coupled RHD, simulations of volumes large enough to fully account for the effects of cosmic variance on 21cm statistics do not yet resolve galaxies \citep[e.g.][]{semelin_lyman-alpha_2007}. Others couple approximate RT methods to the hydrodynamics \citep[e.g.][]{davies_efficient_2023}.

In this paper we aim to strike a middle ground, by relying upon existing large cosmological hydrodynamical simulations of galaxies and by post-processing them to model reionization. Our goal is to compare predictions for the 21cm signal of hydrogen during reionization. In particular, we use the outcome of three cosmological simulations (Illustris, IllustrisTNG and EAGLE, see Section~\ref{sec:mths}) and contrast their predictions to quantify the link between different galaxy formation models and the 21cm signal. These simulations allow us to self-consistently account for the complex interplay between hydrodynamics and galactic feedback processes.

In Sec. \ref{sec:mths} we present the simulations used in this work, our approach for modeling the ionization and heating of the IGM gas during reionization, and finally our assumptions when computing the brightness temperature of the 21cm line, and its power spectrum. Then, in Sec. \ref{sec:results} we showcase the predicted brightness temperatures, starting with maps, moving on to the average global signal, and finally the 21cm power spectra. To explain these findings, we study statistics about the IGM gas and galaxies in Sec. \ref{sec:disc}. We finish with our concluding remarks in Sec. \ref{sec:concl}.


\section{Methods}

\label{sec:mths}

\subsection{Cosmological hydrodynamical simulations of galaxies}

In this paper, we investigate the 21cm signal from three cosmological hydrodynamical galaxy simulations suites: Illustris, IllustrisTNG, and Eagle. We base our analysis on the similar, roughly \SI[exponent-base=100, input-exponent-markers=k]{k3}{\mega\parsec\cubed}, volumes. All three simulation projects feature models that account for the star formation and baryon feedback physics required to form reasonably realistic galaxies by $z=0$. However, significant differences exist in the implementation, parametrization, and outcome of these models. Whereas these have been compared to each other and to observational constraints across a diverse set of galaxy and large-scale structure properties at low redshifts \citep[e.g.][]{somerville_review_2015, vogelsberger_cosmological_2020}, here we focus on the effects of the different underlying galaxy-formation models on the resulting 21cm signals during the epoch of reionization. In fact, differences in the feedback models across these three simulations are expected to impact the temperature and density of gas in the environments of galaxies, in turn altering the 21cm signal from these regions.

\begin{table*}
\centering
\begin{tabular}{l|c|cc|c}
     \multirow{2}{*}{} & \multirow{2}{*}{\textbf{\Large Illustris}} & \multicolumn{2}{c|}{\multirow{2}{*}{\textbf{\Large IllustrisTNG}}} & \multirow{2}{*}{\textbf{\Large Eagle}} \\
     &&&&\\
     & & {\large TNG100} & {\large TNG100-NR} &\\
     \hline 
     &&&&\\
     {\large Numerical code}& \code{AREPO} \citep[][]{springel_arepo_2010} & \multicolumn{2}{c|}{\code{AREPO} \citep[][]{springel_arepo_2010}} & \code{GADGET-3} \citep[][]{springel_gadget2_2005}\\
     
     {\large Volume} & $110.7^3 \rm \, Mpc^3$ &\multicolumn{2}{c|}{$110.7^3 \rm \, Mpc^3$}& $100^3 \rm \, Mpc^3$ \\
     
     {\large Number of Particles/Cells} & $1820^3$ & \multicolumn{2}{c|}{$1820^3$} & $1504^3$ \\
     {\large Baryon mass resolution} & $\rm 1.26 \times 10^6 \,  M_\odot$& \multicolumn{2}{c|}{$\rm 1.4 \times 10^6 \,  M_\odot$} & $\rm 1.81 \times 10^6\,  M_\odot$\\
     
     {\large Gas physics} &&&&\\
     \emph{ \qquad ISM/dense gas model} & two-phase model$^\dagger$ & two-phase model$^\dagger$ & none & equation of state$^\star$\\
     \emph{ \qquad cooling} & primordial + metal & primordial + metal & none & primordial + metal\\
     \emph{ \qquad UVB} & \cite{faucher-giguere_new_2009} & \cite{faucher-giguere_new_2009} & none & \cite{haardt_modelling_2001}\\
    
     {\large Star formation model} &&&&\\
     \emph{ \qquad formation threshold} & $n_{\rm H} < 0.1 \rm  \, cm^{-3}$ & $n_{\rm H} < 0.1 \rm  \, cm^{-3}$& n/a & $n_{\rm H}(Z=0.002) < 0.1 \rm  \, cm^{-3}$$^\star$\\
     \emph{ \qquad IMF} & \cite{chabrier_galactic_2003} & \cite{chabrier_galactic_2003} & n/a & \cite{chabrier_galactic_2003}\\
     
     {\large Stellar feedback} & & & &\\
     
     \emph{ \qquad feedback channel} & kinetic wind & kinetic wind & none & thermal energy$^{\circ}$ \\
     
     {\large SMBHs \& feedback} &&&&\\
     \emph{ \qquad seed mass} & $\rm 10^5 \, M_\odot$ & $8 \times \rm 10^5 \, M_\odot$ & none & $\rm 10^5 \, M_\odot$\\
     \emph{ \qquad halo mass for seeding} & $\rm 5 \times 10^{10} \, M_\odot$ & $\rm 5 \times 10^{10} \, M_\odot$ & n/a & $\rm 10^{10} \, M_\odot$\\
     \emph{ \qquad accretion model} & Bondi-Hoyle & Bondi-Hoyle & n/a & Bondi-Hoyle\\
     \emph{ \qquad feedback channel(s)} & quasar, radio, radiative & quasar, kinetic, radiative & n/a & thermal energy\\
     \large{ Reference(s)} & \cite{vogelsberger_model_2013} & \cite{weinberger_simulating_2017} & --& \cite{crain_eagle_2015}\\
     &\cite{sijacki_illustris_2015}& \cite{pillepich_simulating_2018}&--&\cite{schaye_eagle_2015} \\
     \hline
\end{tabular}
\caption{Summary table of simulation characteristics, key differences in physical model assumptions and methods, and references for the important physical model aspects. $^\dagger$ : \protect\cite{springel_cosmological_2003}, $^\star$ : \protect\cite{crain_eagle_2015}, $^{\circ}$ : \protect\cite{dalla_simulating_2012}}
\end{table*}

\subsubsection{Illustris}

The original Illustris simulation \citep[][]{vogelsberger_introducing_2014,sijacki_illustris_2015,genel_introducing_2014}, encompasses a volume of \SI[exponent-base=110.7, input-exponent-markers=k]{k3}{\mega\parsec\cubed}, resolved by $1820^3$ particles and an equal number of gas cells at the initial conditions. Each dark matter particle (gas cell) has a mass of \SI{6.29e6}{\simsun} (\SI{1.26e6}{\simsun}). Dark matter and stars have $z=0$ gravitational softening lengths of \SI{\sim1.4}{\kilo\parsec}, while the gas softenings are adaptive down to \SI{\sim0.7}{\kilo\parsec}. Illustris was performed with the moving-mesh code \textsc{arepo} \citep{springel_arepo_2010}, which solves the coupled equations of gravity, galaxy astrophysics, and hydrodynamics in expanding universes by adopting, for the former, an N-body Tree-PM scheme and, for the latter, a Riemann solver on an unstructured Voronoi mesh.

The Illustris galaxy formation model \citep{vogelsberger_model_2013, torrey_model_2014} includes the key physics thought to be responsible for galaxy formation, and we here summarise the main aspects, focusing on the implementation of star formation and feedback. Gas cools radiatively, via hydrogen and helium collisional processes, bremsstrahlung, and inverse Compton cooling off the CMB \citep{katz_cosmological_1996}, in addition to metal-line cooling. It is heated by a spatially uniform, though time-variable, background radiation field \citep{faucher-giguere_new_2009}.

The unresolved dense star-forming ISM gas is modeled with a sub-resolution two-phase model, which partitions star-forming cells into cold and hot phases \citep[][]{springel_cosmological_2003}. Stars can form from the cold phase via a stochastic generation of discrete stellar particles, above a threshold density of $n_{H} = 0.1 \, \rm cm^{-3}$. These represent stellar populations that form simultaneously, following a \cite{chabrier_galactic_2003} initial mass function (IMF). Galactic-scale winds due to supernova feedback are modeled with a kinetic decoupled wind scheme \citep{springel_cosmological_2003}, which is effective at regulating the star formation of lower mass galaxies.

SMBHs are seeded in all sufficiently massive dark matter haloes ($ \gtrsim 7 \times 10^{10}\, \rm M_\odot$). They subsequently grow over time via smooth gas accretion and SMBH-SMBH mergers. Feedback from SMBHs is modeled via three channels: thermal (or quasar), mechanical (or radio), and radiative. Thermal or quasar-mode feedback heats the nearby gas, occurs for SMBHs with high accretion rates, and is time continuous. Mechanical or radio-mode feedback takes the form of energy injected into the halo by jet-inflated bubbles, is implemented for SMBHs with low activity states, and is time stochastic. Finally, the radiative feedback directly alters the local radiation field, reducing the effectiveness of cooling in nearby gas.

\subsubsection{IllustrisTNG}
\label{sec:tng}

The three IllustrisTNG simulations, TNG100 and TNG300 \citep[][]{nelson_first_2018,springel_first_2018,marinacci_first_2018,naiman_first_2018,pillepich_first_2018} and TNG50 \citep{pillepich_first_2019, nelson_first_2019}, are the successors to Illustris and are also based on the code \textsc{arepo}. Here we focus on the intermediate volume simulation, TNG100, which spans \SI[exponent-base=110.7, input-exponent-markers=k]{k3}{\mega\parsec\cubed} and which has the same initial conditions random seed as Illustris. TNG100 has $2\times1820^3$ dark matter particles and initial gas cells, with corresponding particle/cell masses of \SI{7.5e6}{\simsun}  and \SI{1.4e6}{\simsun} (slightly different than Illustris because of the different values of the cosmological parameters -- see Section~\ref{sec:cosmologies}). 

The TNG galaxy formation model has several significant changes with respect to Illustris, such as the inclusion of magnetic fields \citep[see][for an overview]{pillepich_simulating_2018}. Here we focus on the model differences that could affect the gas over the large scales pertinent to the study of the 21cm signal.

First, the low accretion state SMBH feedback model is replaced by a more local, kinetic feedback channel. The SMBH seed mass was also increased, compensating for the removal of the boost factor in the Bondi-Hoyle accretion rate  \citep[see][for a full review of the modified AGN feedback and SMBH physics]{weinberger_simulating_2017}. The updated SMBH feedback model produces a more realistic massive galaxy population \citep{donnari_star_2019,rodriguez-gomez_optical_2019,nelson_first_2019}. Second, the galactic winds generated by stellar feedback are updated in a number of aspects, including metallicity-dependent energetics, isotropy of the energy input at the injection scale, and the introduction of a redshift scaling of launch velocity. Overall TNG has stronger and faster winds, particularly at high redshift, producing important heating and enrichment around star-forming galaxies at early times \citep{pillepich_simulating_2018}.

We also analyse ``non-radiative'' runs of the TNG100 simulation in which we deactivate much of the baryonic physics: cooling, star formation and feedback. This TNG100-NR simulation has the same initial conditions and resolution as TNG100, enabling direct comparisons. Its behaviour is closer to the assumptions made in analytical (and some semi-analytical approaches) regarding baryonic matter. It also serves as a control for the physics that we wish to study in this paper. In this simulation, where there are no collisional or radiative cooling or heating processes, we avoid making any assumptions about the temperature of the gas. When required, we use the TNG100 stellar particles as sources in TNG100-NR.

\subsubsection{Eagle}

The primary Eagle simulation \citep[][]{schaye_eagle_2015, crain_eagle_2015} has a volume of \SI[exponent-base=100.0, input-exponent-markers=k]{k3}{\mega\parsec\cubed}, resolved by $ 1504^3$ dark matter and $1504^3$ gas particles. This results in a slightly lower resolution, with dark matter (gas) particles masses of \SI{9.7e6}{\simsun} (\SI{1.81e6}{\simsun}). Eagle uses a modified version of the smoothed particle hydrodynamics code \textsc{gadget-3} \citep{springel_gadget2_2005}.

Though many of the physical ingredients for galaxy formation are largely similar as in Illustris and TNG, their implementations are different. First, there is no sub-grid two-phase ISM model. Instead, star-forming gas follows an effective equation of state. The \cite{chabrier_galactic_2003} IMF is also adopted, while the stochastic star-formation scheme relies on a metallicity dependent density threshold. Third, stellar feedback from Type II SNe is modelled by a stochastic thermal injection scheme as described in \cite{dalla_simulating_2012}, as opposed to the kinetic schemes of the other two simulations. This heats nearby gas to very high temperatures to produce effective outflows. Third, SMBHs are seeded in lower mass ($> 10^{10} \rm M_\odot$) haloes. Fourth, the feedback from SMBHs is also modelled by a single thermal channel (as opposed to three channels of Illustris and TNG), and is time integrated rather than continuous \citep{booth_cosmological_2009}.

\subsubsection{Cosmologies and data structure}
\label{sec:cosmologies}

Illustris assumes cosmological parameters consistent with WMAP 9 \citep[][]{hinshaw_nine-year_2013}: $ \Omega_m=0.2727, \, \Omega_\Lambda=0.7274, \, \Omega_b=0.0456, \, \sigma_8=0.809, \, n_s=0.963, \, H_0=70.4 \, \rm km s^{-1} Mpc^{-1}$. The TNG simulations assume updated cosmological parameters from \citetalias{planck_collaboration_planck_2016}: $ \Omega_m=0.3089, \, \Omega_\Lambda=0.6911, \, \Omega_b=0.0486, \, \sigma_8=0.8159, \, n_s=0.9667, \, H_0=67.74 \, \rm km s^{-1} Mpc^{-1}$.
\citetalias{planck_cosmo_2013} cosmology was assumed for the Eagle simulations, with  $ \Omega_m=0.307, \, \Omega_\Lambda=0.693, \, \Omega_b=0.04825, \, \sigma_8=0.8288, \, n_s=0.9611, \, H_0=67.77 \, \rm km s^{-1} Mpc^{-1}$. In this work, distances are always given in comoving units.

The data of these three simulations, both catalogs and particle-based data, are publicly available and described by \cite{nelson_release_2015}, \cite{eagle_release_2017}, \cite{nelson_release_2019} for Illustris, EAGLE and TNG, respectively. In all three simulations, dark matter haloes and galaxies were identified using, consecutively, a Friends-of-Friends and the \code{SUBFIND} algorithms \citep[][]{springel_gadget_2001}.\footnote{In this work we do not use the original EAGLE particle data, but a version that has been rewritten and re-analysed with the TNG halo finder, enabling an apples-to-apples comparison \citep{nelson_release_2019}.}

\subsection{UV/X-ray background and temperature of the IGM}
\label{sec:TIGM}

Illustris, TNG and EAGLE are {\it not} fully coupled radiation and hydrodynamics (RHD) simulations and therefore they do not explicitly model the heating from ionizing photons (UV and X-ray) that originate from galaxies and SMBHs. Hydrogen reionization is implemented by activating a time-dependent, spatially-uniform ionizing background (UVB). In the case of Eagle, this is adopted from \cite{haardt_modelling_2001} and turned on at $z=11.5$ \citep{schaye_eagle_2015}. Illustris and TNG also include a UVB, from \cite{faucher-giguere_new_2009}, which is only turned on at $z=6$. Therefore, prior to $z=6$, in Illustris and TNG100 the IGM is cold ($ T \ll 10^4 \, \rm K$), whereas any non-negligible UVB would act to heat this gas to $\sim  10^4 \, \rm K$. 

To model more realistic gas temperatures at $z>6$ on the large spatial scales of the IGM, we recalculate gas cell temperatures in post-processing. To do so we re-run the TNG model cooling network for each gas cell, including the self-shielding from \cite{rahmati_impact_2013}, stopping when equilibrium is reached, i.e. when the relative change in temperature between cooling steps is smaller than \SI{e-4}{}. For low density IGM gas, this happens relatively quickly (in $ \lesssim 10 \, \rm Myr$) compared to the time separation of snapshots ($ > 40 \, \rm Myr$), avoiding transitions between snapshots. As the heating is done in post-processing, we cannot consistently account for heating due to gravitational effects, and stellar and AGN feedback. Therefore, in cells that are already hotter than typical temperatures from photo-ionization ($ > 2 \times 10^{4} \, K$), we retain the original simulation temperatures.

We verify this approach with a variation of the TNG100-3 simulation -- a run of the same TNG100 volume with the same initial conditions, unchanged galaxy-formation model but with 64 (4) times worse mass (spatial) resolution -- where the UVB was included for all $z<11$. This resolution is sufficient to capture effects in the low-density IGM, where we modify temperatures. Adopting a power law for the temperature-density relation in low density gas ($ n_{\rm H} <10^{-3} \rm cm^{-3}$), we find $<8\%$ relative difference in temperatures between the full UVB run and our equilibrium approach. Furthermore, a comparison of the average IGM temperature of TNG, Illustris, and Eagle to observational constraints (which we do not show) returns a broad agreement \citep[with e.g.][]{boera_revealing_2019, gaikwad_probing_2020}, which is satisfactory for this comparative study (as it does not focus on modeling the most realistic Reionization).

This post-processing allows us to model the temperature in the IGM when heating has already finished \citep[$z \lesssim 8-10$, see][]{ciardi_probing_2003}. At higher redshifts, our simulations and post-processing cannot account for some of the physical effects that are important in the early heating of the IGM.\footnote{This could be the case, for instance, due to the formation of the first stars in $ \rm H_2$-cooled mini-haloes \citep[see][for an overview of early star formation]{klessen_first_23}. Similarly, the early heating of the IGM by X-ray sources, which is important for modeling the brightness temperature of the 21cm line at higher redshifts \citep[e.g.][]{pritchard_21-cm_2007,fialkov_signature_2019}, can also play a role.} Consequently, throughout this paper, we refrain from making predictions for $z>8$.

\subsection{Brightness temperature of the 21cm line}
\label{sec:21cm}
    
In this work, we derive the differential brightness temperature of the 21cm line with respect to the CMB. The relevant expression is given by \citep[][]{furlanetto_cosmology_2006}:
\begin{equation} \label{eq:dTb}
 \delta T_{\rm b}(\nu) =  \frac{T_{\rm s} - T_{\rm CMB}}{1 + z}\big(1 - e^{-\tau_{\nu_0}}\big)
\end{equation}
\noindent where $ \delta T_{\rm b}$ is the differential brightness temperature (hereafter, simply referred to as the brightness temperature, for brevity), $ T_{\rm s}$ is the spin temperature of the gas, $ T_{\rm CMB}$ is the temperature of the cosmic microwave background, $z$ is the cosmological redshift, and $ \tau_{\nu_0}$ is the optical depth at 21cm line center. $\nu$ is the wavelength of observation, which is cosmologically redshifted (i.e. $\nu = \nu_0 / (1 + z)$).
    
Making several simplifying assumptions \citep[see][for details]{furlanetto_cosmology_2006}, this can be rewritten as \citep[][]{mesinger_21cmfast:_2011}:
    
\begin{equation}
\label{eq:dTb2}
\begin{aligned}
  \delta T_{\rm b} \approx 27 \left( \frac{\Omega_{\rm b} h^2}{0.023} \right) \left(\frac{0.15}{\Omega_{\rm m} h^2} \frac{1+z}{10}\right)^{0.5} x_{\rm HI} \frac{\rho_{\rm H}}{\mean{\rho_{\rm H}}} \\ \times \left(1 - \frac{T_{\rm CMB}}{T_{\rm s}} \right) \left(\frac{H}{\delta_v + H} \right) \: {\rm mK},
\end{aligned}
\end{equation}

\noindent where $ \Omega_b$ is the cosmological baryon density, $ \Omega_m$ is the cosmological density parameter for matter, z is cosmological redshift, \xhi \, is the mass fraction of neutral hydrogen gas, $ \rho_{\rm H}$ is the gas density of hydrogen nuclei (and $ \mean{\rho_{\rm H}}$ its mean), $H$ is the Hubble parameter, and $ \delta_v$ is the gas velocity gradient along the line of sight (LoS).    

We use the cosmological parameters of each simulation. $ \rho_{\rm H}$ is taken from the simulation data. $ T_{\rm CMB}$ is taken to be $ T_{\rm CMB}=2.725 (1 + z)$ \citep[][]{furlanetto_cosmology_2006,kannan_introducing_2021}, and $ \delta_v$ is computed using simulation gas velocities along the x direction. For $ x_{\rm HI}$, we adopt a post-processing strategy described below (Section~\ref{sec:xHI}).

The velocity term of Eq.~\ref{eq:dTb2} can diverge if the LoS gas peculiar velocity gradient is negative and close to the Hubble expansion rate, leading to power spectra of the brightness temperature that can be artificially dominated by velocity effects. A possible remedy is to limit the values of $ \delta_v/H$ \citep[e.g.][]{mesinger_21cmfast:_2011}, though this poses the question of which threshold to pick, and renders the resulting \dtb{} \, more complicated to interpret. Here, we circumvent this issue and adopt an approach similar to \cite{mao_redshift-space_2012} as implemented in the \code{tools21cm} software package \citep[][]{giri_tools21cm_2020}. For every LoS ($x$ direction\footnote{We find a small $\leq 3\%$ mean difference in the final 21cm power spectrum at $z=6$ when switching LoS directions.}),\, \dtb \, in each cell of the uniform grid is sampled by 20 pseudo-particles. These particles are then moved along the LoS direction according to the local gas peculiar velocities (as directly measured from the hydrodynamical simulations) and Hubble expansion. The particles are then binned back onto the uniform grid, smoothing the initial \dtb \, values and mimicking the effects of redshift space distortion (RSD) on \dtb. We discuss these choices further in the section \ref{sec:limits}.
    
A common assumption is that, during reionization, $ T_{\rm s} \gg T_{\rm CMB}$ \citep[e.g.][]{mesinger_21cmfast:_2011,kannan_arepo-rt_2019}, with the effect that the spin temperature term in Eq.~\ref{eq:dTb2} vanishes. However, in this paper, we wish to examine the effect of galaxy formation model choices on the power spectra of the brightness temperature. These can alter the temperature of gas around galaxies, e.g.\ due to stellar or SMBH outflows. This encourages us to use the gas temperature in our model of $ T_{\rm s}$, in order to capture the differences from the different galaxy formation models. $ T_{\rm s}$ can be expressed as follows (Eq.~\ref{eq:Ts}):
    
\begin{equation} \label{eq:Ts}
   T_{\rm s} = \frac{1+x_c+x_\alpha}{T_{\rm CMB}^{-1} + x_c T_{\rm gas}^{-1} + x_\alpha T_{\rm c}^{-1}} \, ,
\end{equation}

\noindent where $ x_c$ is the coupling coefficient for collisions, $ x_\alpha$ is the coupling coefficient for the Wouthuysen-Field (WF) effect, $ T_{\rm gas}$ is the gas temperature, and $ T_{\rm c}$ is the colour temperature of the Lyman-$\alpha$ line. As we cannot compute $ x_{\alpha}$ from first principles, we limit our study to $z \lesssim 8$, when the heating of the IGM is finished, $ x_\alpha \gg x_c \approx 1$, and $ T_{\rm c} \approx T_{\rm gas}$, so that $T_{\rm s} \approx T_{\rm gas}$ \citep[e.g.][]{ciardi_probing_2003, furlanetto_cosmology_2006}. We checked these assumptions by computing $x_c$, $x_\alpha$, and $T_{\rm s}$ as in \cite{gillet_21cm_2021}, and found negligible differences in our results. In TNG100-NR, we always assume $ T_{\rm s} \gg T_{\rm CMB}$.

\subsection{Deriving \xhi~ in post-processing}
\label{sec:xHI}

As mentioned in Section~\ref{sec:intro} and described in Section~\ref{sec:21cm}, the 21 cm brightness temperature directly depends on the fraction of neutral hydrogen gas and thus on its density. In TNG, Illustris, and Eagle, the fraction of neutral hydrogen gas in each resolution element is determined by a balance between collisional processes computed from the gas state, and photo-heating and photo-ionisation processes computed using a uniform ultra-violet background (UVB). This UVB represents the combined emission of all ionising sources.

In practice, the UVB is applied to all gas in a spatially uniform manner. Thus, in these simulations, the IGM is progressively ionized in a rather homogeneous way, while in reality the IGM reionizes due to nearby sources. Therefore, the first areas to reionize are situated close to the overdensities where galaxies form. To accurately model the 21cm signal during the EoR, we must take this into account, and post-process the simulations so that the ionized fraction of hydrogen gas reflects the source density. We carry out this post-processing with the \code{CIFOG} code \citep[][]{hutter_accuracy_2018}.

\subsubsection{CIFOG}
\label{sec:cifog}    

\code{CIFOG} \citep[][]{hutter_accuracy_2018} is a tool for post-processing simulations in order to determine the ionization fractions of hydrogen and helium as they evolve throughout reionization. It is based on the excursion set formalism widely used in 21cm predictions \citep[see e.g.][]{furlanetto_cosmology_2006, mesinger_21cmfast:_2011}. However, \code{CIFOG} allows the input of grids of gas density and ionizing photons rates as directly predicted or derived from hydrodynamical galaxy simulations, whilst retaining a low computational expense when compared to full radiative transfer. For our use case \code{CIFOG} allows post-processing of the simulations to capture varying star formation and galaxy formation models in order to assess their imprint on the 21cm hydrogen emission line. For precise details about \code{CIFOG}'s design and examples of its use, we refer the reader to \cite{hutter_accuracy_2018} and \citet{hutter_astraeus_2020}.
        
With \code{CIFOG} we generate outputs at the same redshifts for all three simulations. Since the 21cm signal is proportional to the neutral fraction of gas, and because the average neutral gas fraction evolves by several orders of magnitude during the final $\approx 100$ Myr of reionization, we take \code{CIFOG} outputs separated by $ \approx 20 \rm \, Myr$ to sample rapid changes in the 21cm signal.

\subsubsection{Eulerian grid} \label{sec:grid}

\code{CIFOG} only supports Eulerian input grids of size $2^n$ (with $n$ an integer). We also require an evenly sampled grid to compute the power spectrum of the simulation fields. We therefore construct a set of Eulerian grids for the density, temperature, and line of sight velocity fields of each snapshot, for each simulation. For the bulk of the analysis, we choose a grid of 1024$^3$ cells, giving a spatial resolution of roughly $ \Delta_x \approx 100 \, \rm kpc$ (depending on the simulation). This spatial resolution is adequate for resolving the large spatial scales that can be constrained by current or near-term future observations of the 21cm emission line of hydrogen \citep[see e.g.][]{koopmans_cosmic_2015, kolopanis_simplified_2019, kolopanis_new_2022, hera_2022}.

We use the standard cubic-spline kernel with an adaptive size and deposit gas mass and mass-weighted gas properties accordingly into the Eulerian grid \citep[following][]{nelson_halo_2016}. For TNG and Illustris the kernel size is the spherical volume-equivalent radius of each Voronoi cell. For Eagle, we use the SPH adaptive smoothing length. Rarely, low density void regions are not sampled, and in these cases we infill the grid values from nearest neighbor values.\footnote{In the Eagle SPH simulation, this occurs in regions not within the hydrodynamical softening length of any particle. In the AREPO Voronoi mesh simulations, this happens in cells between and around Voronoi cells who's geometries are not well described by a sphere, as is assumed when using the cubic-spline kernel. In these cases we fill empty grid cells with the values from the nearest gas cell. Note that this procedure introduces a small increase in mass ($<0.3\%$ in Eagle, where the problem is the most prevalent).}

With \code{CIFOG} we save regularly spaced outputs of the \xhi{} field, obtaining more \code{CIFOG} outputs than simulation snapshots. This is particularly true for $z>6$ and in Eagle, for which we only have 3 snapshots covering the $ 6 \leq z \leq 8$ interval. When necessary we associate each \xhi{} field to the closest in redshift snapshot. Where possible, we verified that our conclusions persist if we limit ourselves to the study of redshifts where we have a snapshot. This is possible because the evolution of \xhi{} is the main driver of large scale evolution of the brightness temperature and power spectrum of the 21cm line during the epoch we study \citep[e.g. ][]{furlanetto_cosmology_2006}.
            
\subsubsection{Sources of ionizing radiation and escape fraction}
\label{ssec:src}

We model the sources of ionizing radiation from star forming regions as follows, neglecting throughout radiation from SMBHs.\footnote{The ionizing radiation from AGN is only thought to become important at lower redshifts, based on their low density during reionization \citep[e.g.][]{kulkarni_evolution_2019}, and results from simulations \citep[][]{trebitsch_obelisk_2021}.}

To determine the hydrogen ionizing luminosities of stellar particles we use the BPASS v2.2.1 \citep[][]{eldridge_population_2020} stellar evolution model to tabulate ionizing luminosity per stellar mass as a function of stellar particle age and birth metallicity, including the effects of binaries. We include all ionizing wavelengths from the BPASS v2.2.1 stellar spectra, spanning $1.2397 \times 10^4 \, {\rm eV} \geq h \nu \geq 13.6 \rm \, eV$.

Due to the coarse resolution of our Eulerian grid, and the approximate treatment of radiative transfer performed by the \code{CIFOG} code, we must account for the absorption of ionizing photons by neutral HI gas as they travel from stellar sources to the IGM heuristically. We take the simplest choice and assume a constant fraction of all produced photons are absorbed by dense neutral galactic gas. The corresponding escape fraction is therefore an effective value, that does not evolve in time, or account for any dependencies on galactic properties \citep[as increasingly suggested by simulations, for instance:][]{rosdahl_lyc_2022,kostyuk_ionizing_2022}. Nevertheless, similar models are widely used, particularly in analytical and semi-analytical approaches \citep[e.g.][]{dayal_reionization_2020}. We calibrate the escape fraction (\fesc) using the timing of reionization in TNG100, finding a reasonable result for \fesc$=0.2$ (\fescglob).

\subsubsection{Fiducial post-processing model}
\label{sec:fiducial}

Putting together the methodologies described above, throughout this paper we adopt the following fiducial choices to obtain a prediction of the 21cm brightness temperature from the Illustris, TNG100 and Eagle simulations. We assume that the spin temperature ($T_{\rm s}$) is coupled to the gas temperature ($T_{\rm gas}$). We adopt ionizing luminosities of stellar populations from the binary BPASS V2.2.1 stellar evolution model \citep[][]{eldridge_population_2020}. We assume a fixed escape fraction of 20\%. We then derive the mass fraction of ionized gas in each cell, using the source and hydrogen density distributions from each simulation, with \code{CIFOG}. Our fiducial model for the brightness temperature of the 21cm line accounts for redshift space distortion (RSD) along the line of sight.

In Appendix \ref{app:model_var} we examine the sensitivity of our 21cm brightness temperature calculation to all different model choices and assumptions discussed above.

\subsection{Computing power spectra}    
\label{sec:pk}

Beyond its spatial average, the most informative and accessible summary statistic of the 21cm brightness temperature is its 2-point correlation function, or power spectrum.    

To compute the power spectrum of the brightness temperature (\dtbPS{}, or 21cm power spectrum hereafter), we first determine the mean of the perturbation subtracted \dtb. We then use the {\textsc tools21cm} package \citep{giri_tools21cm_2020} to compute \dtbPS{} in Fourier space.

The smallest scales we probe are limited by the grid resolution, while the largest are limited by the simulation box sizes. In our case this corresponds to the interval $ k \in [\approx 0.06~\rm cMpc^{-1}, \approx 58.49~ \rm cMpc^{-1}]$. However, the power spectra at the largest scales (smallest k) suffer from under-sampling, and those at the smallest scales (largest k) are susceptible to aliasing. Throughout the paper we plot a conservative range, where we have more than eight resolution elements contributing to the radially averaged power spectrum. This corresponds to $ k \in [\approx 0.11~ \rm cMpc^{-1}, \approx 29.25~ \rm cMpc^{-1}]$.

In order to investigate the differences between \dtbPS{} across simulations, we also decompose the dimensionless power spectra according to the individual terms of Eq.~\ref{eq:dTb2}:
\begin{equation}
\begin{array}{l l}
     \Delta^2_{21} = &  \Delta^2_{\rho_{\rm H},\rho_{\rm H}} + \Delta^2_{x_{\rm HI},x_{\rm HI}} \\
    & + \Delta^2_{T_{\rm s},T_{\rm s}} + 2\Delta^2_{\rho_{\rm H},x_{\rm HI}} + 2\Delta^2_{\rho_{\rm H},T_{\rm s}} + 2\Delta^2_{T_{\rm s},x_{\rm HI}} \\
    & + \Delta^2_{\rm higher \, order}
\end{array}
\end{equation}
where $ \Delta^2_{X,X}$ is the power spectrum of the field X, and $ \Delta^2_{X,Y}$ is the cross power spectrum between fields X and Y.
This allows us to compare the contribution of different simulation fields to the total \dtbPS, with two caveats. First, we do not compute 3rd order and higher terms, which can be important at small scales \citep[e.g.][]{furlanetto_cosmology_2006, georgiev_large-scale_2022}. Second, since the velocity effects are not directly included in the computation of \dtb \, but accounted for by smearing \dtbPS, the contribution from the velocity term (i.e. redshift space distortion) is derived by a simple subtraction between the final \dtbPS \, and the \dtbPS \, before the correction for velocity effects is made. Thus, we do not separate between different velocity and cross-correlation velocity terms.

In Appendices \ref{app:model_var} and \ref{app:fesc} we examine the sensitivity of our power spectra calculations to different model choices and assumptions. Further, in Appendix \ref{app:conv} we study the convergence of our fiducial \dtbPS{} predictions with respect to the resolution of our Eulerian grid.
 

\section{Results}

\label{sec:results}

\subsection{Maps of the 21cm brightness temperature}

\begin{figure*}
    \includegraphics[width=\textwidth]{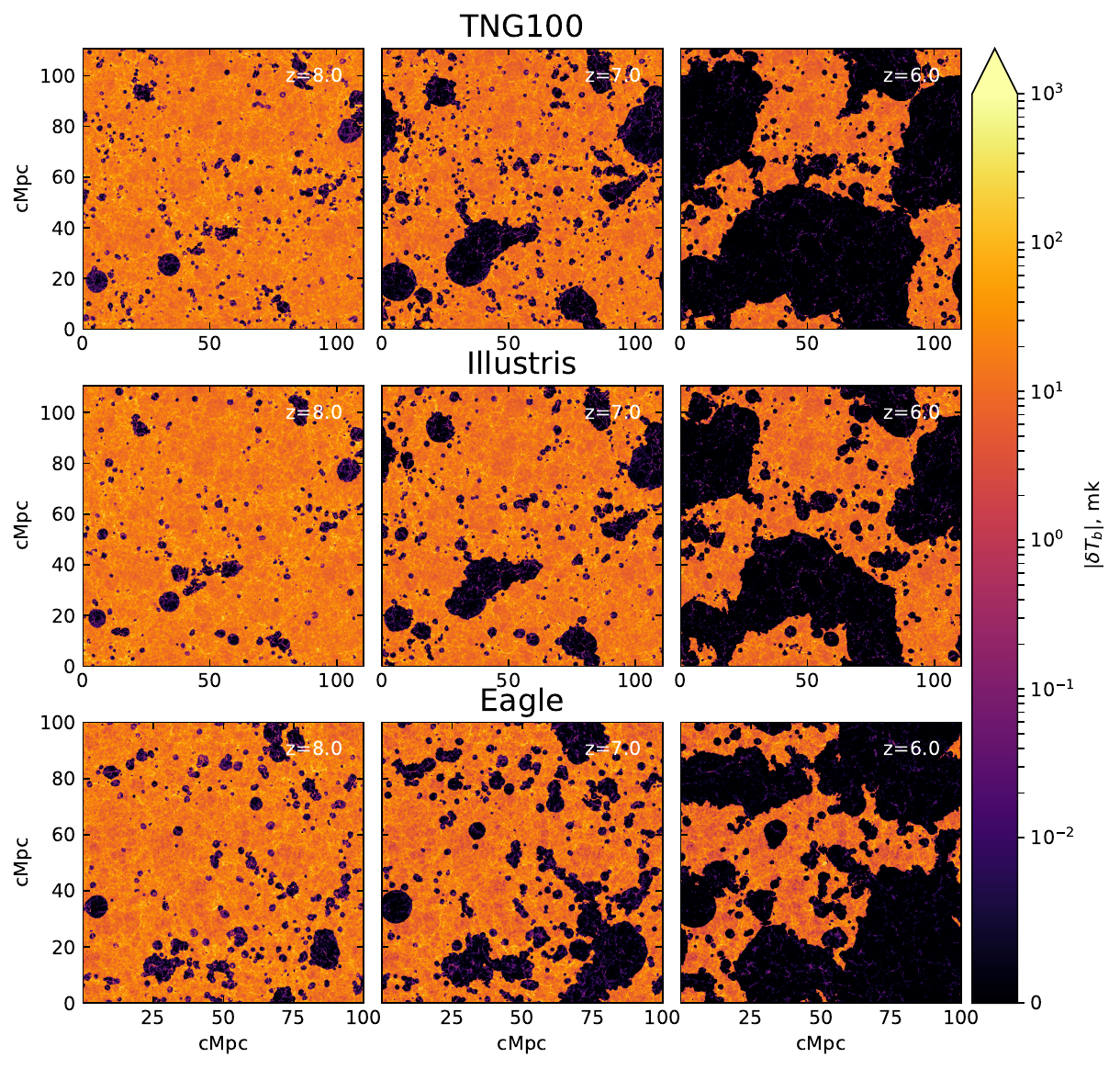}
    \caption{Evolution of the absolute differential brightness temperature of the 21cm line, for the TNG100, Illustris and Eagle simulations (top to bottom), from $z=8$ (left) to $z=6$ (right). All are computed with our fiducial post-processing model. Hydrogen becomes progressively more ionized in a spatially localised and heterogeneous manner, reflecting the distribution of ionizing photon sources. This causes the brightness temperature values (\dtb) to decrease in correspondingly overdense regions. Within ionized regions, the remaining neutral gas in self-shielded areas emits a weak signal. Despite Illustris and TNG100 sharing the same initial conditions, their maps differ, showcasing the impact of the galaxy formation physics and feedback models in determining \dtb{}.}
    \label{fig:maps}
\end{figure*}

Figure \ref{fig:maps} shows slices of differential brightness temperature $|$\dtb{}$|$, where each panel is \SI{\approx 100}{\mega\parsec} a side, at $z=8,7,6$ in TNG100, Illustris, and Eagle. Emission peaks occur at the overdensities in filaments and galaxies. As time progresses, more and larger regions of low \dtb{} appear. These are mainly driven by the formation and growth of ionized bubbles around collapsed structures.

There are many similarities among the simulations, particularly between TNG100 and Illustris as they share the same initial conditions. However, differences due to star formation and gas distributions are also apparent. The TNG100 and Eagle maps have more, smaller bubbles of low \dtb{} (ionized HI bubbles), when compared to Illustris, but also more bubbles overall. This suggests that the distribution of star formation amongst haloes differs strongly at these high redshifts. Within these bubbles, one can distinguish faint filament-like emission features that correspond to dense underlying gas structures.

\subsection{Evolution of the average 21cm brightness temperature}

\begin{figure*}
  \includegraphics[width=0.3322\textwidth]{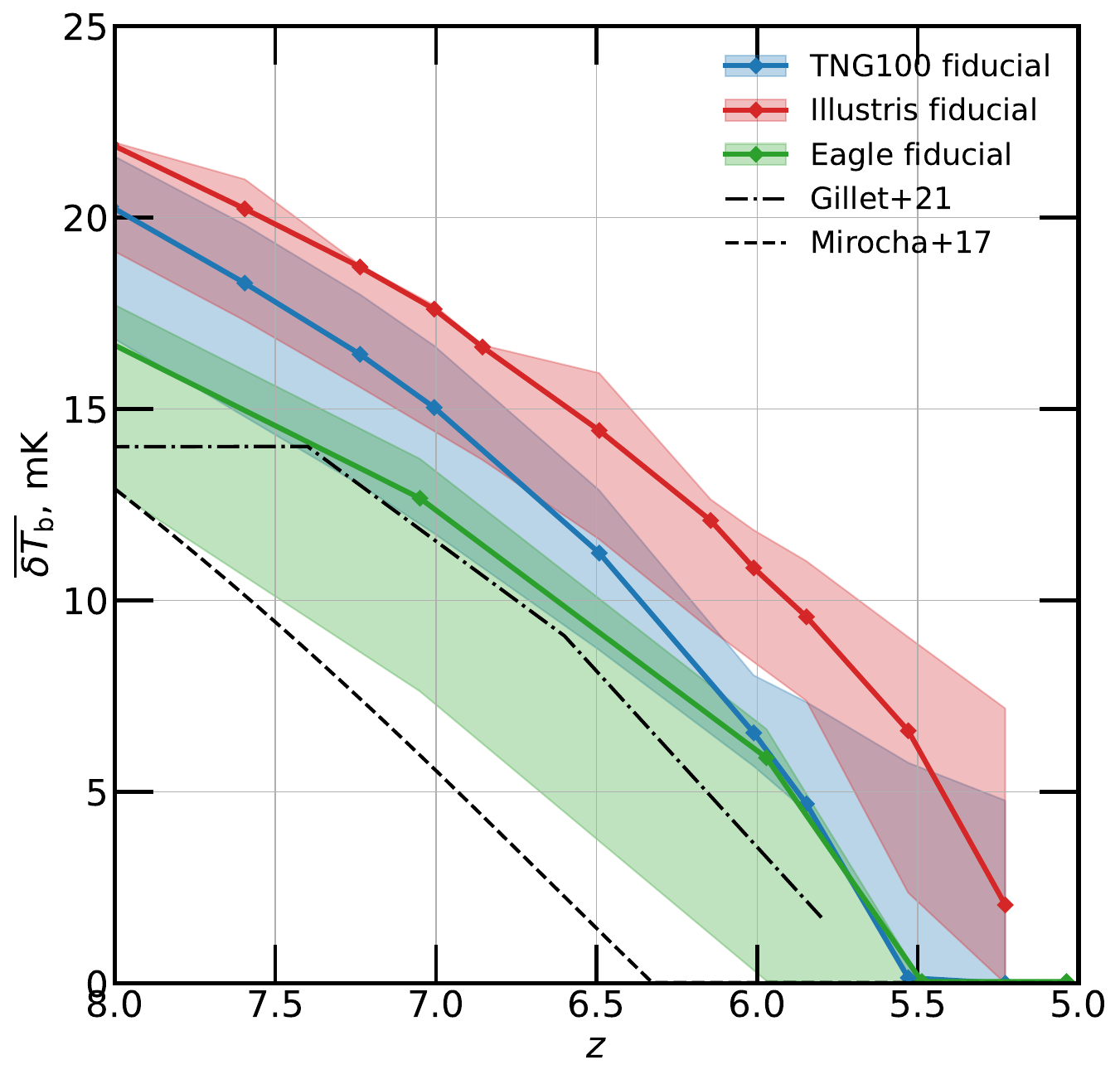} 
  \includegraphics[width=0.3322\textwidth]{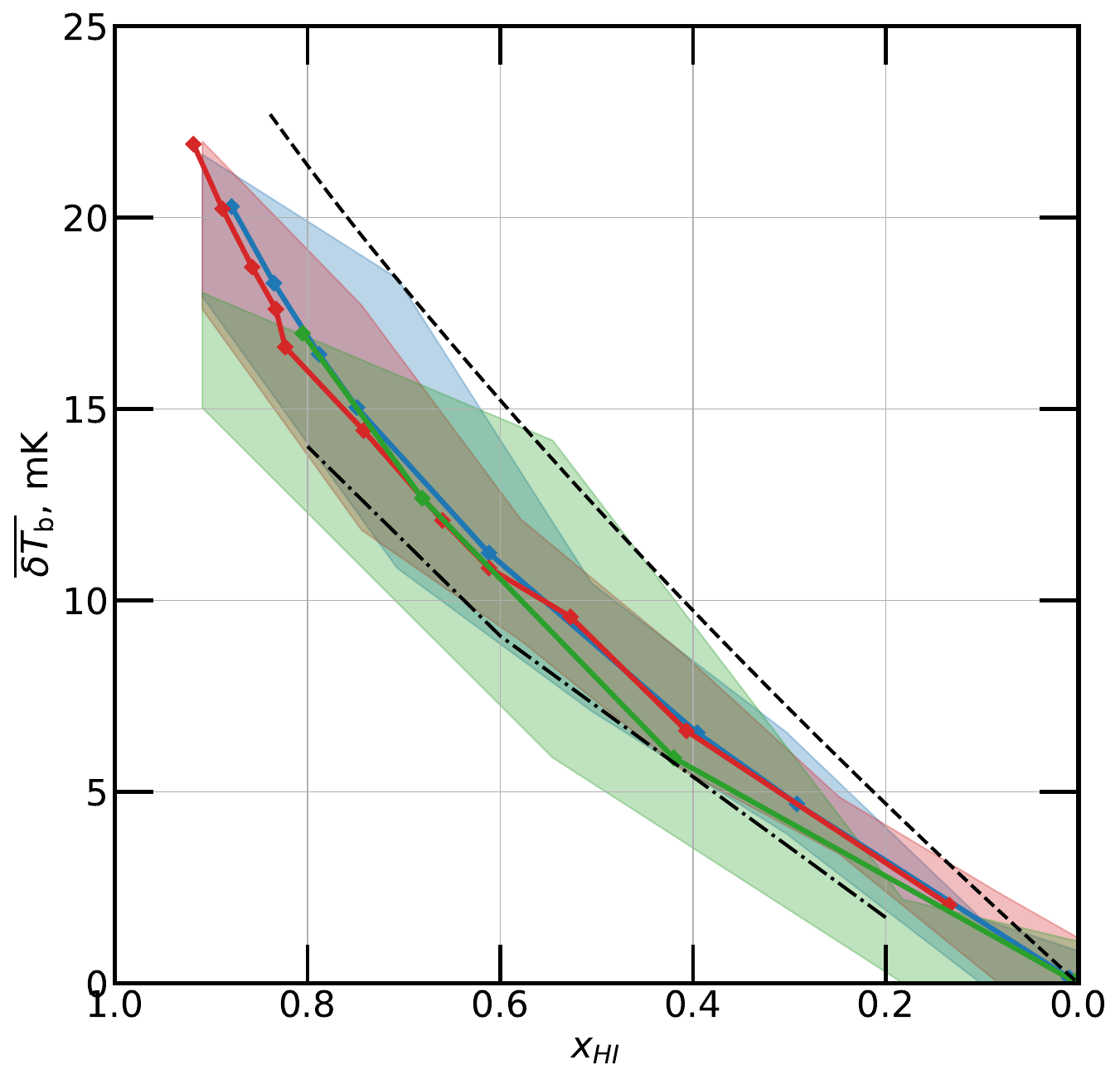} 
  \includegraphics[width=0.325\textwidth]{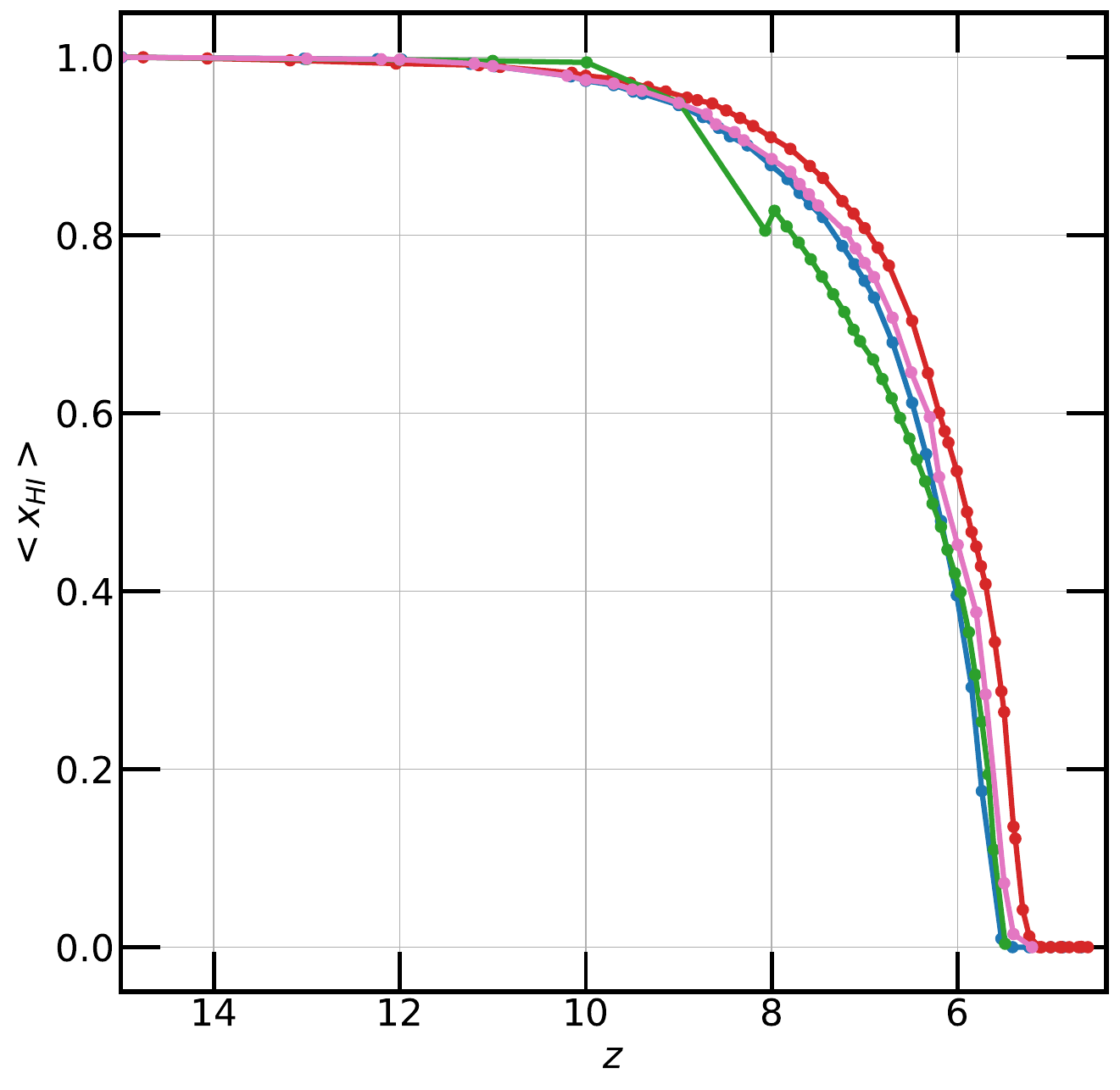}
  \caption{\emph{Left: }Time evolution of the mean differential brightness temperature of the 21cm line for TNG100 (blue), Illustris (red), and Eagle (green). Diamonds show our fiducial post-processing model, whereas the shaded areas encompass all our explored models. Also shown are the analytical prediction of \protect \cite{mirocha_global_2017} in the saturated case, and the predicted mean signal from the full radiation and hydrodynamics simulation of \protect \cite{gillet_21cm_2021}. \emph{Middle: } Average brightness temperature as a function of average \xhi{} (used as a proxy for the progress of reionization; right). \emph{Right: } Average \xhi{} as a function of redshift in the fiducial model. The three simulations predict very similar \avgdtb{} at fixed \xhi{}, whose evolution is largely set by the decrease of average \xhi{} during reionization. In this figure, the results from other works appear truncated so as to represent a similar redshift range as in the left hand panel.}
  \label{fig:mean_dtb}
\end{figure*}

The left panel of \figref{fig:mean_dtb} shows the redshift evolution of the average brightness temperature in TNG100, Illustris, and Eagle. The markers and thin lines show the evolution of \dtb{} according to our fiducial post-processing model (Section~\ref{sec:fiducial}). For all three simulations, \dtb{} decreases over time as the volume filling factor of ionized hydrogen increases, and the total volume where \dtb{}$ \approx 0$ increases. 

We predict a lower signal in Eagle across all redshifts. Illustris and TNG100 are similar at $z=8$, while the signal decreases faster in TNG100 and is closer to the Eagle result for $z \lesssim 6$. Differences in \avgdtb{} between TNG100 and Illustris arise from their unique feedback prescriptions. This is likely also the case for differences in the \avgdtb{} of Eagle, as they cannot be explained by differences in large scale density (closer than $2\%$). The shaded regions show a level of modeling uncertainty representing the breadth of all explored post-processing models (described in Appendix~\ref{app:models}). Broadly, the Illustris and TNG100 models are the closest, whereas Eagle-based models of \dtb{} show the largest scatter in values, lying below the other two simulations for most model combinations. 

The middle panel of \figref{fig:mean_dtb}, shows the evolution of the average \dtb{} as a function of the spatially-averaged neutral fraction (average \xhi{}). This accounts for differences in reionization histories across the simulations and source models. As expected, \avgdtb{} decreases rapidly with decreasing average \xhi{}, and the three predictions of \avgdtb{} are strikingly close at fixed \xhi{}, irrespective of the underlying hydrodynamical galaxy simulation. The Eagle model variants are again the most diverse from the fiducial model.

\cite{gillet_21cm_2021} perform full RHD simulations, with sub-grid models for star formation and luminosities based on smaller, more resolved RHD simulations. They report a similar evolution of the brightness temperature as ours from Illustris, TNG100 and Eagle, particularly when compared to Eagle. On the other hand, we find significantly higher temperatures  than the analytical result (in the saturated case) from \cite{mirocha_global_2017}. However, when accounting for our different reionization histories, our results are close to their predictions near $z=6$, and they are even closer to the RHD simulation results of \cite{gillet_21cm_2021}.

\subsection{Power spectra of the 21cm brightness temperature}

\begin{figure*}
    \centering
    \includegraphics[width=0.49\textwidth]{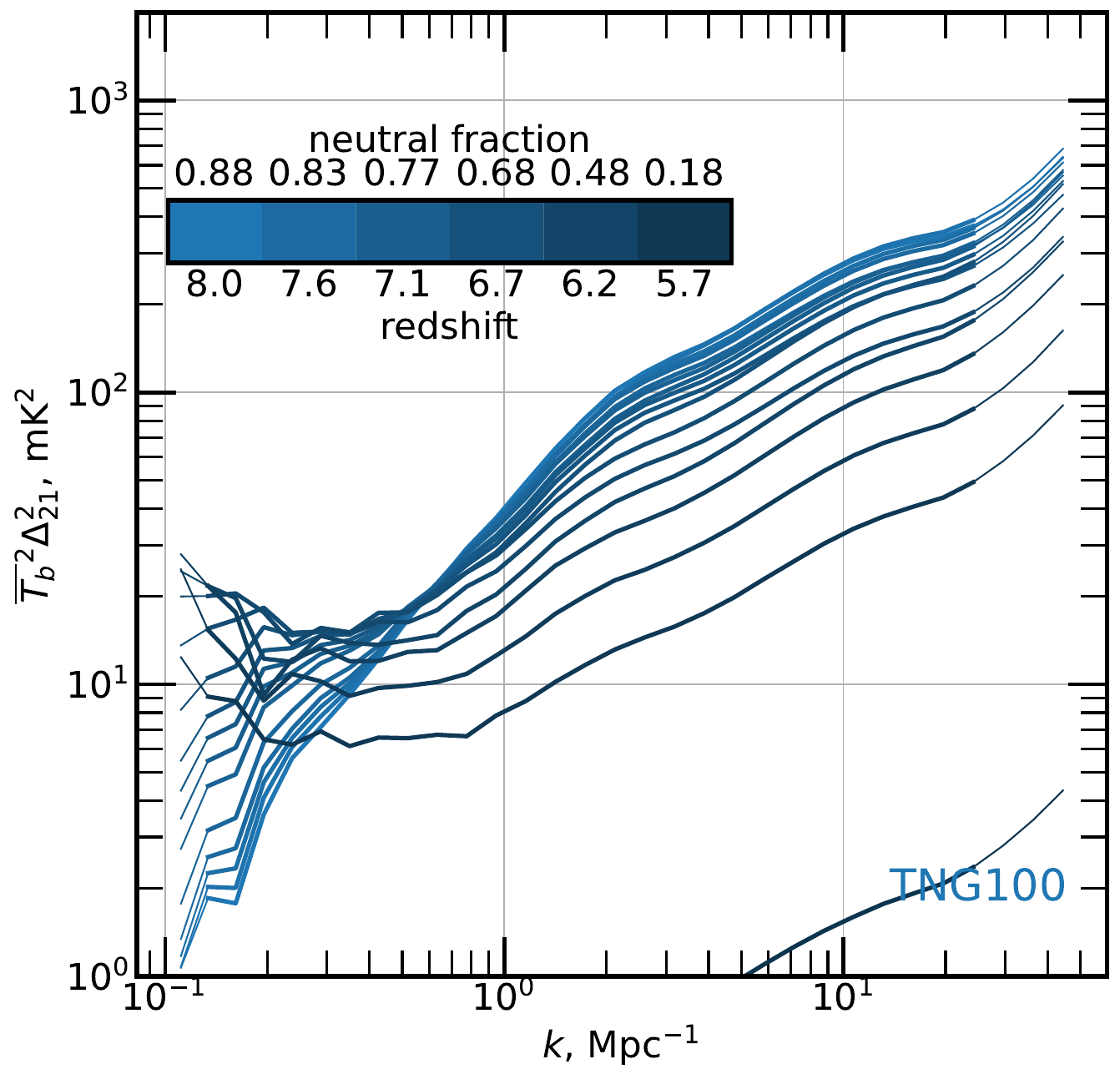}
    \includegraphics[width=0.49\textwidth]{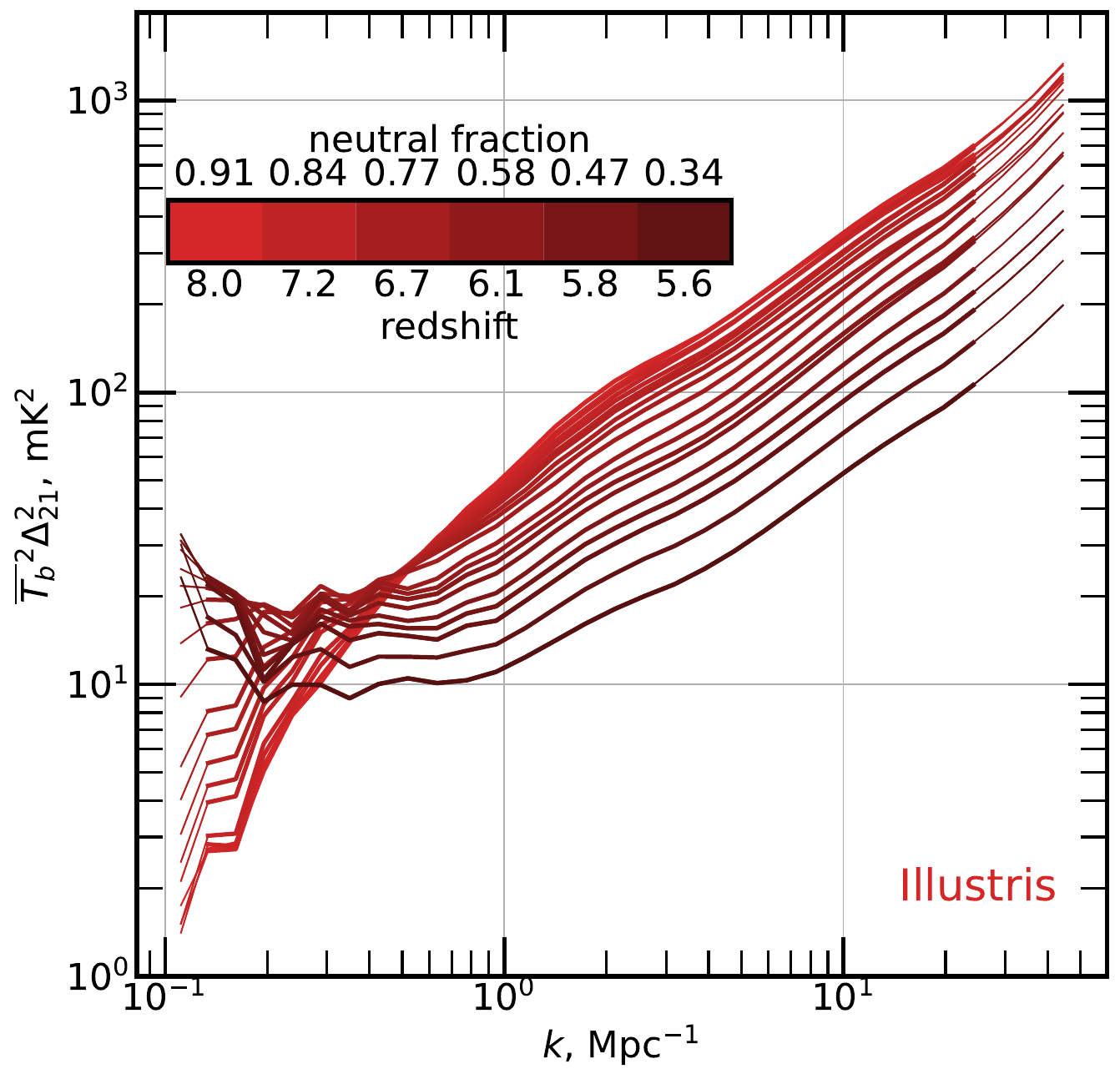}\\
    \includegraphics[width=0.49\textwidth]{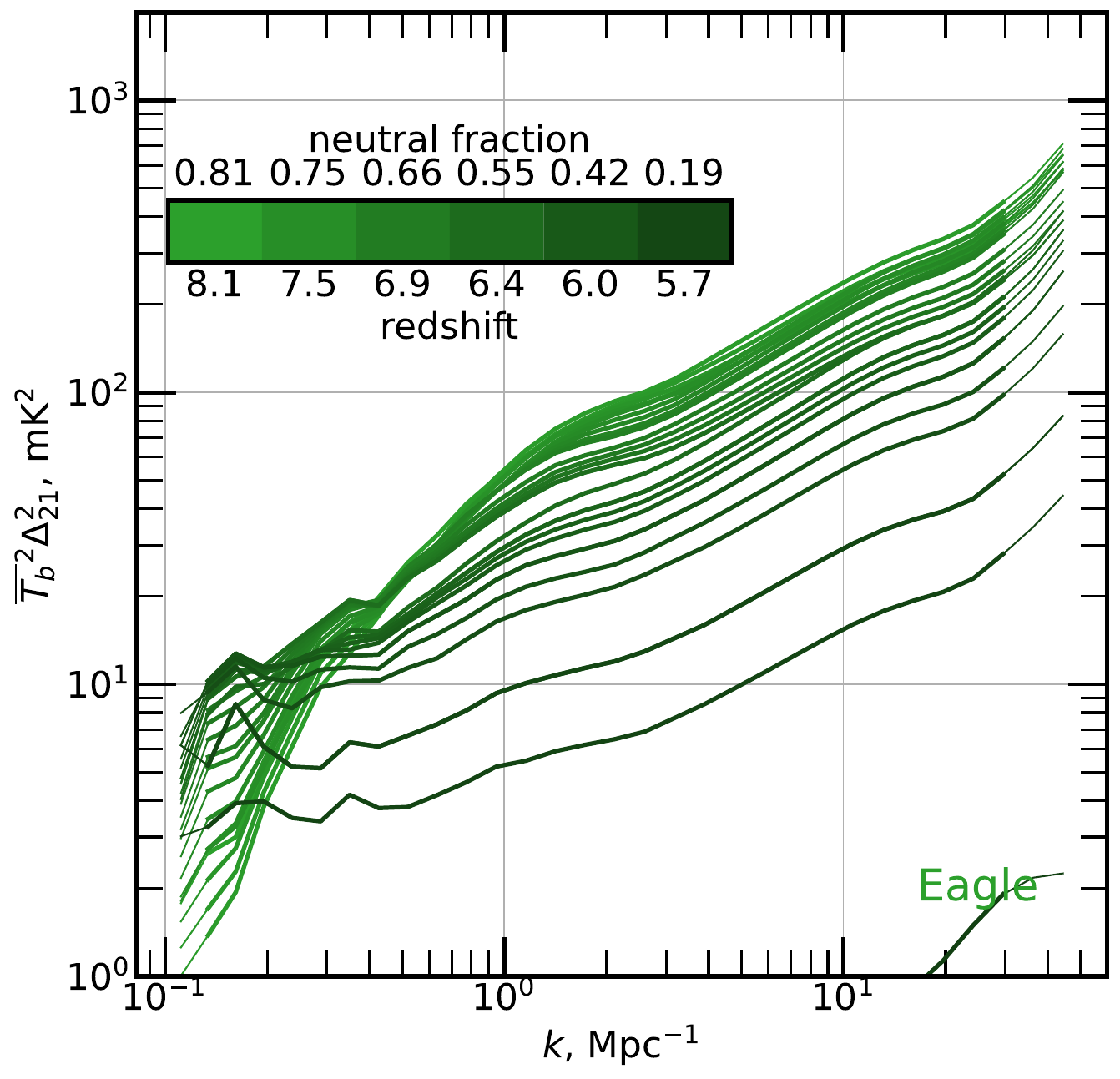}   
    \includegraphics[width=0.49\textwidth]{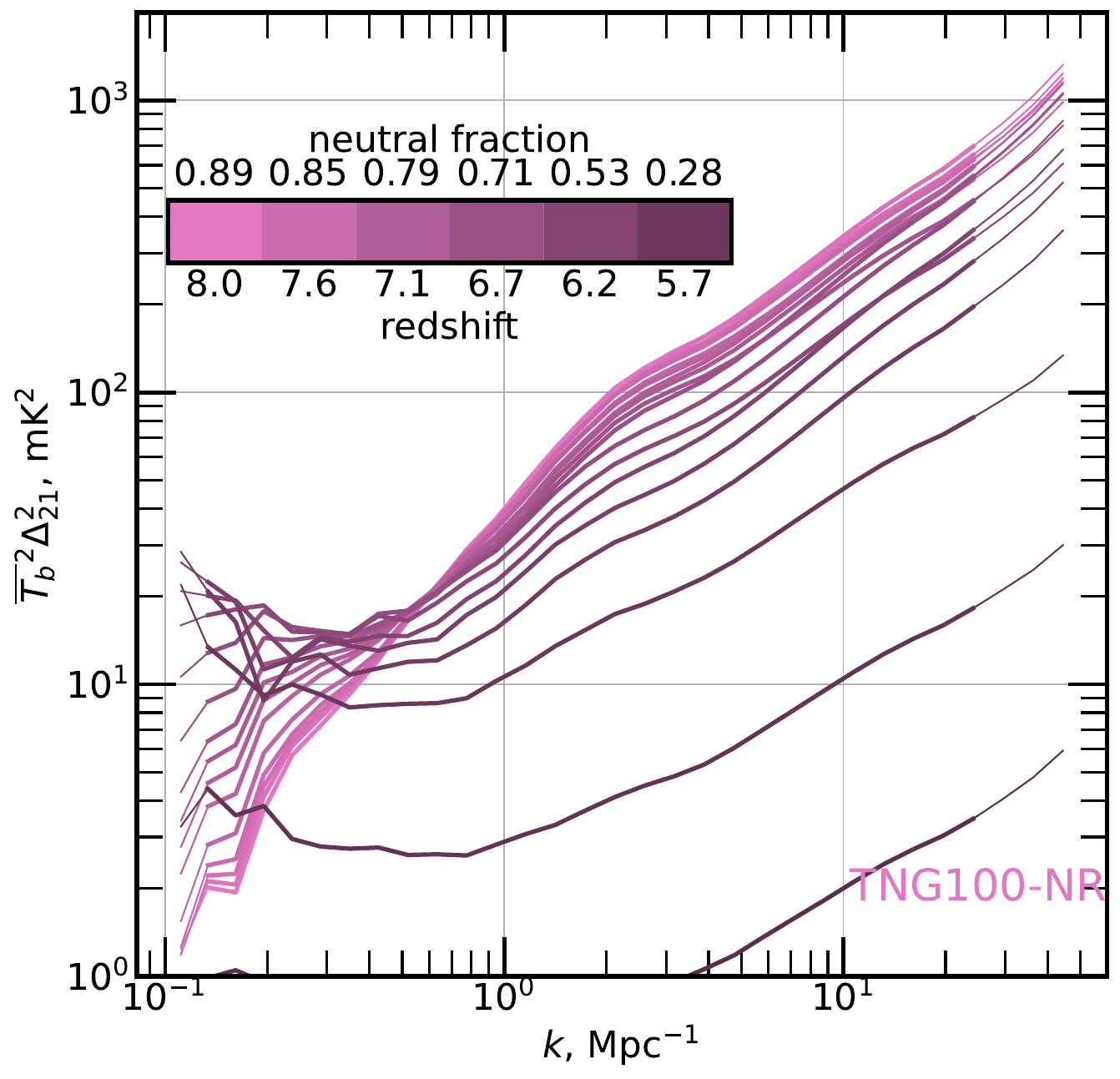}
    \caption{From left to right and top to bottom: evolution of the power spectrum of the 21cm brightness temperature, \dtbPS{}, according to TNG100, Illustris, Eagle, and TNG100-NR in our fiducial post-processing model. The colours denote the progression of the power spectra from $z=8$ (light) to $z\leq 5.7$ (dark). The associated colour bars also show the average neutral fraction at each corresponding redshift. Curves are thinner on scales that are compromised by aliasing or box size effects. For all simulations, the 21cm power spectrum first increases with time at large spatial scales; then, as a greater fraction of the volume is ionized, it decreases homogeneously over all scales.}
    \label{fig:PS_presentation}
\end{figure*}

\figref{fig:PS_presentation} quantifies the fiducial power spectra of the 21cm brightness temperature, \dtbPS{}, and the fiducial reionization histories from TNG100, Illustris, Eagle, and TNG100-NR (see Section~\ref{sec:tng}). We focus on the epoch between $z=8$ and $z\sim6$. For all simulations and times, the 21cm power spectrum increases with decreasing spatial scale, reflecting the underlying matter density power spectrum and the high variance in the density of baryons at small scales.

Broadly, all the predicted 21cm power spectra show the same time evolution: as the first large (>\SI{1}{\mega\parsec}) ionized bubbles form between $z=8$ and $z=7$, the large scale variance in the brightness temperature increases, driving an increase in the power of \dtb{} at large spatial scales, which reaches a maximum somewhere near $z \approx 6.5$ ($ x_{\rm HI} \approx 0.65$). Beyond this point, the power drops at all scales as the volume filling fraction of ionized hydrogen climbs to unity \citep[see e.g.][for recent similar findings]{kannan_introducing_2021}.

Beyond this similar evolution of the 21cm power spectra across different simulations, the values across scales also appear relatively consistent with one exception: both TNG100 and Eagle have significantly less small scale power than Illustris and TNG100-NR. Finally, although the reionization histories of the simulations are similar ($  5.3 < z(x_{\rm HI}=0.1) \leq 6$ for the fiducial model), they are not identical. For instance, Eagle starts reionization earlier than the other boxes under our fiducial assumptions. This can been seen in the very low normalisation of 21cm power spectrum at $z=5.7$ in Eagle with respect to the other simulations at the same redshift. 

\begin{figure*}
    \centering
    \includegraphics[width=0.33\textwidth]{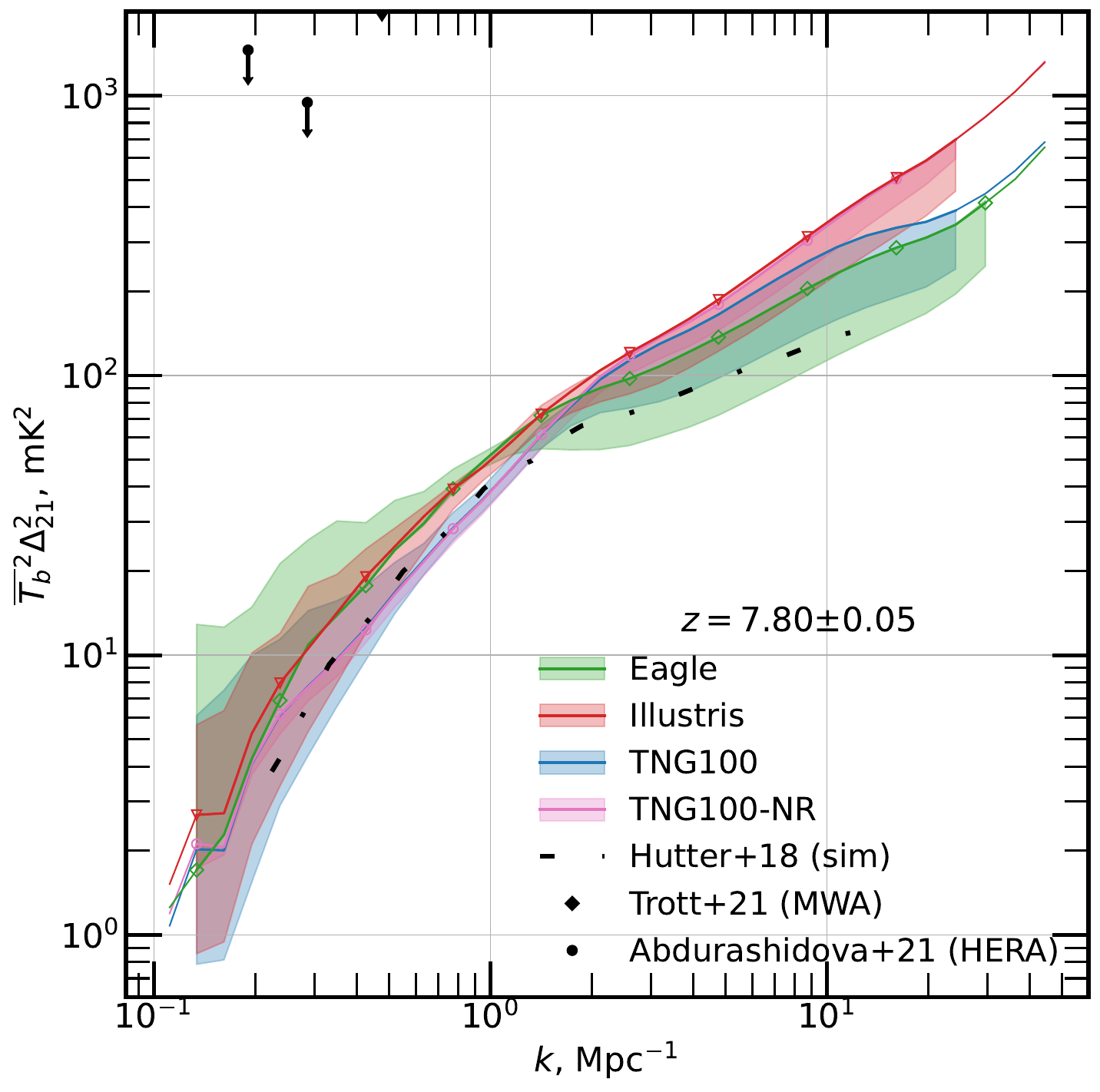}
    \includegraphics[width=0.33\textwidth]{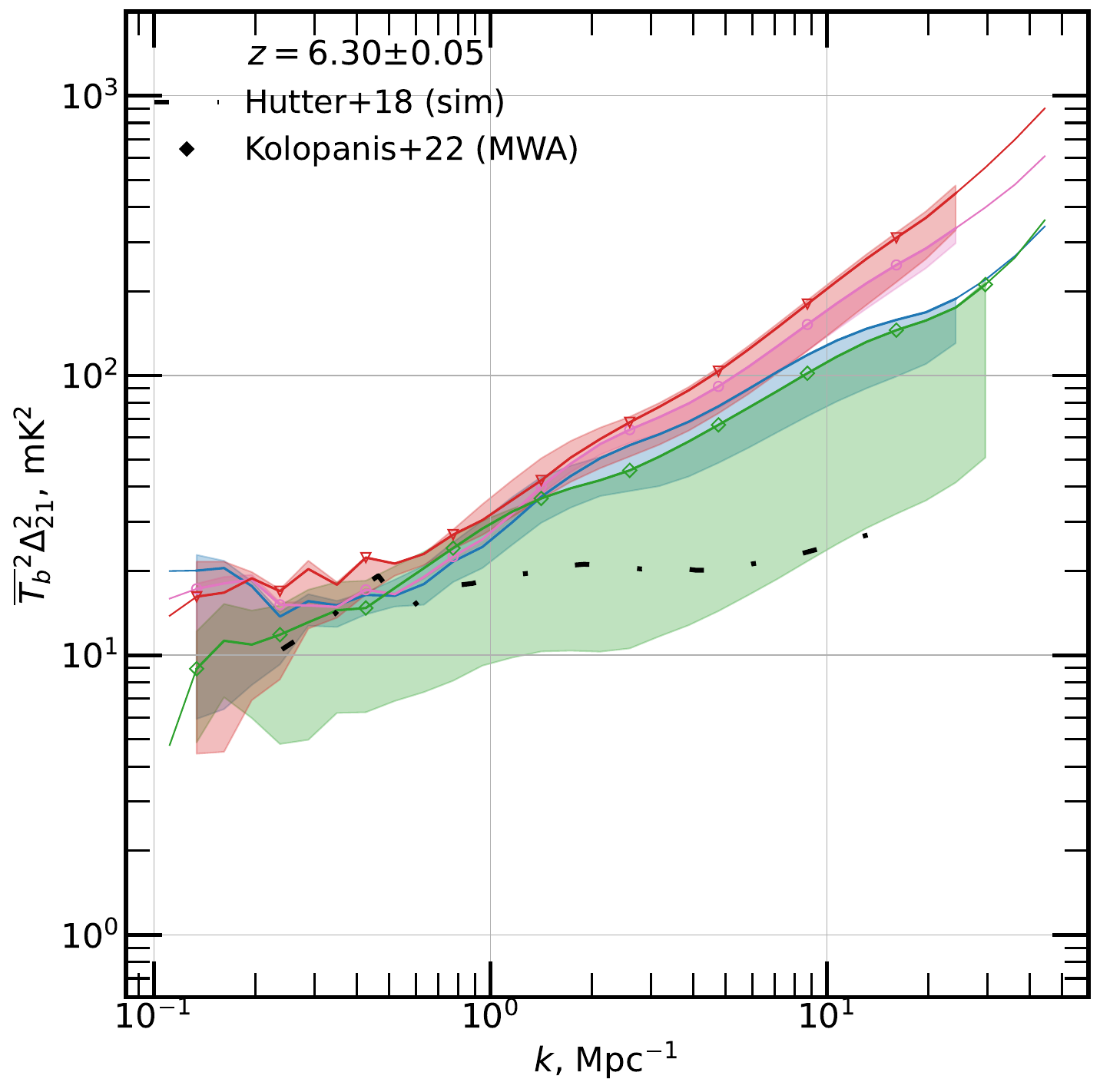}
    \includegraphics[width=0.33\textwidth]{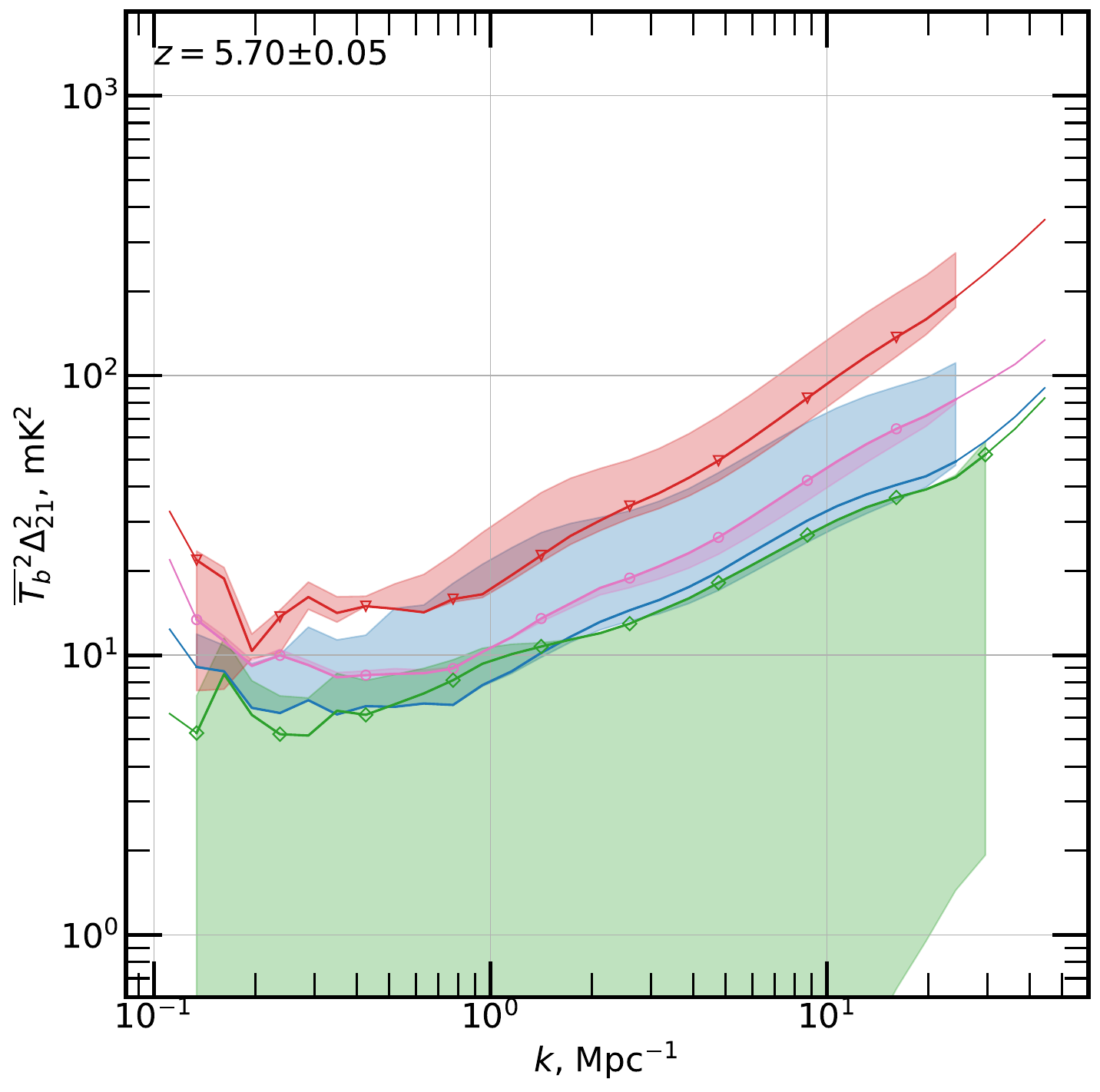}

    \includegraphics[width=0.33\textwidth]{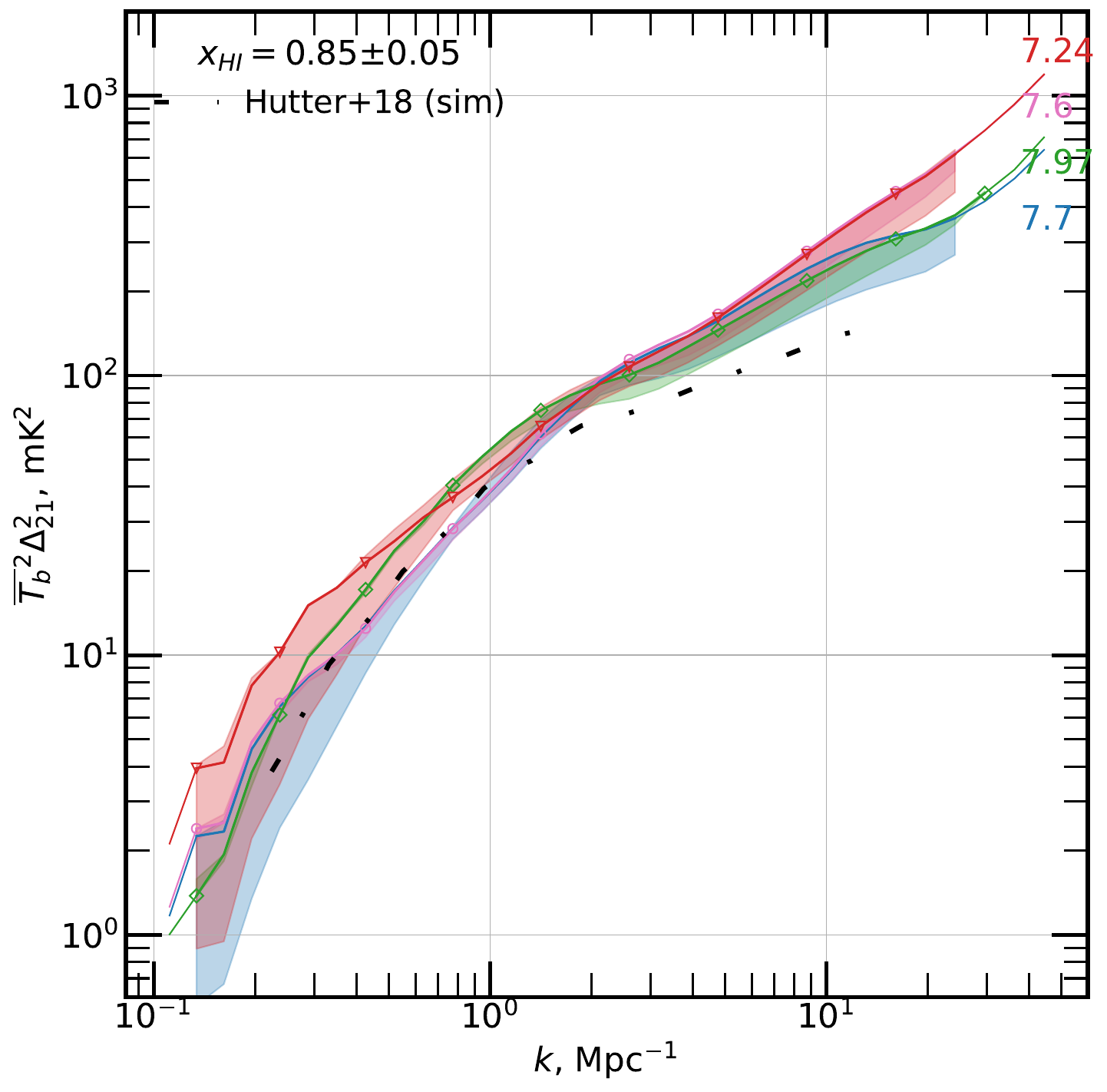}
    \includegraphics[width=0.33\textwidth]{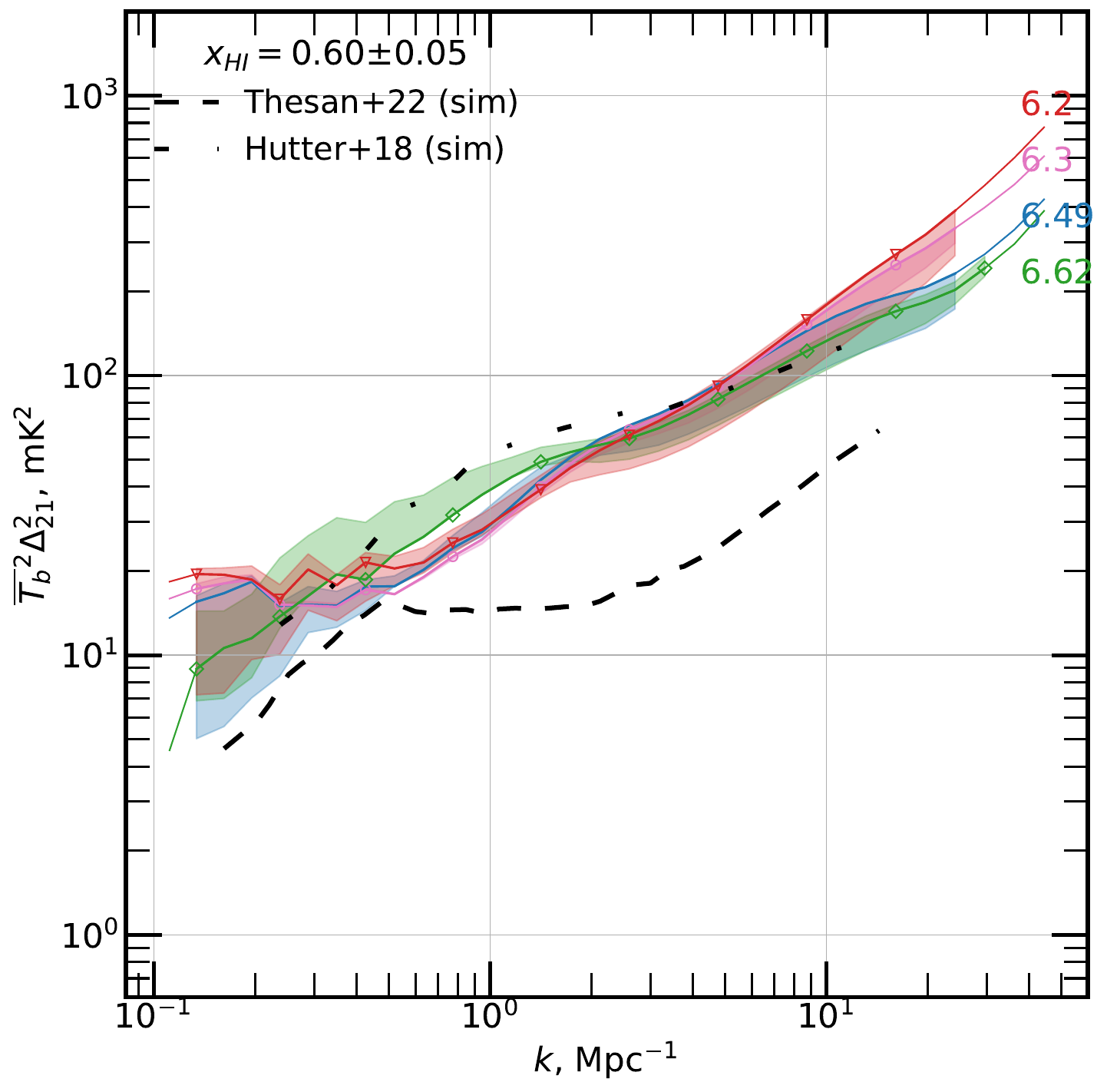} 
    \includegraphics[width=0.33\textwidth]{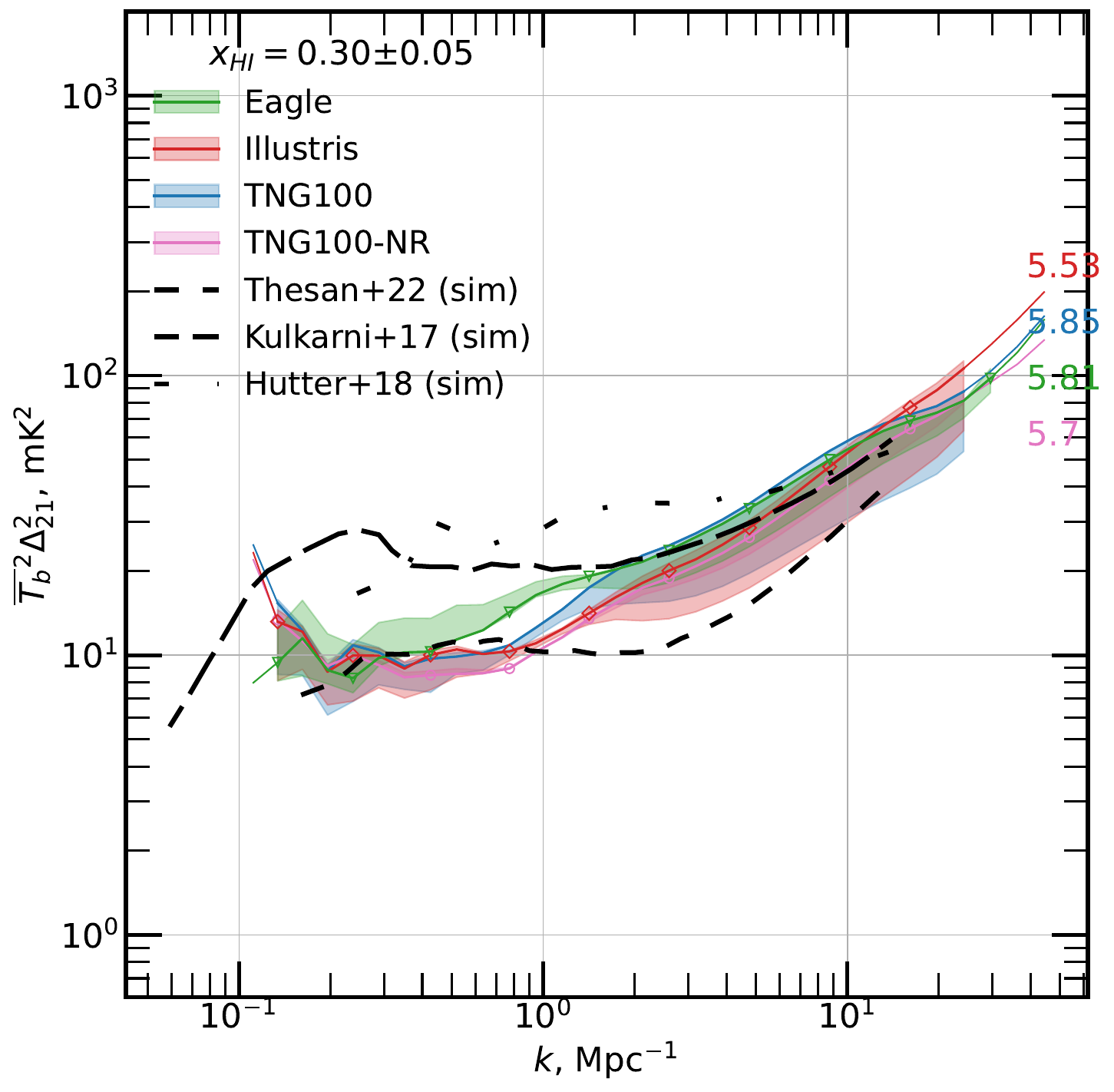}  

    \includegraphics[width=0.33\textwidth]{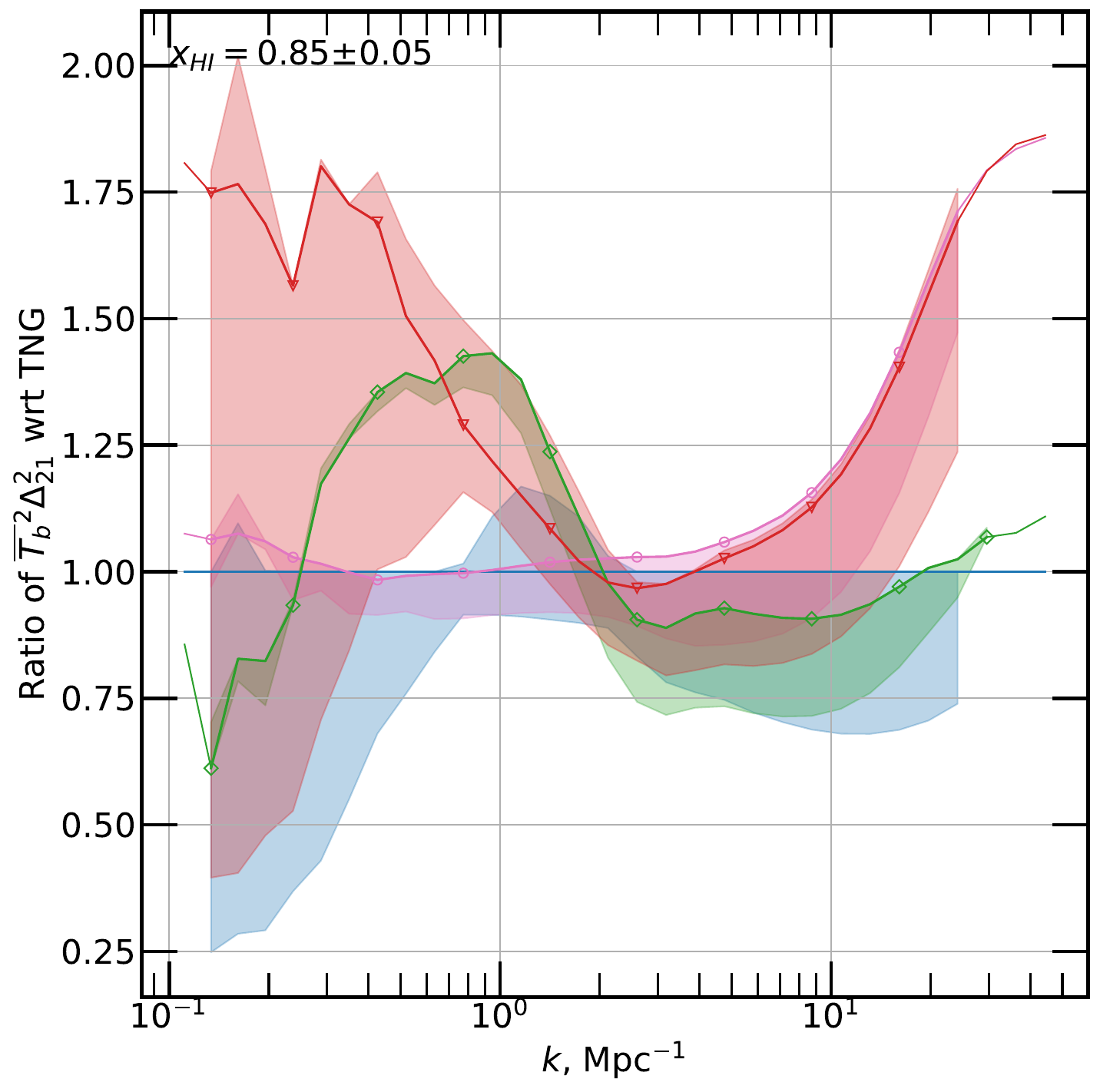}
    \includegraphics[width=0.33\textwidth]{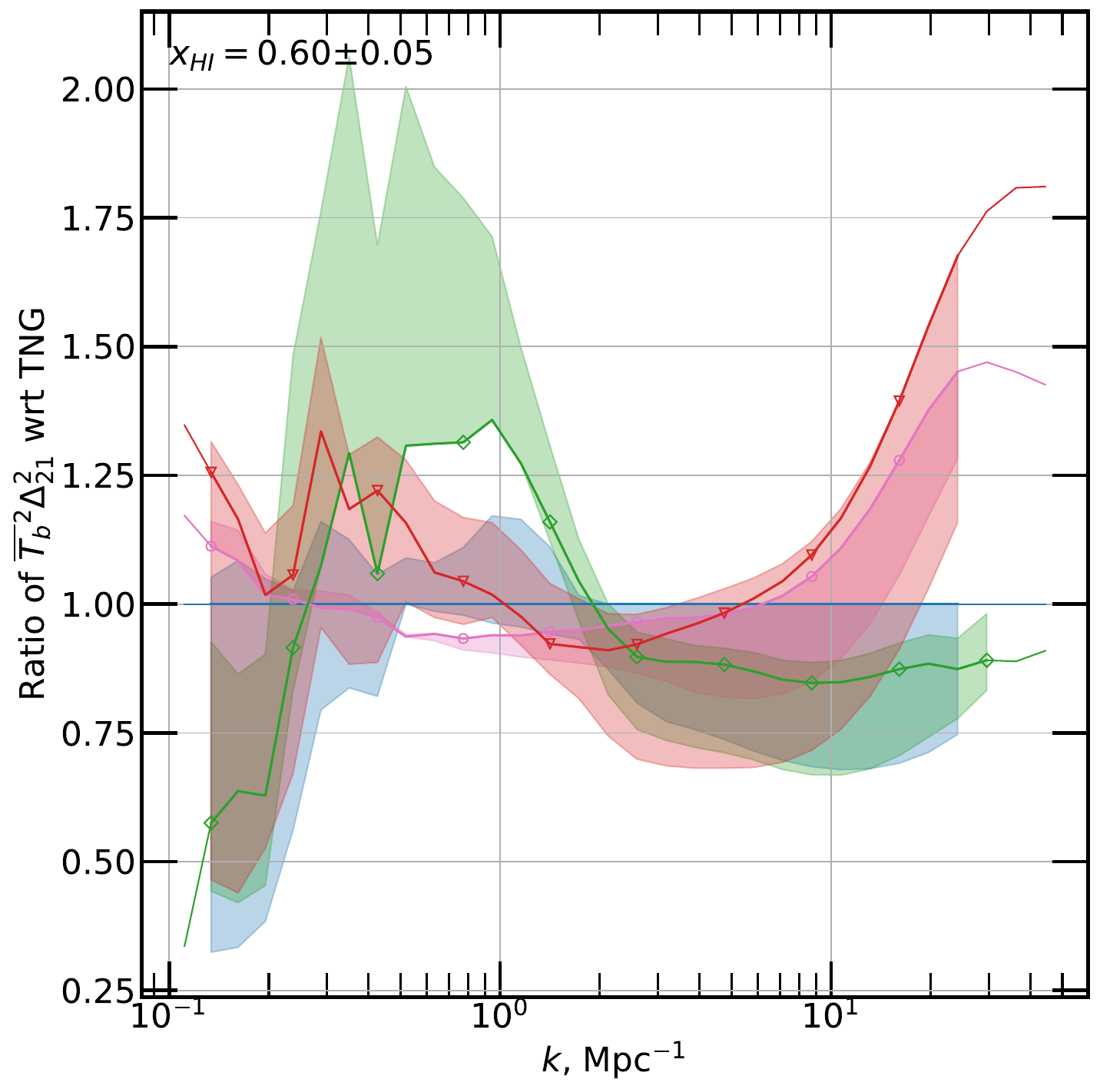} 
    \includegraphics[width=0.33\textwidth]{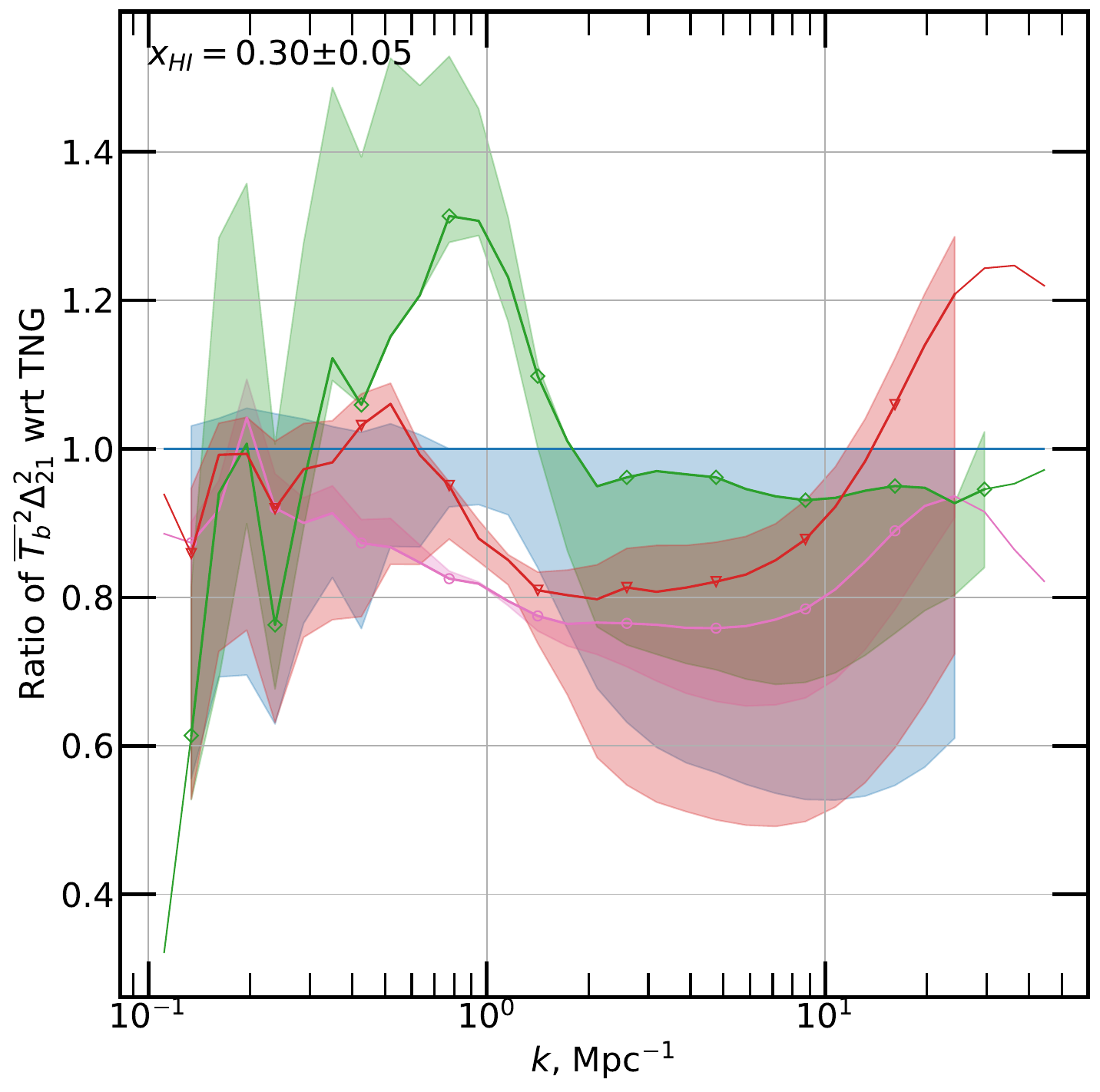}  
    
    \caption{Direct comparison of the power spectra of the 21cm brightness temperature, \dtbPS{}, from TNG100, Illustris, Eagle, and TNG100-NR across time. Coloured curves show the fiducial results, whereas the shaded areas indicate the spread of results using different post-processing models for \dtb{} (Section~\ref{sec:fiducial} and Appendix~\ref{app:model_var}). Thinner curves denote scales where the predictions are affected by resolution or box size. \textbf{Top row:} power spectra for all four simulations at fixed redshifts. \textbf{Middle row:} power spectra grouped so that the average neutral fraction (\xhi{}) in every panel is similar. \textbf{Bottom row:} Same as the middle row, but normalising all power spectra by the TNG100 result to quantify the level of (dis)agreement.}
    \label{fig:PS_compare}
\end{figure*}

To directly compare the three simulations, the top row of \figref{fig:PS_compare} shows the 21cm power spectra at three redshifts (left to right), with all simulations in each panel (TNG100, Illustris, Eagle, and TNG100-NR). Overall and as qualitatively seen above, the 21cm power spectra are very similar across the board. Indeed, the changes produced by our various post-processing models and choices are of the same order or larger than the differences across simulations at fixed redshifts. Across scales and for the first two studied epochs (top left two panels), the different predictions for each simulation show similar trends and normalisation. In fact, the fiducial 21cm power spectra of TNG100 and Eagle are similar for most small scales. This is somewhat surprising as, although they are both recent galaxy formation simulations, they are only calibrated to produce realistic galaxy populations at $z=0$. In addition, the small-scale signals of Illustris and TNG100-NR are similar. Both predict consistently higher 21cm power spectra for all models at the smallest scales we study. Since TNG100, TNG100-NR, and Illustris were run with the same initial conditions, these changes stem solely from the galaxy formation physical models and the choices made for their parameters.

The higher power on small spatial scales, according to Illustris and TNG100-NR, likely arises due to differences in the matter power spectrum of gas, that is the main contributing term to the 21cm power spectra at small scales \citep{furlanetto_cosmology_2006}. Feedback simultaneously smooths out baryons at small scales within galaxies, while pushing a fraction of the smallest scale power to larger scales. Since the small-scale differences between TNG100 and Illustris must arise from the different prescriptions for star formation, and stellar and SMBH feedback, and since TNG100-NR was run without star formation or feedback and is very close at small scales to Illustris, then the high-redshift feedback in Illustris is clearly weaker than in TNG. These conclusions are consistent with the changes to the stellar feedback model between the Illustris and IllustrisTNG models. In particular, high-redshift galactic winds have higher initial velocities in TNG100 than in the original Illustris model \citep[see][for more details]{pillepich_simulating_2018}. 

In the top right panel ($z=5.7$), reionization is almost complete in all the models: any large changes in the power spectra are set by different average \xhi{}. Modulo this difference, which mostly affects the overall normalisation, the shape of the power spectra are similar.

We find our results to be quite similar across all scales to those of \cite{hutter_accuracy_2018} at $z=7.8$, that use \code{CIFOG} to post-process N-body dark matter simulations. However, their fiducial prediction is closer to some of our model variants than to our fiducial model. At $z=6.3$ the \cite{hutter_accuracy_2018} curve is much flatter than our predictions, so that we only agree at large scales. On the other hand, all of our predictions are compatible with the most recent observational upper limits (black arrows). However, with \dtbPS{}$< 10^3 {\rm mK^2}$ \citep[][]{abdurashidova_hera_2022}, current data is not yet constraining. Significant observational advances are required in order to discriminate across our results.

To account for reionization history differences among the simulations and models, we turn to the $ 2^{nd}$ row of \figref{fig:PS_compare} where the 21cm power spectra in each panel are shown at epochs with similar average neutral fractions (average \xhi{}). For all simulations and times, the predictions are closer, and the shaded areas representing our other post-processing models are much tighter and closer to our fiducial results. To enable quantitative comparisons, we show the same results in the $ 3^{rd}$ row, but normalised by the TNG100 outcome. 

We find that, overall, our various predictions are similar: all models are within less than a factor of two from the fiducial TNG100 predictions. TNG100 and Eagle are very similar at small scales, as commented upon above. In particular, the fiducial predictions from Eagle and TNG100 are roughly within 15\% for $k \geq \,$\SI{2}{\per\mega\parsec}, i.e. at scales smaller than about $500 \rm kpc$. At the same scales, and before \xhi{}$\,= 0.3$, Illustris predictions are almost identical to TNG100-NR, and both have more power than Eagle and TNG100, by an amount that increases up to 200 \% by $k \approx \,$\SI{20}{\per\mega\parsec}. 

Near $k \approx \,$\SI{0.8}{\per\mega\parsec}, our fiducial Eagle predictions are about 30 \% higher than those based on TNG100, TNG100-NR, or Illustris (when \xhi{}$\, \lesssim 0.6$). Depending on our post-processing model combinations, this difference is less or more pronounced. At the largest spatial scales, all simulations agree to within 50 \%, except Illustris when \xhi{}$\,=0.85$, which has far more power than TNG100 and whose predictions are very dependent on our post-processing modeling choices -- the shaded area covers $50-180$ \%. TNG100 and TNG100-NR are quantitatively similar at large scales.

When accounting for the reionization history, we find that our predictions are the most similar across all scales when average \xhi{}$=0.3$, with the majority of results lying within $\approx 40\%$ of the fiducial TNG100 result. 

\begin{figure*}
  \includegraphics[width=0.33\textwidth]{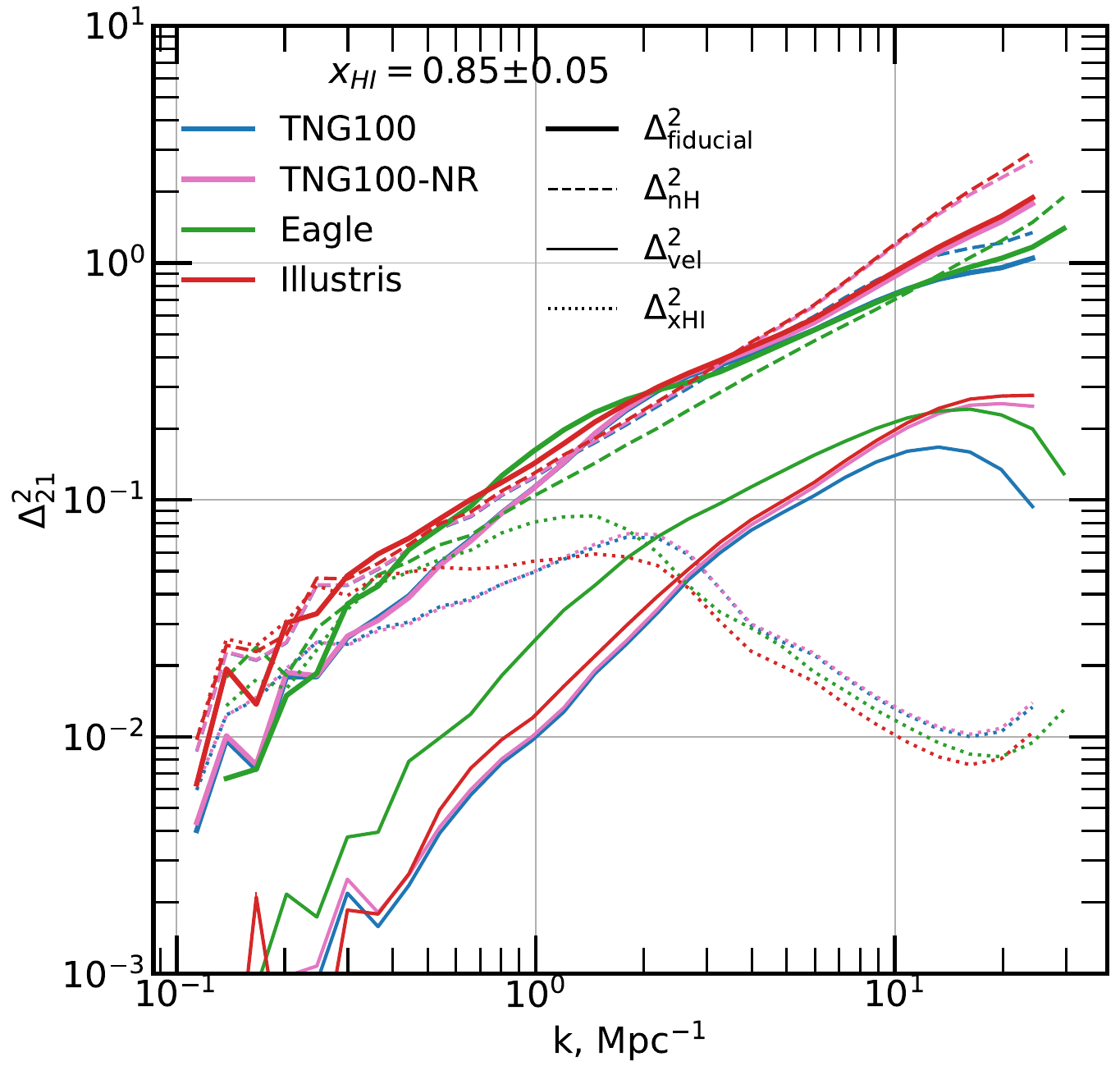}
  \includegraphics[width=0.33\textwidth]{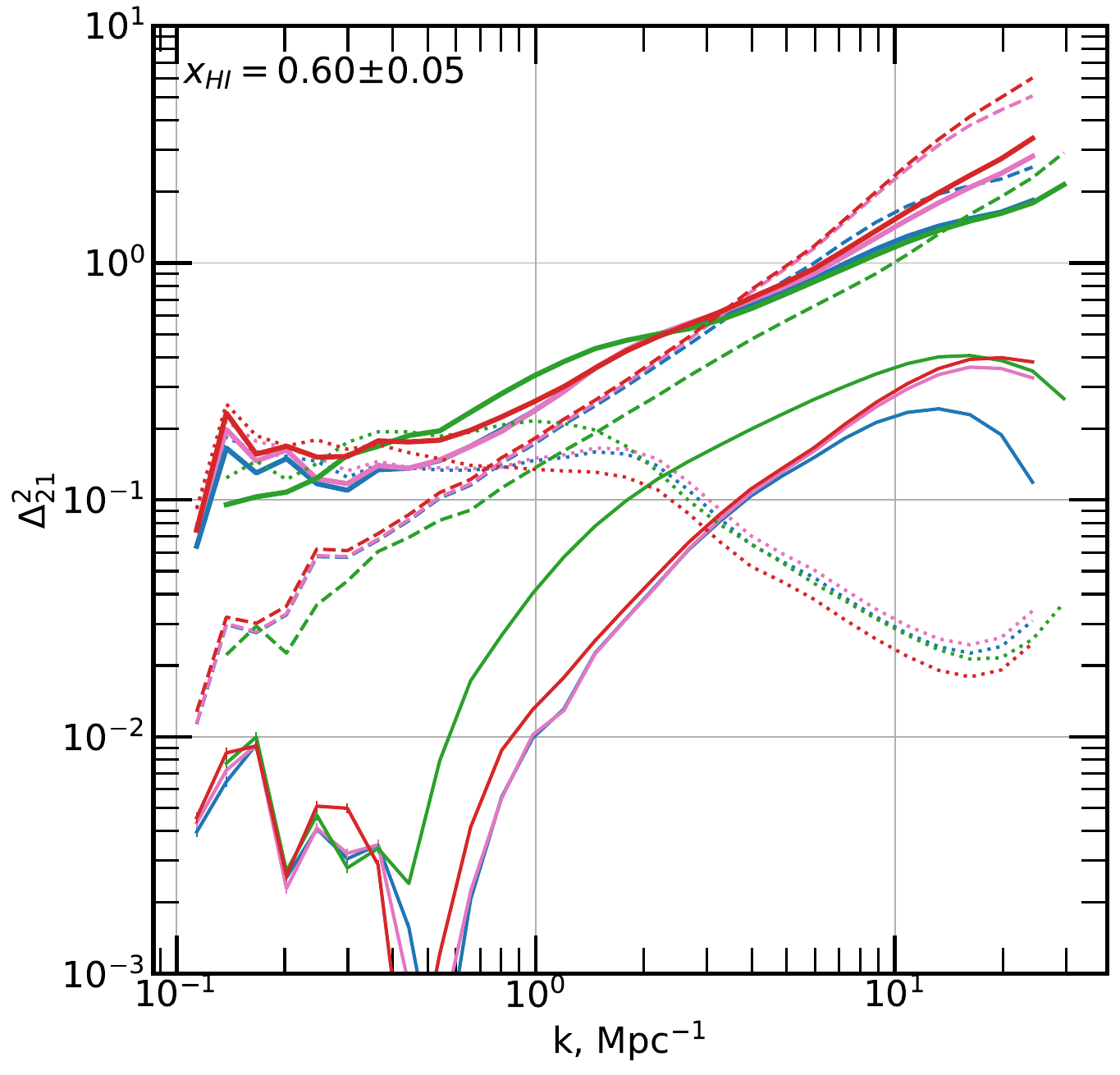} 
  \includegraphics[width=0.33\textwidth]{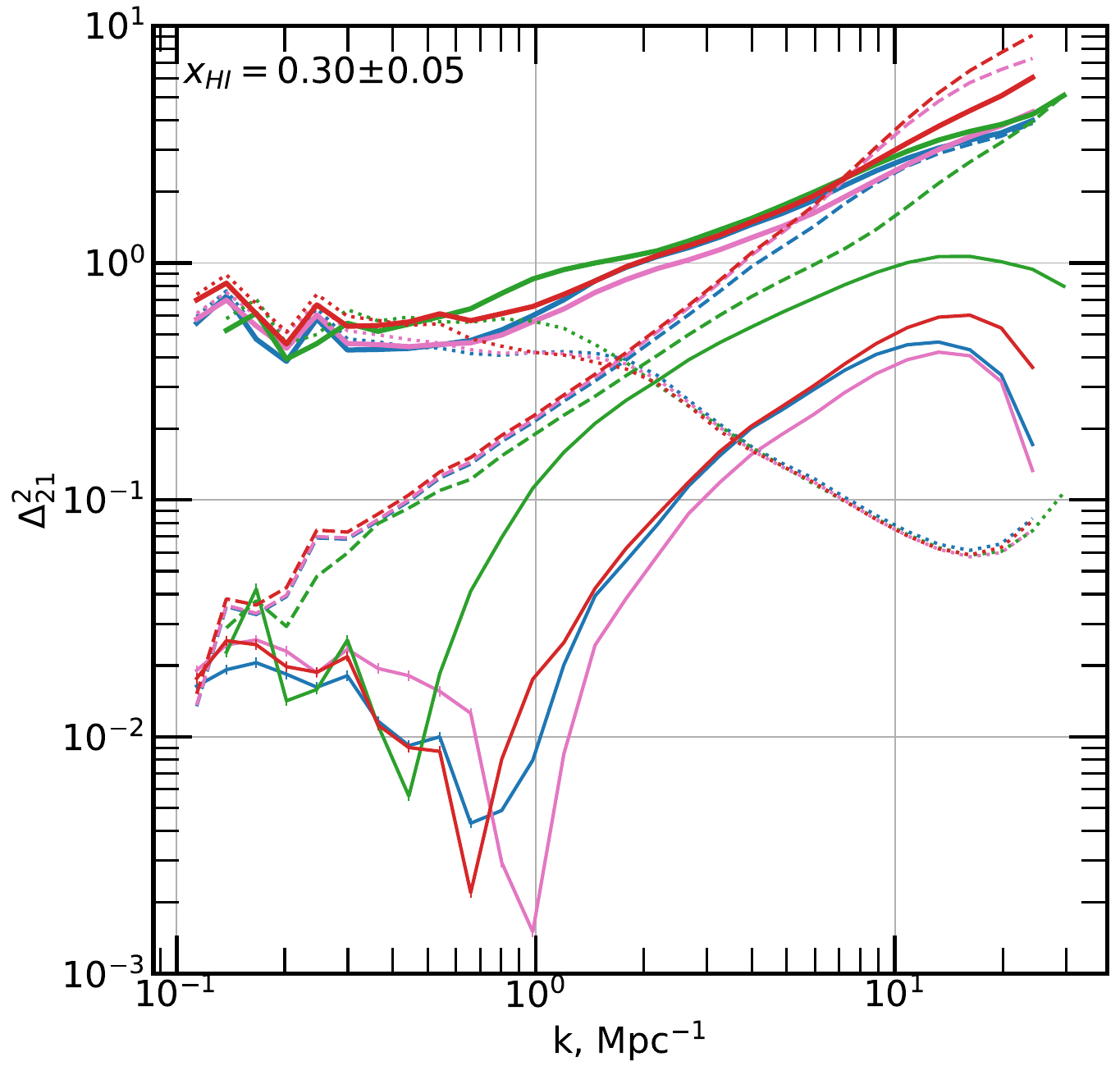}
  \caption{Normalized 21cm power spectrum, $ \Delta^2_{21}$, in our fiducial post-processing model and its contributing terms (as detailed in Eq.~\ref{eq:dTb2}), for all simulations. To simplify this plot, we do not show any constituent cross terms of $ \Delta^2_{21}$, which can however contribute significantly at some scales and ionization levels \protect\citep[e.g.][]{georgiev_large-scale_2022}. We have grouped the simulations at redshifts when their average ionization fractions are similar. Solid thick curves show the fiducial results. Dashed curves denote the approximate contribution to $ \Delta^2_{21}$ from the gas density term, whereas dotted and solid thin curves represent the \xhi{} term and the approximate contribution of the redshift-space velocity distortions (velocity term), respectively.}
  \label{fig:ps_ingredients}
\end{figure*}

Comparing to previous models, the predictions from other works are within a factor two of our results. The agreement with the \cite{hutter_accuracy_2018} predictions is improved by comparing results at fixed average \xhi{}. Since the THESAN \citep[][]{kannan_introducing_2021} results are based on similar physical prescriptions as TNG100 (with the addition of fully coupled radiative transfer), the comparison between the THESAN and TNG100 predictions is informative: see the dashed black curves in the middle panels of \figref{fig:PS_compare}. At average \xhi{}$\,=0.6$ the lowest TNG100 prediction is close to the THESAN outcome on large spatial scales. However, at progressively smaller spatial scales (larger $k$), the THESAN result is systematically 50 \% lower than TNG100. At average \xhi{}$\,=0.3$, the agreement is better: at large scales the results between TNG100 and THESAN are interchangeable, whereas at the smallest scales the THESAN outcome is close to the lowest TNG100 prediction. However, near $k \approx \,$\SI{2}{\per\mega\parsec}, the THESAN predictions are again roughly 50 \% lower. This larger power in TNG100 versus THESAN may be due to our use of \code{CIFOG} in lieu of a full radiative transfer (RT) method. Indeed \cite{hutter_accuracy_2018} shows differences of a similar degree between their \code{CIFOG} scheme and a full RT scheme at these scales and larger. However, coupling the TNG100 models to RT could also have significant effects on galaxy formation and the sources of reionization, potentially increasing the differences with respect to TNG100. We further expand on the comparison to other simulation-based models and observations in the Section \ref{sec:disc}.

Overall, our results paint a picture of relatively close predictions for the 21cm signals at $z=6-8$ from recent cosmological hydrodynamical non-RT galaxy simulations. This is particularly true for Eagle and TNG100 at small scales, and in comparison to other theoretical works. At the same time, unique features from individual simulations are present (e.g.\ the bump at $k \approx \,$\SI{0.8}{\per\mega\parsec} in Eagle with respect to the other boxes), which are robust to our post-processing modeling choices. In the next section, we discuss these findings, in terms of the power spectra of the constituent fields of \dtb{}, with respect to galaxy properties, and as a result of our post-processing choices.


\section{Discussion}
\label{sec:disc}

\subsection{Physical interpretation}

\subsubsection{Contributions to the 21cm power spectrum}

\figref{fig:ps_ingredients} shows the largest contributing terms of the dimensionless power spectrum of the 21cm brightness temperature, \PS{}, that we compute (as per Eq.~\ref{eq:dTb2}) alongside the fiducial predictions from each simulation, at three key moments of reionization. As before, in all four simulations, the fiducial \PS{} (solid thick curves) increases with time at large scales. Here, we directly see the rise of the \xhi{} term (dotted) at large scales between average \xhi{}$ \,=0.85$ and \xhi{}$\,=0.3$, which drives a corresponding increase for the total 21cm power. 

It has been extensively shown that the two most important components to the 21cm spatial correlations are the gas density power spectrum and the neutral fraction power spectrum \citep[][]{furlanetto_cosmology_2006}. Our results echo these findings: for all simulations and considering all cosmic epochs, the \xhi{} term (dotted) is the most important at large scales ($k < \,$\SI{2}{\per\mega\parsec}) whereas the gas density term (dashed) dominates at smaller scales, especially towards the end of reionization. We find that the next most important contribution to our fiducial result comes from RSD (i.e. the velocity term; solid thin curves), which can contribute close to 10 \% of the total power between $ k=1$ and \SI{20}{\per\mega\parsec}, depending on the simulation. 

This decomposition is only precisely valid if the correlations between the constituent fields of \dtb{} are null, which is the not the case. This is visible at certain scales, where the sum of the depicted components does not equal the fiducial result. Particularly at the smallest scales, the difference between the gas density power spectrum and \PS{} is made up by the missing (negative) contribution from the higher order terms, that we omit in this work. However, this simplification does not affect a relative comparison between the simulations.

At $k \approx \,$\SI{1}{\per\mega\parsec}, there is more \xhi{} power in the Eagle simulation than in TNG100 and Illustris: this contributes to the bump in the 21cm power spectrum in Eagle with respect to TNG100 near $k \approx \,$\SI{1}{\per\mega\parsec}. At the largest scales, and at average \xhi{}$\,=0.85$, the Illustris \xhi{} power is greater than in the other simulations, potentially resulting in the correspondingly larger \dtbPS{} in Illustris.

At smaller spatial scales, when the gas density power spectrum is largest, we find that differences in the \PS{} across simulations map to differences in the gas density power spectrum, at least for average \xhi{}$\, \geq .35$. Indeed, the spatial correlations of gas density and velocity in Illustris and TNG100-NR are similar over the scales where the simulations have comparable normalised 21cm power spectra. At the same time, the apparent agreement between the 21cm power spectrum from Eagle and TNG100 is actually a coincidence that arises from a lower gas density, but higher velocity, power spectrum in Eagle. 

The significant differences (up to a factor of 2) in the gas density and velocity power spectra at $k \geq \,$\SI{10}{\per\mega\parsec} in TNG100 and Illustris (and TNG100-NR), that share the same initial conditions, can only arise from the underlying differences in the baryonic and feedback physics models. The similarity of the Illustris and TNG100-NR results suggests that feedback in Illustris has a negligible impact on gas at these scales and high redshifts ($z\gtrsim 6$). With respect to TNG100 and Eagle, stellar feedback in Eagle may be more effective at ejecting matter from the central regions of galaxies, with even higher velocity winds than in TNG. This would explain the lower matter density and higher velocity power spectra in Eagle. In fact, for $k \leq \,$\SI{8}{\per\mega\parsec}, and particularly as reionization completes, the velocity term from Eagle has the most power (by a factor of $ \lesssim 2$), indicating the greatest variation in line of sight velocity gradients.

\begin{figure}
    \centering
    \includegraphics[width=0.45\textwidth]{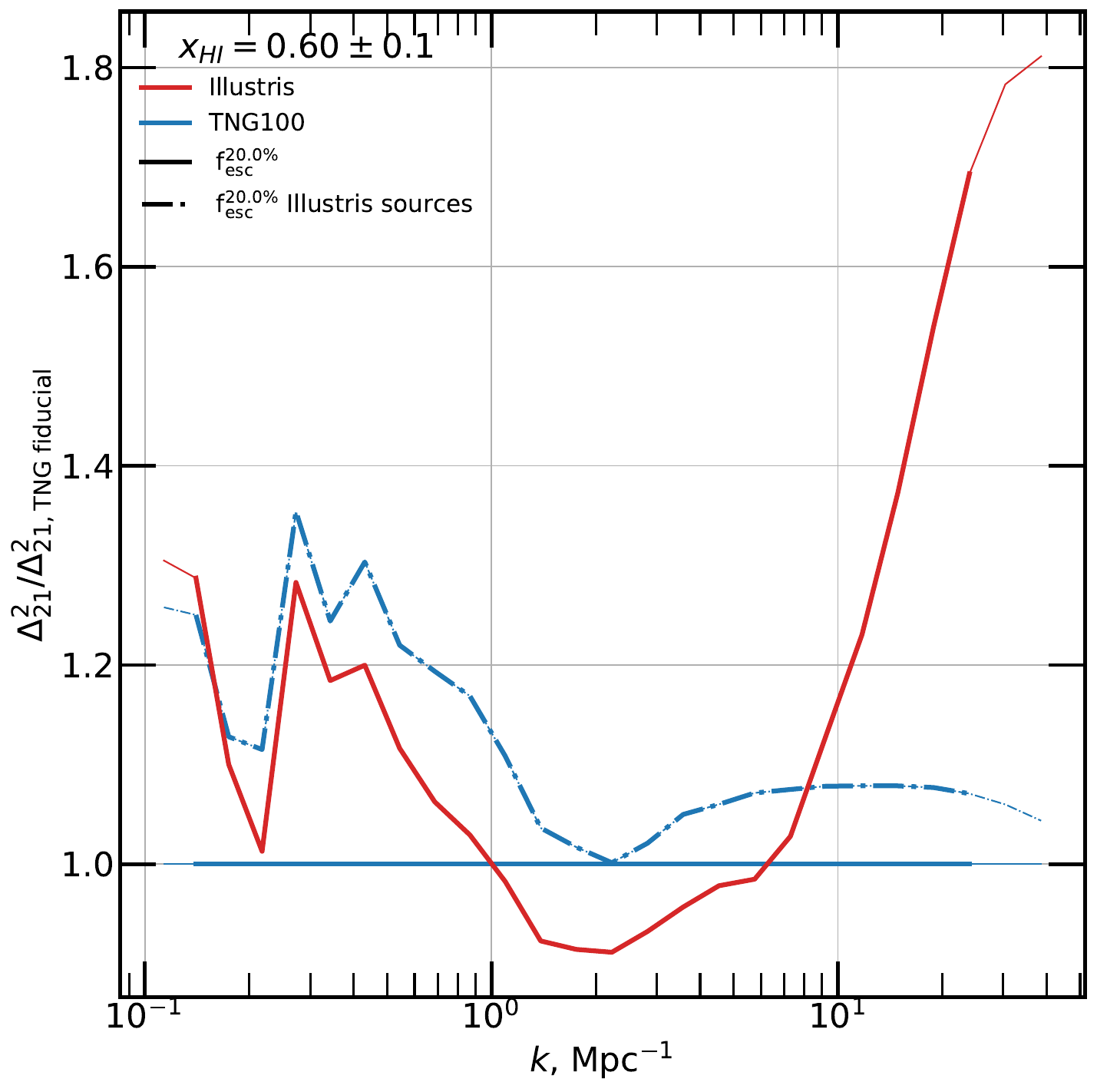}
    \caption{Relative 21cm power spectra for an experimental analysis of the TNG100 simulation, keeping all fiducial choices except we use the sources of ionizing radiation as predicted by Illustris. This test run is called ``Illustris sources'' (dashed dotted curves) and is compared to the fiducial Illustris and TNG100 predictions via the ratio of the 21cm power spectrum to TNG100. Each power spectrum is taken at an epoch where average \xhi{}$\, \approx 0.6$. The differences in the 21cm power spectrum at $k \leq \,$\SI{7}{\per\mega\parsec} in Illustris vs. TNG100 are caused by differences in the stellar populations rather than in the gas state. Thinner curves denote scales where the predictions are affected by resolution or box size.}
    \label{fig:ill_src_ps}
\end{figure}

Finally, we can confirm that the different predictions of the large-scale 21cm power spectrum from Illustris and TNG100 are due to differences in the large-scale ionization fraction. To directly compare the stellar source populations, we use the shared initial conditions of TNG100 and Illustris to perform another \code{CIFOG} run using TNG100 and our fiducial assumptions, but with the sources of ionizing radiation as predicted by Illustris (see Section~\ref{ssec:src}). \figref{fig:ill_src_ps} shows the ratio of the 21cm power spectrum from this new run (named ``Illustris sources'') and from our fiducial TNG100 and Illustris predictions to the TNG100 result, when average \xhi{}$\, \approx 0.6$. The outcome of this test run is strikingly close to those from Illustris for scales $k\leq \,$\SI{7}{\per\mega\parsec}, demonstrating that it is the differences in the source population rather than in the physical gas state that drive the most important differences in the 21cm power spectrum, for most scales. 

\begin{figure*}
    \centering
    \includegraphics[width=\textwidth]{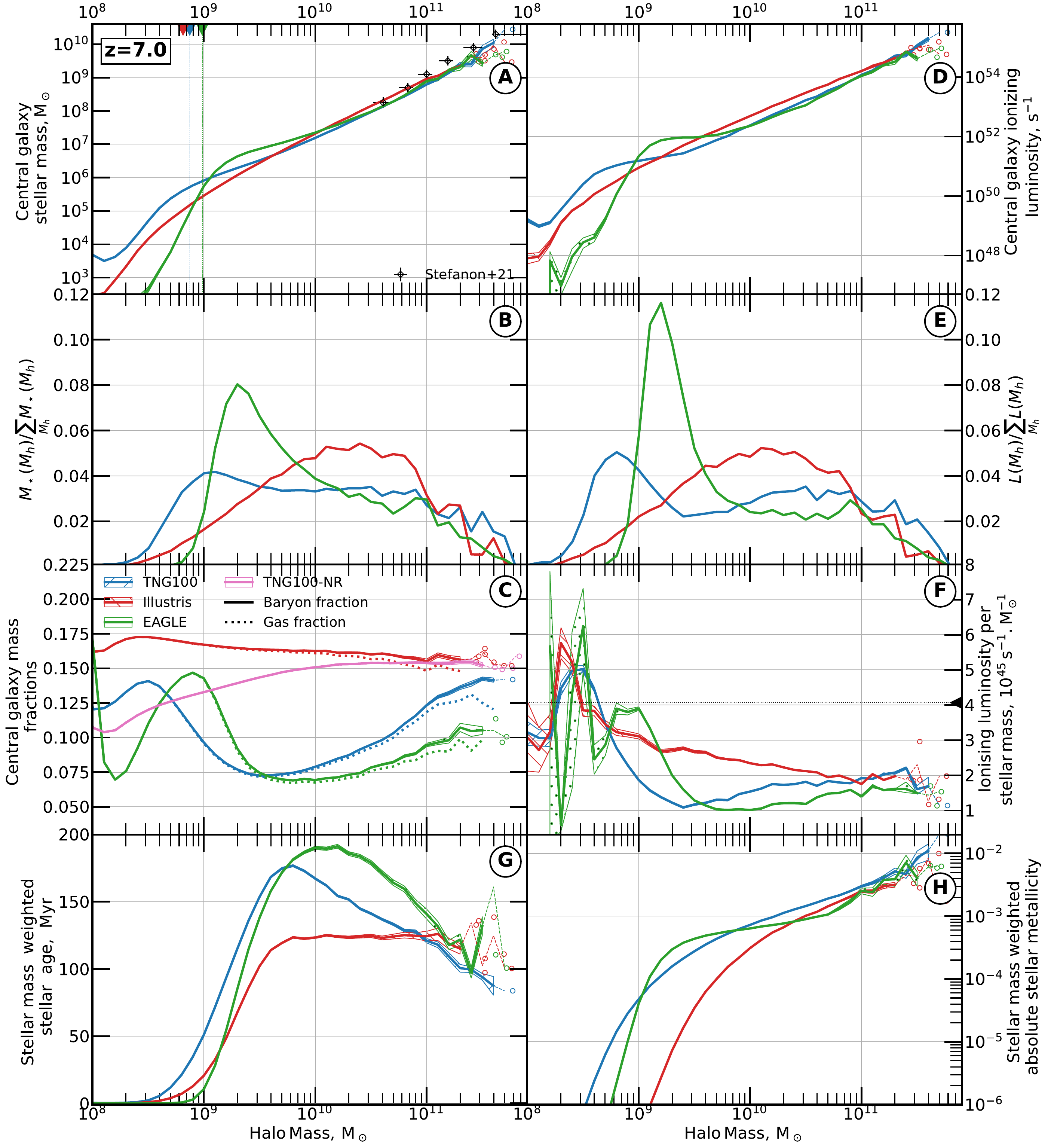}
    \caption{
    Average properties of the central galaxies in TNG100, Illustris, and Eagle as a function of halo mass, at $z=7$. Thinner curves show the means in bins with fewer than two galaxies (in these bins, the individual galaxies are also shown). The areas around curves shows the central 90 \% result obtained from bootstrapping the sample of galaxies. Starting from the top left:
    \textbf{Panel A}: Galaxy stellar mass -- The vertical coloured lines show the minimum possible mass of a halo resolved by 100 dark matter particles in each simulation.
    \textbf{Panel B}: Ionizing photon luminosity from the stellar particles of central galaxies.
    \textbf{Panel C}: Fraction of total stellar mass in the central galaxy haloes.
    \textbf{Panel D}: Fraction of total ionizing luminosity from central galaxies.
    \textbf{Panel E}: Baryon, gas, and stellar mass fractions to total mass.
    \textbf{Panel F}: Ionizing photon luminosity per stellar mass -- The black triangular marker shows the ionizing luminosity in our modeling of a young ($\rm 1 Myr$), low metallicity ($Z_{\rm absolute}=10^{-3}$) stellar population of mass $ 10^6 \rm M_\odot$.
    \textbf{Panel G}: Mass-weighted stellar ages, i.e. average galaxy-wide stellar ages weighted by the stellar-particle masses.
    \textbf{Panel H}: Mass-weighted stellar metallicity.}
    \label{fig:gal_pties}
\end{figure*}

Despite allowing for differences in the simulations' gas temperatures stemming from choices in the implementation of feedback, we find that accounting for temperature in our modeling does not significantly distinguish the 21cm power predictions of the simulations. This resemblance has two likely explanations: First, and foremost, we have limited ourselves to the post-heating epoch, so that the IGM gas in all three simulations is similarly hot. Second, it may be that we do not consider small enough scales for the temperature of the coolest, densest neutral clumps and the hottest outflows to significantly impact the 21cm power.

\subsubsection{Galaxy properties predicted by the different simulations}

To understand the role of different source populations in creating scale-dependent differences in the 21cm power spectra from simulation to simulation, we consider the galactic properties predicted by Illustris, TNG100 and Eagle and the underlying star formation and feedback models at play. Figure \ref{fig:gal_pties} shows several properties of galaxies as a function of dark matter halo mass at a representative redshift of $z=7$. 

First, we examine the stellar mass of galaxies (top left panel, panel A). Overall and for all simulations, the stellar mass increases between $ M_{\rm h}=10^8$ and \SI{3e11}{\simsun}. This is expected as the more massive haloes are able to accrete and concentrate more gas in spite of stellar feedback, permitting more star formation. For the highest mass haloes ($ M_{\rm h} \gtrsim 2\times 10^{10} \rm M_\odot$), \mstel{} is a power law function of $ M_{\rm h}$. At these masses, the TNG100 and Eagle simulations are very similar, whereas Illustris galaxies have on average somewhat more stellar mass at fixed halo mass. For lower masses, stellar feedback is efficient at suppressing star formation, and this results in a knee in the average stellar mass to halo mass relation, giving way to a stronger decrease of stellar mass with decreasing halo mass. The mass at which this knee manifests itself arises from details in the physical models of each simulation.\footnote{The relationship between stellar mass and halo mass is also complicated by the mass resolution of the stellar particles and the stochastic nature of star formation in these simulations. Indeed, between \SI{e8}{\simsun} and $ \lesssim 2\times 10^9 \rm M_\odot$ haloes transition from rarely forming a single stellar particle, to systematically forming particles, leading to an artificially rapid jump in mean \mstel{}.}

Panel B shows the fraction of the total stellar mass contained in each halo mass bin. In other words, it shows which halo masses contain the most stellar mass. This is governed by two factors: the abundance of galaxies decreases with increasing mass, and stellar mass increases with halo mass. These competing trends give rise to a halo mass interval that forms the highest fraction of all stars. The resulting curves are strikingly different for each simulation. For instance, stars in Illustris are more likely to form in $ >10^{10} \rm M_\odot$ haloes than in the other two simulations. Indeed, the halo mass bin that contains the highest fraction of stellar mass is \SI{e9}{\simsun} in TNG versus \SI{2e9}{\simsun} in Eagle and \SI{e10}{\simsun} in Illustris. Further, the most star forming halo masses contain somewhat different fractions of the total stellar mass: $ \approx 5$ \% in TNG100 and Illustris, but 8 \% in Eagle. Despite the differences between TNG100 and Eagle, they have close fractions of stellar mass for $ M_{\rm h} \gtrsim 10^{10} \rm M_\odot$, and only differ for lower masses. Stars in TNG easily form in $\lesssim 10^9  \rm M_\odot$ haloes, while in Eagle these haloes form very little stellar mass, potentially related to resolution differences. For the highest mass haloes, all three simulations are almost identical.

Panel C shows the average mass fractions for gas (solid lines) and baryons (dotted lines). Each mass fraction is defined as the ratio between the gas mass (or baryon mass) and the total mass of all particles and cells associated with the \code{SUBFIND} sub-haloes (i.e. gravitationally-bound material). As before, TNG100 and Eagle are more similar to each other than to Illustris: the average baryon and gas fractions in Illustris haloes is nearly constant with halo mass. In contrast, the TNG100 and Eagle fractions are lower for all halo masses, and decrease until a few \SI{e9}{\simsun}, above which they rise with increasing halo mass. These differences between Illustris and TNG100 haloes are driven by physical models changes, including the non-monotonicity. In fact, there is no corresponding drop in the total halo gas fraction in TNG100-NR. This indicates that the halo-mass trends of panel C for TNG100 and Eagle are associated with cooling, star formation, and feedback. The latter can expel gas from the TNG100 and Eagle haloes, whilst heating their surroundings and slowing accretion of new gas from the IGM -- predominantly at lower redshifts \citep[e.g.][]{pillepich_simulating_2018}. 

The differing effectiveness of feedback between the simulations explains why the haloes that contain the most stellar mass are more massive in Illustris. For $>$\SI{3e9}{\simsun}, haloes in Eagle have the lowest average baryon and gas fractions, suggesting the strongest feedback effects. TNG100-NR actually has a lower gas fraction than TNG100, despite its absence of feedback. This may be due to the lack of gas cooling in TNG100-NR, slowing the accretion of gas into galaxies. These differences are reflected in the 21cm power spectra, e.g. in \figref{fig:ps_ingredients}. The simulations with the highest gas fractions (Illustris and TNG100-NR) also have the largest small-scale power in the spectra of gas density. In TNG100 and Eagle, where the gas fractions are lower, the gas density power is also lower at small scales \citep[][]{schneider_new_2015, van_daalen_contributions_2015}. We infer that stellar feedback in TNG100 and Eagle is capable of evacuating gas from galaxies out to $\leq \,$\SI{}{\mega\parsec} scales \citep[see][for a study of large-scale feedback-driven gas redistribution]{ayromlou_feedback_2022}.

\subsubsection{Connecting galaxy properties to the ionizing luminosity}

We now move to connect the galactic stellar masses to their ionizing luminosities, in the right hand panels of \figref{fig:gal_pties}. 

Panel D shows the average galaxy ionizing luminosity (or \lintr{}). In our source modeling (Section~\ref{ssec:src}), at fixed stellar population age and metallicitiy, the ionizing luminosity scales linearly with stellar mass. Thus, the average \lintr{} increases with halo mass, roughly as \mstel{}. However, the conversion is not straightforward, because of the dependence on stellar ages and metallicities, which are affected by the detailed star formation and enrichment histories of individual galaxies. Indeed, the highest mass haloes in Illustris appear to have somewhat higher average \lintr{}, with respect to the other simulations, than their \mstel{} alone suggests.

Panel E shows the fraction of the total ionizing luminosity produced in each halo mass bin. Qualitatively, this metric closely resembles the stellar mass counterpart (panel to the left). However, the halo masses that contain the largest fraction of overall stellar mass contain even larger fractions of the total ionizing luminosity. Further, these masses are slightly lower for the luminosity curves than for the stellar mass curves.  As a result, the ionizing stellar photons overwhelmingly (roughly 50 \%) come from \SI{\approx e9}{\simsun} haloes in Eagle, whereas in TNG100 and Illustris the ionizing sources are more evenly distributed across halo masses. This implies different typical sizes and spatial distribution of ionized bubbles in Eagle with respect to the other two simulations, and could be associated with the higher \xhi{} power seen in Eagle near $k \approx \,$\SI{ 0.8}{\per\mega\parsec} (\figref{fig:ps_ingredients}).

To understand this difference with respect to the stellar mass fractions, we show the total halo ionizing luminosity per stellar mass (in units of number of photons with $h\nu \geq 13.6 \, {\rm eV}\, \rm{s^{-1}} \, \rm M_\odot^{-1}$), that we will call specific luminosity, in panel F. Across all three simulations, the specific luminosities of the lowest mass central galaxies are roughly \SI{3e45}{\per\second\per\simsun}. Between \SI{3e9}{\simsun} and \SI{3e10}{\simsun}, the specific luminosities of TNG100 and Eagle decrease to \SI{e45}{\per\second\per\simsun} at \SI{2e9}{\simsun} and \SI{8e9}{\simsun} respectively. In this halo mass range, Illustris has significantly ($ \gtrsim 0.4 \, \rm dex$) higher specific luminosity than the other two simulations. At higher mass, the TNG100 and Eagle specific luminosities increase to approximately \SI{1.7e45}{\per\second\per\simsun} at \SI{e11}{\simsun}. All this suggests that the different predictions of the 21cm power spectrum from the various simulations arise not just because of differences in the galactic stellar masses available to emit ionizing photons, but also from the properties of the stellar populations that have (on average) different specific luminosity. 

Panels G and H quantify the stellar mass-weighted, average stellar ages and metallicities of Illustris, TNG100 and Eagle galaxies. In all three simulations, galaxies are young, and have recently formed in the lowest mass haloes, while ages increase to a maximum near \SI{e10}{\simsun}, after which the ages either plateau (in Illustris) or decrease with increasing halo mass (TNG and Eagle), reaching a rough agreement for the highest mass haloes. Since the ionizing emissivity decreases rapidly with age, this propagates into differences in specific ionizing luminosity (panel F). In fact, the ordering of the age curves roughly gives the ordering of the specific ionizing luminosity curves, showing that stellar age is the main driver of differences in the ionizing luminosities of galaxies at fixed stellar mass. 

The average age of TNG100 galaxies increases more rapidly at lower masses than in the other two simulations, leading TNG100 to have the lowest specific ionizing luminosities in low mass haloes. Between \SI{e9}{\simsun} and \SI{5e9}{\simsun}, the average stellar age increases rapidly in all simulations. This is most pronounced in Eagle, leading its ionizing luminosities to drop. Beyond this point, the typical ages stagnate in Illustris, continue to rise in Eagle, and decrease in TNG100. Consequently the specific luminosity in TNG100 increases. For \SI{e10}{\simsun} and beyond, the specific luminosities in Eagle decrease rapidly, converging with the other simulations for massive haloes. These age comparisons map well to the differences between baryon fractions: the oldest populations have the lowest baryon fractions. When feedback suppresses star formation in a galaxy, the ionizing luminosity decreases via two channels. First, there is simply less stellar mass. Second, the missing stellar mass would have been young stars efficiently producing ionizing photons.

Panel H shows that the mass-weighted stellar metallicities of galaxies always increase with halo mass. This is readily understood as they accrue more metals through successive stellar generations. For low and intermediate mass haloes ($M_{\rm halo} \lesssim 10^{10} \, \rm M_\odot$), there are significant differences in the stellar metallicities. Illustris galaxies are more metal poor than in TNG100 and Eagle. However, the majority of stars in this galaxy mass range are metal poor ($Z_{\rm absolute}<10^{-3}$), falling below the minimum metallicities for which our stellar evolution model makes ionizing luminosity predictions. These differences therefore do not affect the ionizing emissivities of galaxies. At the same time, the metallicities of the higher mass haloes are fairly similar. Overall, this confirms that the differences in specific ionizing luminosity are mainly driven by differences in stellar population ages, as a result of differing stellar feedback models.

\begin{figure}
    \centering
    \includegraphics[width=0.45\textwidth]{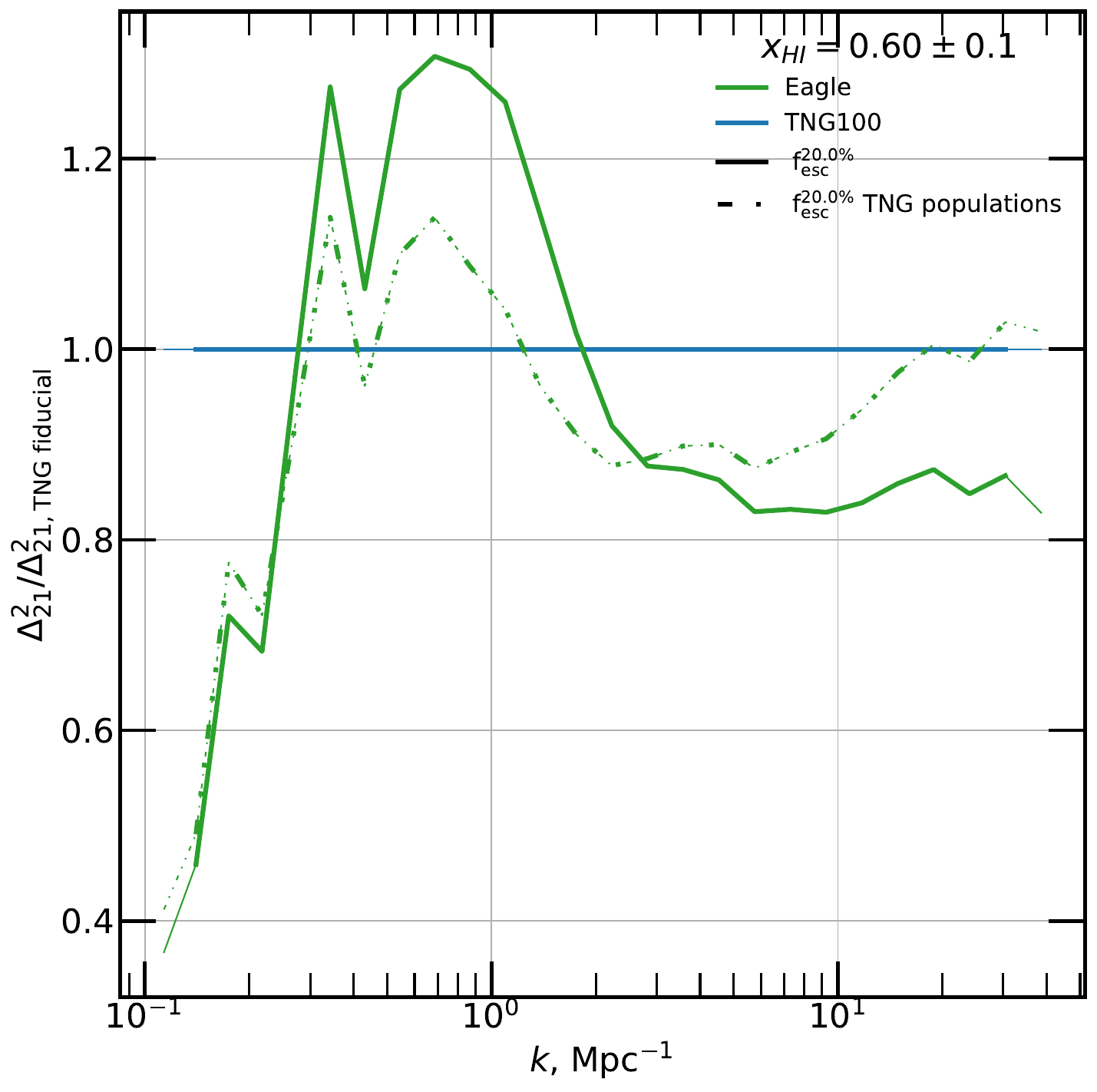}
    \caption{Relative 21cm power spectra, comparing several numerical experiments. We compare our fiducial Eagle simulation result (solid green) to a case where we rescale the ionizing luminosity of the stellar particles as a function of galaxy halo mass, so that the average ionizing luminosity per stellar mass is the same as in TNG100. This test run is called ``TNG100 populations (dashed dotted curves) and its 21cm power spectrum is compared to the predictions from the fiducial Eagle and TNG100 runs via a ratio, as in Fig.~\ref{fig:ill_src_ps}. Each power spectrum is taken at an epoch where average \xhi{}$\, \approx 0.6$. The higher 21cm power spectrum at $\approx \,$\SI{0.8}{\per\mega\parsec} of Eagle is partially explained by its older, redder, and more metal rich stellar populations. Thinner curves denote scales where the predictions are affected by resolution or box size.}
    \label{fig:tng_pop}
\end{figure}

\subsubsection{The impact of different stellar populations}

To further investigate the impact of the stellar ages and metallicities for the 21cm power spectrum, we perform the following test. We make a variant of the fiducial model and \code{CIFOG} run for Eagle, where the ionizing luminosity of the stellar particles is scaled as a function of galaxy halo mass, so that the average ionizing luminosity per stellar mass is the same as in TNG100. In other words, we compensate for the luminosity-weighted effects of the differences in stellar population ages and metallicities between TNG100 and Eagle.

In \figref{fig:tng_pop}, we compare this ``TNG100 populations'' run to the fiducial TNG100 and Eagle predictions. We find that the higher power bump in the power spectrum predicted by Eagle around $\approx \,$\SI{0.8}{\per\mega\parsec} is roughly half as strong in the ``TNG100 populations'' test. At the same time, it has roughly the same power as TNG100 at the smallest scales, whereas elsewhere it is interchangeable with the fiducial Eagle run. This confirms that scale-specific differences among our 21cm power spectrum predictions arise not only from differences in the stellar mass available to produce ionizing photons, but also in the typical ages (and metallicities) of the stellar populations in the different simulations. It also further shows that differences between the Eagle and TNG100 fiducial 21cm power spectra at \SI{\approx 0.8}{\per\mega\parsec} arise from underlying choices in the baryon and stellar feedback models. 

\subsection{Comparison to previous work}

\begin{figure}
    \centering
    \includegraphics[width=0.47\textwidth]{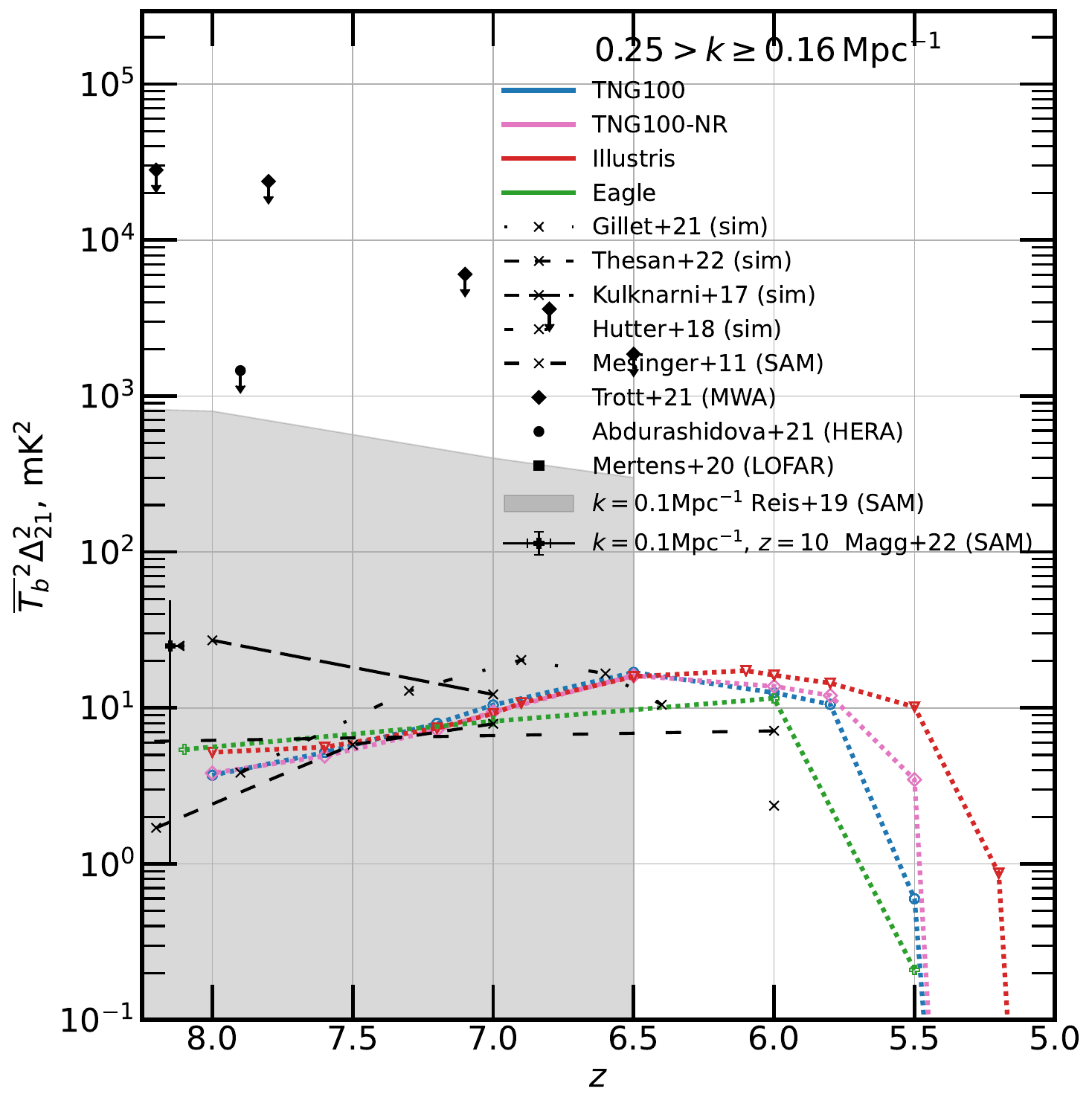}
    \caption{Redshift evolution of the 21cm power spectrum at large scales ($ 0.2 ^{+0.05}_{-0.04} \, \rm Mpc^{-1}$) in TNG100, TNG100-NR, Illustris, and Eagle. We also show relevant upper limits from observations (black markers) and other hydrodynamical simulation-based predictions from the literature (black lines). The predictions of \protect \cite{reis_mapping_2020} are represented roughly as a grey polygon. Simulation-based predictions lie within roughly 1 dex, but differ up to factors of several, depending on scale.}
    \label{fig:large_scale_evol}
\end{figure}

Figure \ref{fig:large_scale_evol} shows the redshift evolution of the 21cm power spectrum, \dtbPS{}, at large scales predicted in our fiducial modeling by Illustris, TNG100, TNG100-NR and Eagle. Despite significant differences among the simulations and models, the large scale evolution of the 21cm power spectrum is similar across the board at $z>6$. As time progresses there is a gentle rise caused by the formation of large ionized regions increasing the large scale variance of the 21cm signal. After $z\approx 5.7$ the curves drop as the volumes become predominantly ionized. The large-scale 21cm signal from Eagle is slightly lower than in the other simulations. In Eagle and TNG100 reionization also occurs earlier than in Illustris and TNG100-NR. As our predictions lie at least 2 dex below the most constraining observational upper limits, the differences are currently too small to differentiate against.

Other results that use hydrodynamical simulations, either via post-processing or by using fully-coupled RHD, find similar evolution of the large-scale 21cm power spectrum. In fact, all of these predictions can be confined in a area occupying less that one dex in 21cm power. This is striking. First, such simulations all adopt different implementations of baryonic and galactic astrophysics, and yet converge on what is to be expected on large spatial scales for the 21cm clustering. Second, this is not the case for other types of modeling, such as the semi-analytic works of \cite{reis_mapping_2020}, who find a much greater diversity in large scale power, spanning close to 6 dex. In fact, the extremely high and low predicted large-scale 21cm power spectra arise when considering very late heating and very early reionization of the IGM (respectively), which are disfavoured if taking into account other constraints on the IGM and reionization, e.g. the electron scattering depth, and the optical depth in the Lyman-$\alpha$ forest. In our case, we avoid early reionization scenarios by calibrating our source models for late reionization (as suggested by recent data e.g.\ \citealt{kulkarni_large_2019, bosman_hydrogen_2021}) and adopt similar heating histories through our choices of UVB.

\subsection{Implications}

In this work we have quantified the differences among the predictions from state-of-the-art cosmological hydrodynamical simulations of galaxies. A general result arises: across all scales and considered epochs, our predictions for the global 21cm signal and the 21cm power spectra are within a factor of a few, even without accounting for the different reionization histories of the simulations. This is all the more impressive, as we have highlighted substantial differences amongst the galaxy populations and drivers of reionization in the different simulation volumes. When one considers the huge diversity of predictions from SAMs, that span many orders \citep[e.g.][]{reis_mapping_2020}, our results appear to imply that many of these models lie outside of possible outcomes from current cosmological hydrodynamical simulations of galaxy formation. However, a large fraction of this diversity stems from different assumptions regarding the reionization and heating history of the Universe. Namely, our results are arguably in strong disagreement with models that heat up late, and/or reionize early. In addition, no observations at these high redshifts have been used to constrain the underlying galaxy-formation and feedback models adopted e.g. in Illustris, TNG and Eagle.

The largest differences among the simulations occur, not surprisingly, at small spatial scales (i.e. for $k$\SI{\geq 10}{\per\mega\parsec}) where the strength of supernova feedback models imprints a signature into the gas. At these scales, the signature of the escape fraction model (see App. \ref{app:fesc}) in the 21cm power is significantly smaller. As a result, future upper limits on the 21cm power spectrum at these scales can constrain high-redshift feedback models. Such effects are challenging to account for in the many analytical or semi-analytical works in the literature. In our approach, the resulting predicted differences in the 21cm signals are significant (up to a factor of two). However, they are also relatively modest compared to the large range of theoretical predictions. Thus, a significant leap in observations is needed to directly constrain these differences \citep[e.g.][]{abdurashidova_first_2022, koopmans_cosmic_2015, kolopanis_new_2022, trott_constraining_2021}. Furthermore, at such small scales, it is more challenging to impose upper limits on \dtbPS{} \citep[due to larger thermal noise at small scales; e.g.][]{abdurashidova_hera_2022}, limiting the feasibility of such a constraint in the near future.

We also identify an excess of 21cm power near $k \approx \,$\SI{ 0.8}{\per\mega\parsec} in Eagle with respect to the other simulations, due to topological differences in the \xhi{} field. This arises from a large difference in the ionizing photon budget of haloes. In Eagle, most ionizing photons are produced in \SI{\approx e9}{\simsun} haloes at $z=7$, in much greater proportion than in Illustris or TNG. The result is differences in the typical ionized bubble sizes. We attribute these to feedback model choices, as Eagle galaxies have lower baryon fractions, and older stellar populations than TNG100 (the closest comparable simulation). Indeed the more efficient feedback in massive Eagle galaxies makes their stellar populations older, redder, and less ionizing. In fact, this latter manifestation of feedback appears to be the main driver of the $k \approx \,$\SI{ 0.8}{\per\mega\parsec} bump in 21cm power in Eagle. Although the relative difference in the resulting power spectra between the simulations is small (at most 1.5 times greater in Eagle), these scales are easier to observe.
However, as we show in Appendix~\ref{app:fesc}, changes to the escape fraction (and its dependence on halo mass) can also manifest similarly in the 21cm power. This complicates any interpretation of the 21cm signal at these scales. Indeed, the ionizing luminosities of galaxies are the product of stellar mass, ionizing efficiency, and escape fraction. Thus, a low escape fraction could be misconstrued as less efficient star formation, or stellar populations with low ionizing efficiencies. Fitting observations with a model that parameterizes all three of these aspects would therefore be challenging, as they are highly degenerate. At the same time, this highlights the potential importance of accounting for these ionizing efficiency effects: they may be key in providing more accurate ionizing escape fraction estimations from 21cm observations.

\subsection{Limitations of the adopted methodology}
\label{sec:limits}

Illustris, TNG100, and Eagle all have volumes of roughly $100^3$\si{\mega\parsec\cubed}. Although these are large volumes for galaxy formation simulations, it has been shown \citep[][]{iliev_simulating_2014} that volumes of at least $\approx 200^3$\si{\mega\parsec\cubed} are required to achieve convergence of ionized region sizes, and large-scale 21cm power spectrum statistics. On top of this, such volumes would allow us to predict measurements at larger scales, where upper limits on the 21cm power spectrum are more constraining. Although these larger volumes would be desirable for our study, we highlight the need to compromise between volume and resolution in order to be able to model galaxies realistically, in order to accurately capture the impacts of feedback.

At the same time, our works relies on post-processing the simulation-predicted fields, such as gas density and gas temperature, to determine the ionized fraction of hydrogen gas. This does not allow us to account for the reaction of the gas to photo-ionization and photo-heating in a self-consistent way, which has been shown to have an important impact on the IGM \citep[][]{daloisio_hydrodynamic_2020}, and on the star formation of galaxies \citep[][]{dawoodbhoy_suppression_2018, wu_simulating_2019, borrow_how_2022}. Our approach does not rely on solving the radiative transfer equations, but uses an approximate description using excursion set formalism with \code{CIFOG} \citep{hutter_accuracy_2018}. This method does not explicitly conserve the number of photons, leading to excessive ionization. Moreover, we do not track the evolving temperature of the gas, leading to inconsistencies. In particular, \dtb{} in neutral regions could be artificially low, due to heating that only occurs once the region is ionized (since the emission of the ionized regions is $\approx 0$, the reverse situation is not problematic). In the future, full RT post-processing will enable additional physical effects to be captured, at the cost of additional computational expense.

Finally, in this work, we have focused exclusively on the ionizing radiation from stellar sources, which are thought to be the dominant drivers of reionization. As such, we do not model any ionizing contribution from AGN, although ionizing photons from AGN are accounted for in the simulation UVBs. However, they may play a role, especially in later reionization scenarios, when a higher density of luminous AGN are present.

We have also limited ourselves to times when the IGM has already been heated by a UV and X-ray background ($z\lesssim 8$). The prior-heating stage has a profound impact on the 21cm signal \citep[e.g.][]{pritchard_21-cm_2007, fialkov_complete_2014, ross_radiative_2017}, and extending to such times would require explicitly implementing and exploring separate X-ray source models, for AGN and X-ray binaries, for reionization and patchy (i.e. source driven) heating. This would imply handling X-ray ionization independently with their own source field and ionization cross sections, complicating our approach \citep[for instance, see][ who study impact of X-ray binaries]{sartorio_population_2023}. Additionally, when moving to higher redshifts, the masses of the galaxies producing the most of the UV and X-ray emission decrease, a resolution challenge for the simulations considered here. Expanding to higher redshifts, when the heating of the IGM is not yet complete, may reveal other differences among the simulations.

Finally, in our current approach, we have also refrained from computing full light cones \citep[as in][]{chapman_full_2019}: our approach ignores scattering events, breaks down for large optical depths, and does not account for the width of the 21cm line. Further, our treatment of the gas peculiar velocity gradient remains approximate.That being said, we show that the velocity effects are comparatively small, and do not differ greatly between simulations, meaning that our conclusions regarding the velocities are unlikely to change considerably with a more realistic treatment of the velocities. Computing full light cones would also allow the creation of mock observations in a more straightforward manner. Full forward modeling including foreground effects would also allow stronger conclusions about the true feasibility of observing the features we discuss in this paper. For now, we reserve these steps for future studies.


\section{Summary and conclusions}
\label{sec:concl}

In this work we have taken advantage of the rich baryonic physics of large cosmological hydrodynamical simulations of galaxies to predict the emission from the 21cm line of neutral hydrogen in the post-heating era ($z\approx 6-8$). In particular, we have post-processed the outcome of three simulations -- Illustris, IllustrisTNG (specifically TNG100), and Eagle in order to compare and contrast their predictions for 21cm observables. Our key results are as follows:

\begin{itemize}
    \item The predicted global 21cm signal and 21cm power spectra of Illustris, IllustrisTNG, and Eagle are similar, with differences up to a factor of a few at most. These differences are of the same order as the uncertainties in the post-processing modeling, even on small scales ($k \geq \,$\SI{ 5}{\per\mega\parsec}).
    \item Variations between the predicted power spectra of the 21cm brightness temperature at $z\lesssim 8$ are mainly driven by large-scale differences in the progress of reionization.
    \item On the smallest spatial scales, $k \geq \,$\SI{ 10}{\per\mega\parsec}, the three simulations have more significant differences, due to their different models for baryonic and feedback physics. Illustris predicts significantly more power than TNG or Eagle. This occurs due to stellar feedback differences, particularly the higher galactic wind velocities of TNG versus Illustris. The resulting outflows impact the gas densities and velocities around galaxies. At the same scales, the TNG100 and Eagle 21cm power spectra are similar.
    \item On intermediate spatial scales, $k \approx \,$\SI{0.8}{\per\mega\parsec}, we find that key aspects of the sources of ionizing radiation impart features. Eagle predicts a higher 21cm power spectrum, driven by more power in the \xhi{} field. Compared to the other simulations, Eagle galaxies in haloes of \SI{\approx e9}{\simsun} produce a larger fraction ($\approx 50$ \% at $z=7$) of the total ionizing radiation, influencing the topology of reionization. This is partly due to the stellar feedback in Eagle, which has overall stronger impacts than in TNG in the most massive haloes, leading to lower star formation and older, redder stellar populations. These 21cm power spectrum features are similar to those imprinted by different choices of the escape fraction.
    \item As a result, future measurements of the 21cm signal, in addition to other reionization observables, could constrain the details of feedback and the astrophysics of galaxy formation, at small scales ($\geq$\SI{ 10}{\per\mega\parsec}). At larger scales ($\approx \,$\SI{ 0.8}{\per\mega\parsec}), the signature of feedback is degenerate with the escape fraction. For a given cosmology, data therefore jointly constrains the source model, feedback, and other galactic astrophysics at these scales.
\end{itemize}

Our findings suggest that current state-of-the-art cosmological hydrodynamical simulations of galaxies, coupled with post-processing modeling, are viable and compelling tools to study the 21cm signal at large scales during the epoch of reionization. In fact, the impact of stellar feedback on the ages and metallicities, and thus ionizing emissivities, of stellar populations is often omitted in the modeling of the 21cm signal, which instead, for a given predicted 21cm signal, we find to be largely degenerate with the ionizing escape fraction model. Our findings therefore highlight the need for properly accounting for these effects in order to e.g. constrain the ionizing escape fraction from the current observed limits on 21cm power. On the other hand, future studies with fully coupled radiative transfer, as well as higher resolution hydrodynamical simulations with resolved interstellar medium and supernova feedback physics, will enable more quantitative predictions for the \textit{astrophysics} within the 21cm signal.

\section*{Acknowledgements}

This study is supported by the DFG via the Heidelberg Cluster of Excellence (EXC 2181 - 390900948) “STRUCTURES: A unifying approach to emergent phenomena in the physical world, mathematics, and complex data”, funded by the German Excellence Strategy. The authors thank Anne Hutter for their help in using CIFOG. DN acknowledges funding from the Deutsche Forschungsgemeinschaft (DFG) through an Emmy Noether Research Group (grant number NE 2441/1-1). The primary TNG simulations were carried out with compute time granted by the Gauss Centre for Supercomputing (GCS) under Large-Scale Projects GCS-ILLU and GCS-DWAR on the GCS share of the supercomputer Hazel Hen at the High Performance Computing Center Stuttgart (HLRS). Part of this research was carried out using the High Performance Computing resources at the Max Planck Computing and Data Facility (MPCDF) in Garching, operated by the Max Planck Society (MPG). The authors also acknowledge support by the state of Baden-W\"{u}rttemberg through bwHPC. This work is partly supported by the European Research Council via the ERC Synergy Grant “ECOGAL – Understanding our Galactic ecosystem: From the disk of the Milky Way to the formation sites of stars and planets” (project ID 855130) and via the ERC Advanced Grant “STARLIGHT: Formation of the First Stars” (project ID 339177). We also acknowledge funding from the Deutsche Forschungsgemeinschaft (DFG) via the Collaborative Research Center (SFB 881 – 138713538) “The Milky Way System” (subprojects A1, B1, B2 and B8). We also thank the German Ministry for Economic Affairs and Climate Action for funding in the project “MAINN – Machine learning in astronomy: understanding the physics of stellar birth with invertible neural networks” (funding ID 50OO2206). Finally, we thank for computing resources provided by the Ministry of Science, Research and the Arts (MWK) of \emph{The L\"{a}nd} through bwHPC and DFG through grant INST 35/1134-1 FUGG.

\section*{Data Availability}

Data directly related to this publication and its figures will be made available on request from the corresponding author. The Eagle, Illustris and TNG simulations are publicly available \citep{nelson_release_2015, mcalpine_eagle_2016, nelson_first_2019}. Illustris and TNG are accessible in their entirety at \href{www.tng-project.org/data}{www.tng-project.org/data}. 


\bibliographystyle{mnras}
\bibliography{21cm_art}


\appendix

\section{Model variants and convergence}
\label{app:models}

\subsection{Source model variants}
\label{app:fesc}

\begin{figure*}
  \includegraphics[width=0.33\textwidth]{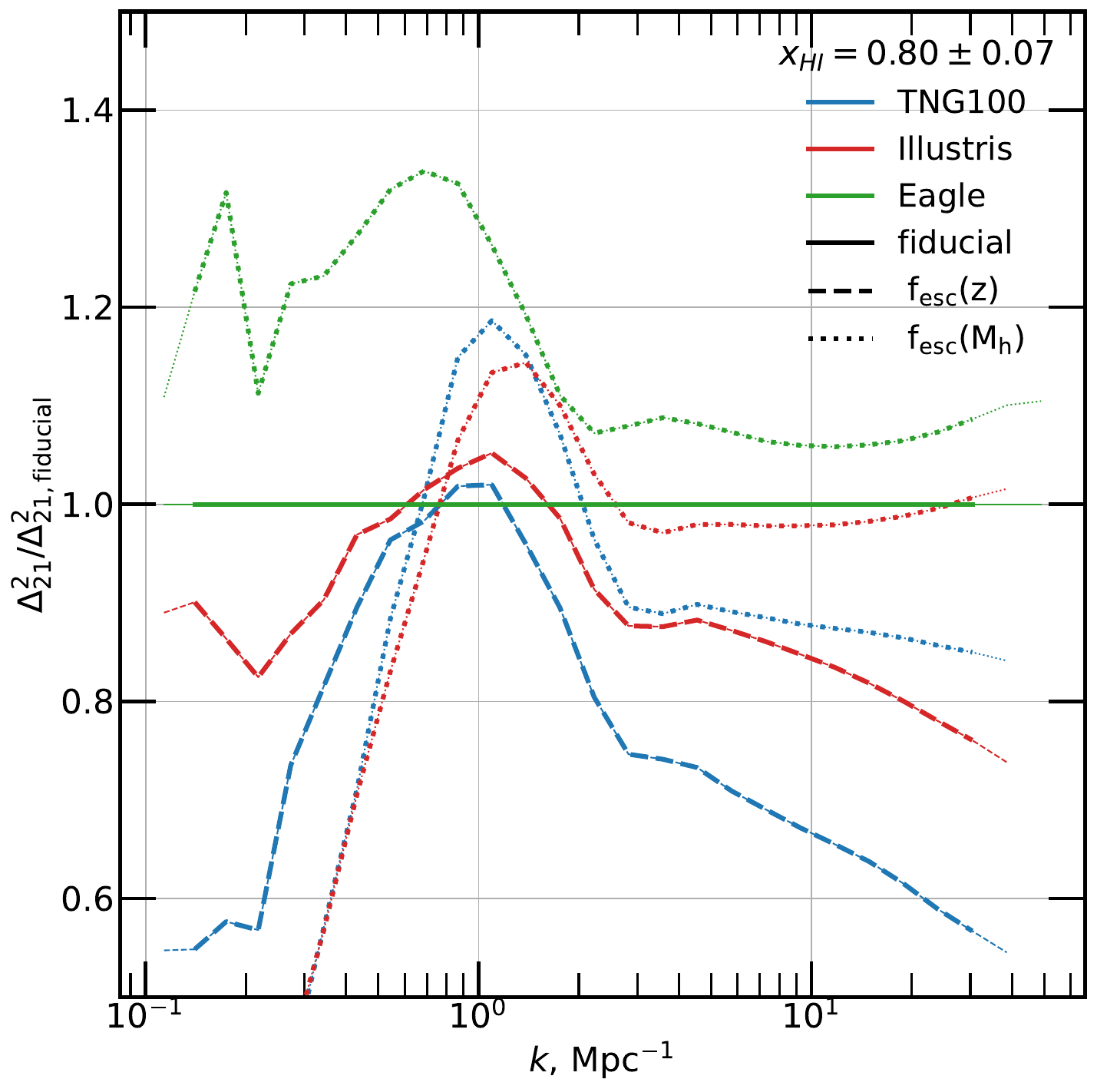}
  \includegraphics[width=0.33\textwidth]{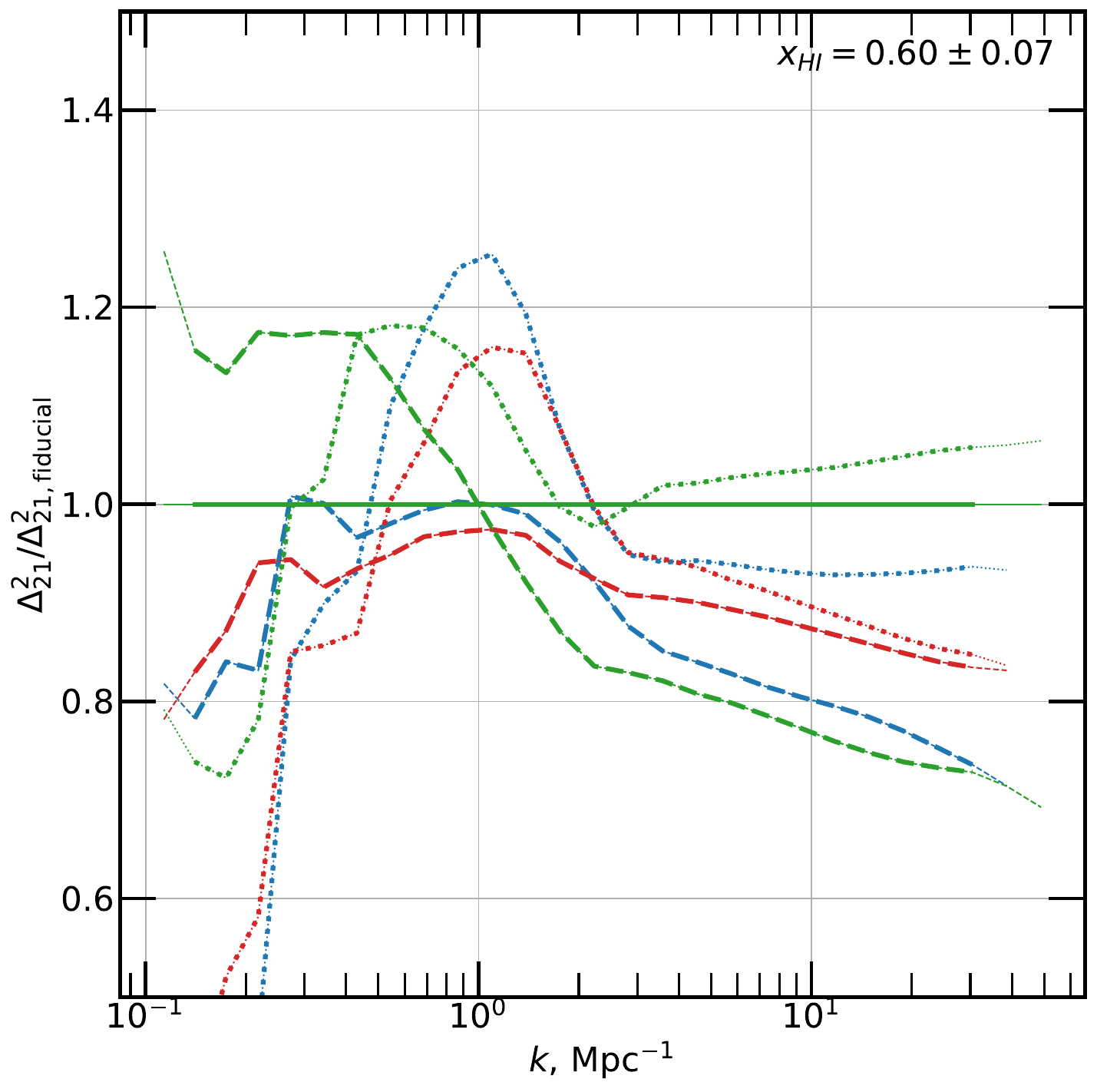}
  \includegraphics[width=0.33\textwidth]{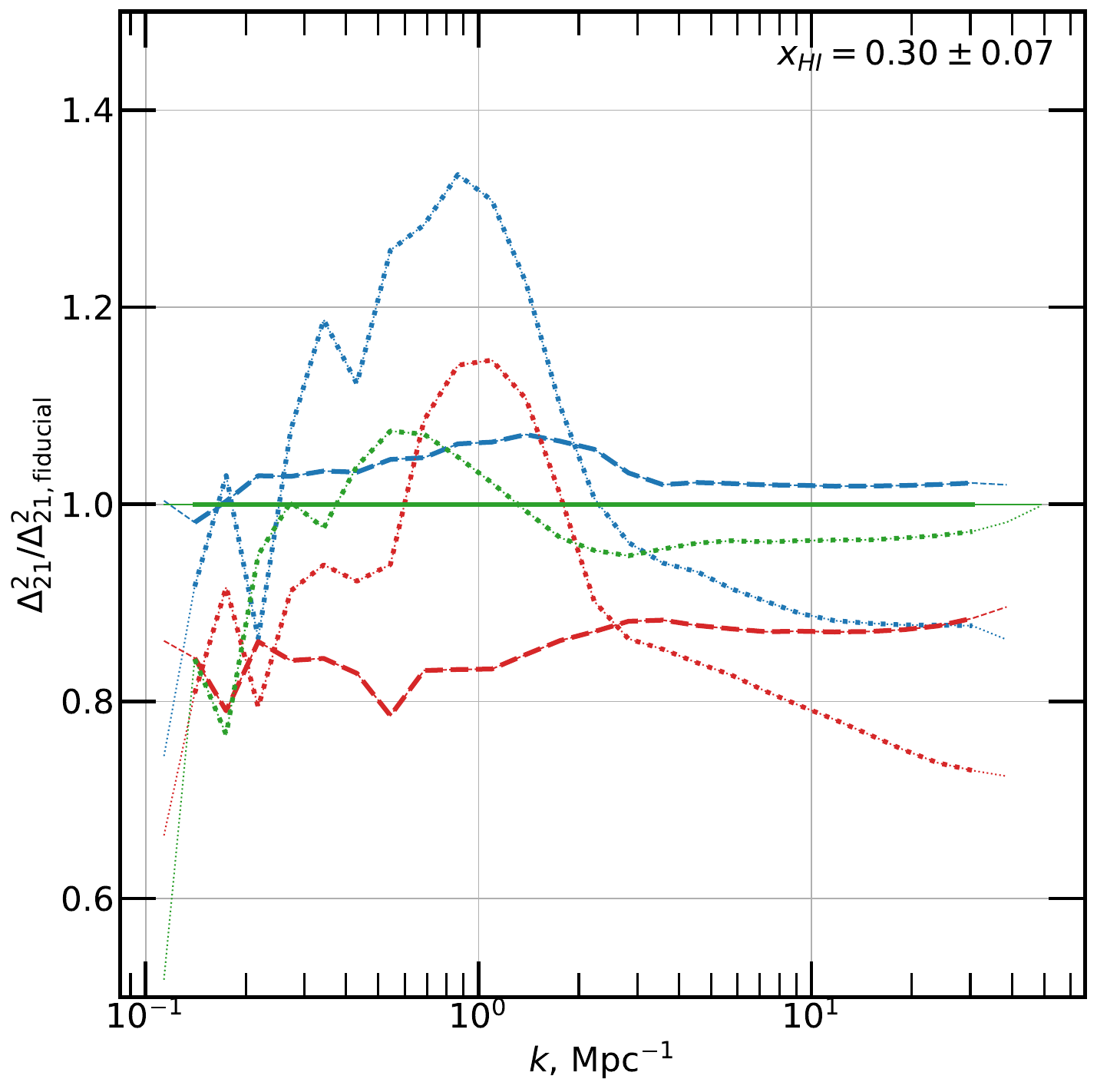}
  \caption{Different escape fraction models, and their relative impact on the $ \Delta^2_{21}$ with respect to our fiducial model (with a constant escape fraction of 20 \%).}
  \label{fig:ps_fesc_model_variants}
\end{figure*}

\begin{figure*}
  \includegraphics[width=0.33\textwidth]{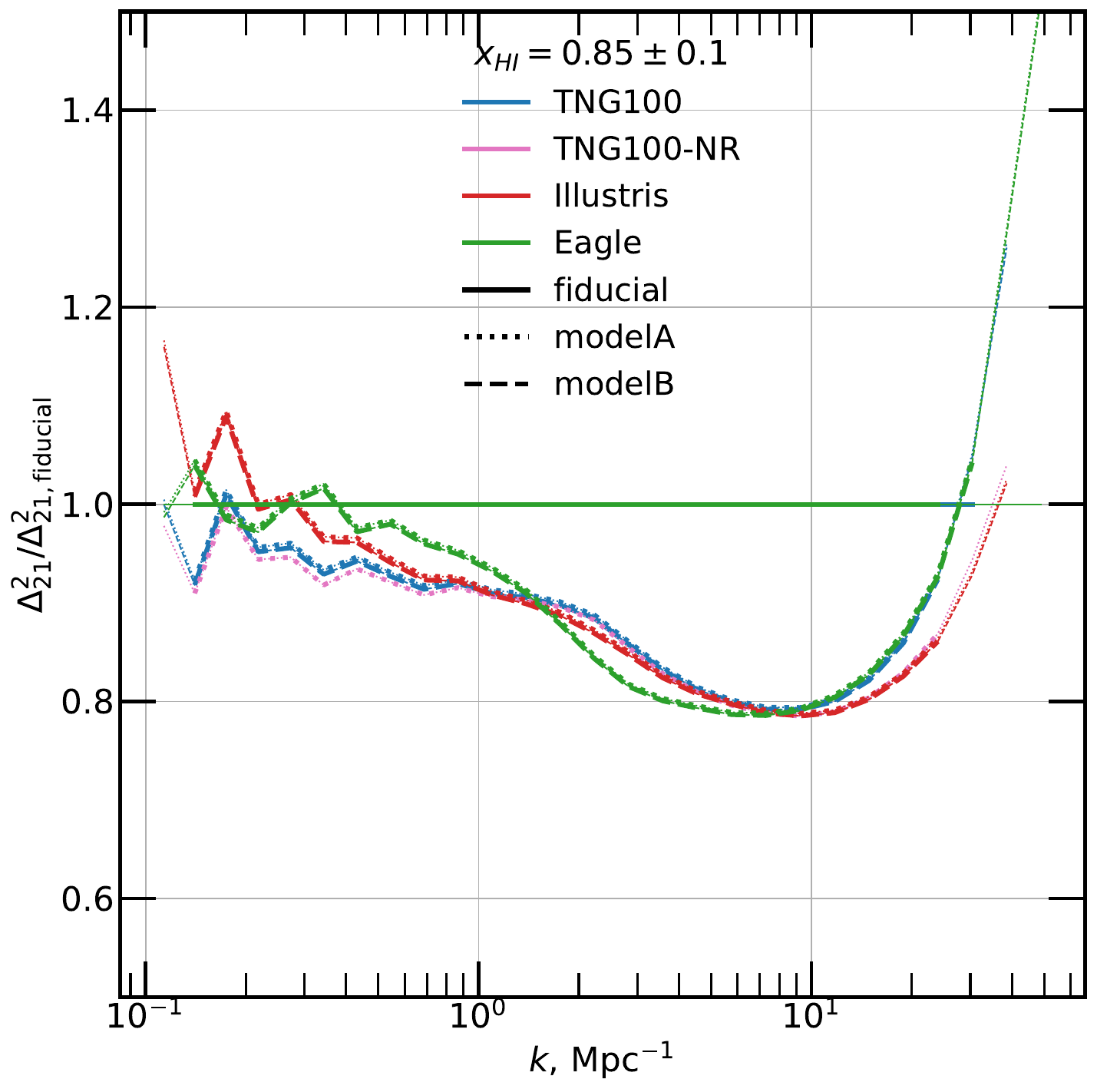}
  \includegraphics[width=0.33\textwidth]{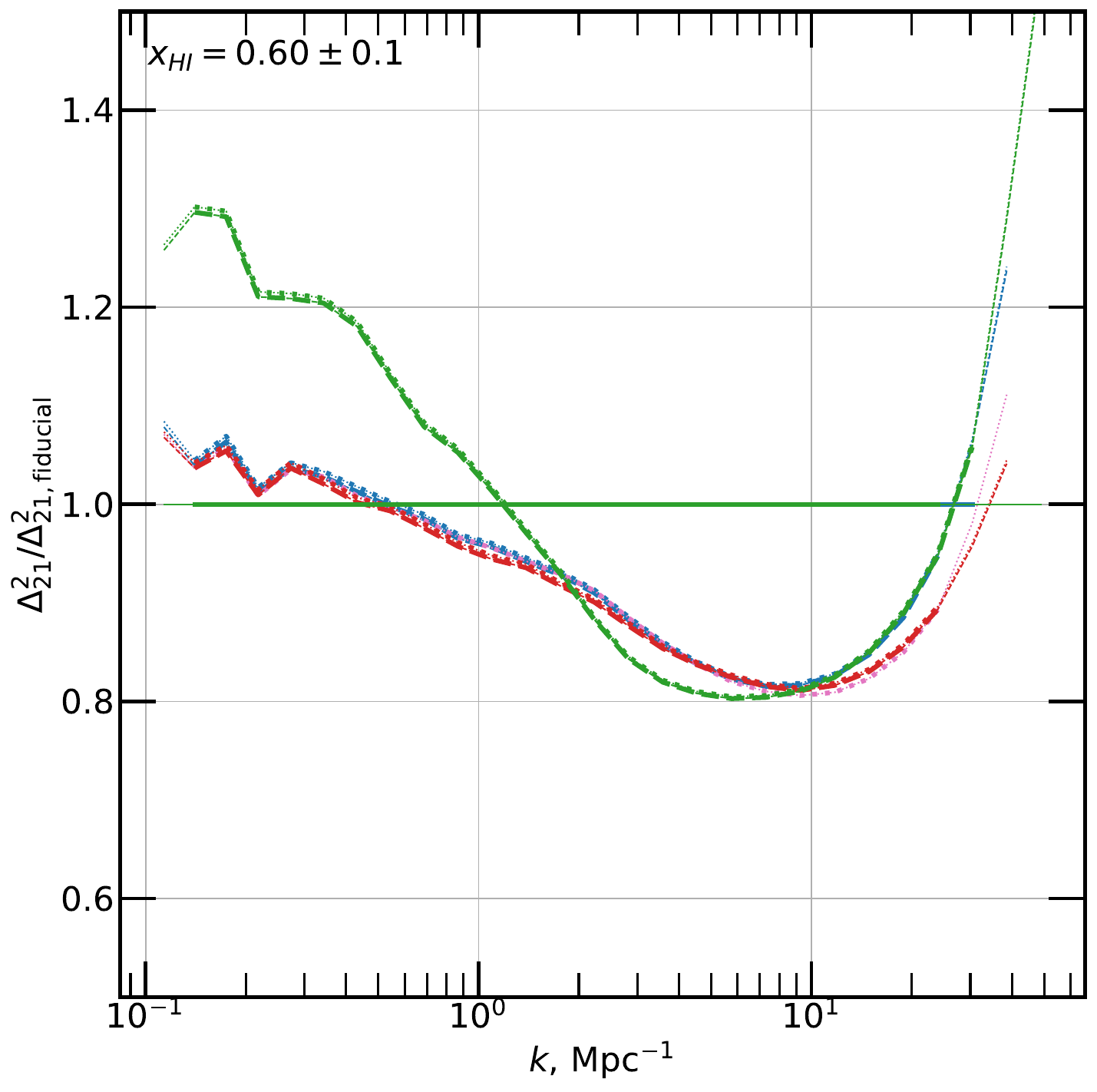}
  \includegraphics[width=0.33\textwidth]{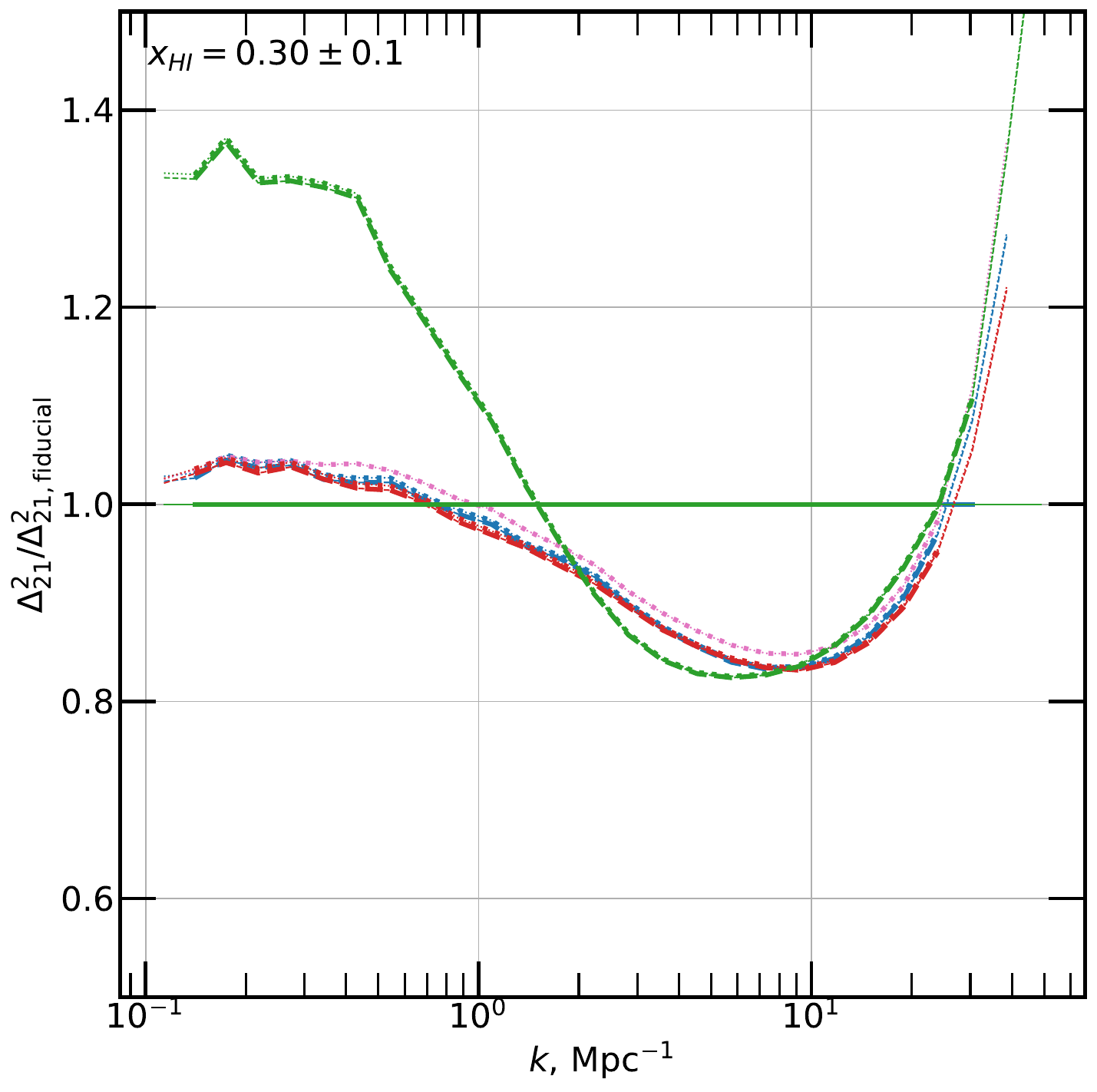}
  \caption{Different \dtb{} models and their relative impact on the \dtbPS{} with respect to our fiducial model. Including redshift space distortion effects (the difference between model B and the fiducial model), has a substantial effect on the PS at large scales in Eagle and \SI{\approx 1}{\per\mega\parsec} scales in all the simulations. }
  \label{fig:ps_dtb_model_variants}
\end{figure*}

We assess the changes in \dtbPS \, due to differences in the three galaxy formation models. By comparing \dtbPS \, in each simulation with several models for the unresolved escape fraction, we can evaluate their relative importance. To this end, we consider three typical models for the ionizing escape fractions of galaxies:
            
\begin{itemize}
  \item Our simplest model considers that a constant fraction of 20 \% of all ionizing photons produced are able to escape from galaxies. We refer to this model as the `constant model' or \fescglob.
                
  \item Next we also investigate a model where the photon escape fraction evolves as a power-law function of redshift. We call this model `evolving' or \fescevol. This is inspired by many papers that connect low measurements of \fesc{} at low redshift and the requirement of high effective \fesc{} at high redshift \citep[e.g. in][]{dayal_reionization_2020, puchwein_consistent_2019}. In our implementation \fescevol \, evolves linearly from 0.4 at $z\geq 10$, to 0.1 at $ 5 \leq z$.
                
  \item Finally we also consider a model in which sources associated with the central galaxy of detected dark matter haloes are the only contributors of photons towards reionization. We assume that the photon escape fraction for these galaxies decreases with increasing host halo dark matter mass as reported by some recent simulations \citep[see, for instance,][]{lewis_galactic_2020, rosdahl_lyc_2022}. Our model sets \fesc{} for central galaxies in haloes of $ M_{\rm h} \leq 5 \times 10^9 M_\odot$ to 0.3. Beyond this threshold, \fesc{} decreases with increasing halo mass according to a power law of exponent $-0.5$. This is the `halo \fesc' model or \feschalo .
\end{itemize}
            
In each of these cases, we calibrated the different \fesc{} models so as to yield a similar (within $\approx 50\%$) cumulative total emission of ionizing photons from $z=15$ till $z=5.5$ in TNG100. We then test these configurations by running them with CIFOG on TNG100 to ensure that their reionization redshift is similar. Finally, we apply the calibrated models to all the studied simulations.

Figure \ref{fig:ps_fesc_model_variants} shows the ratio of $ \Delta^2_{21}$ for the different considered \fesc{} models in TNG, TNG100-NR, Illustris, and Eagle with respect to the fiducial choice (\fescglob{}). Overall, the choice of source model affects the final 21cm power spectrum by roughly 40\% or less, with the most dramatic changes occurring at larges scales. The amplitude of these variations is similar to the large scale differences we observe between simulations in Sec. \ref{sec:21cm}.

\begin{figure*}
    \centering
    \includegraphics[width=0.32\textwidth]{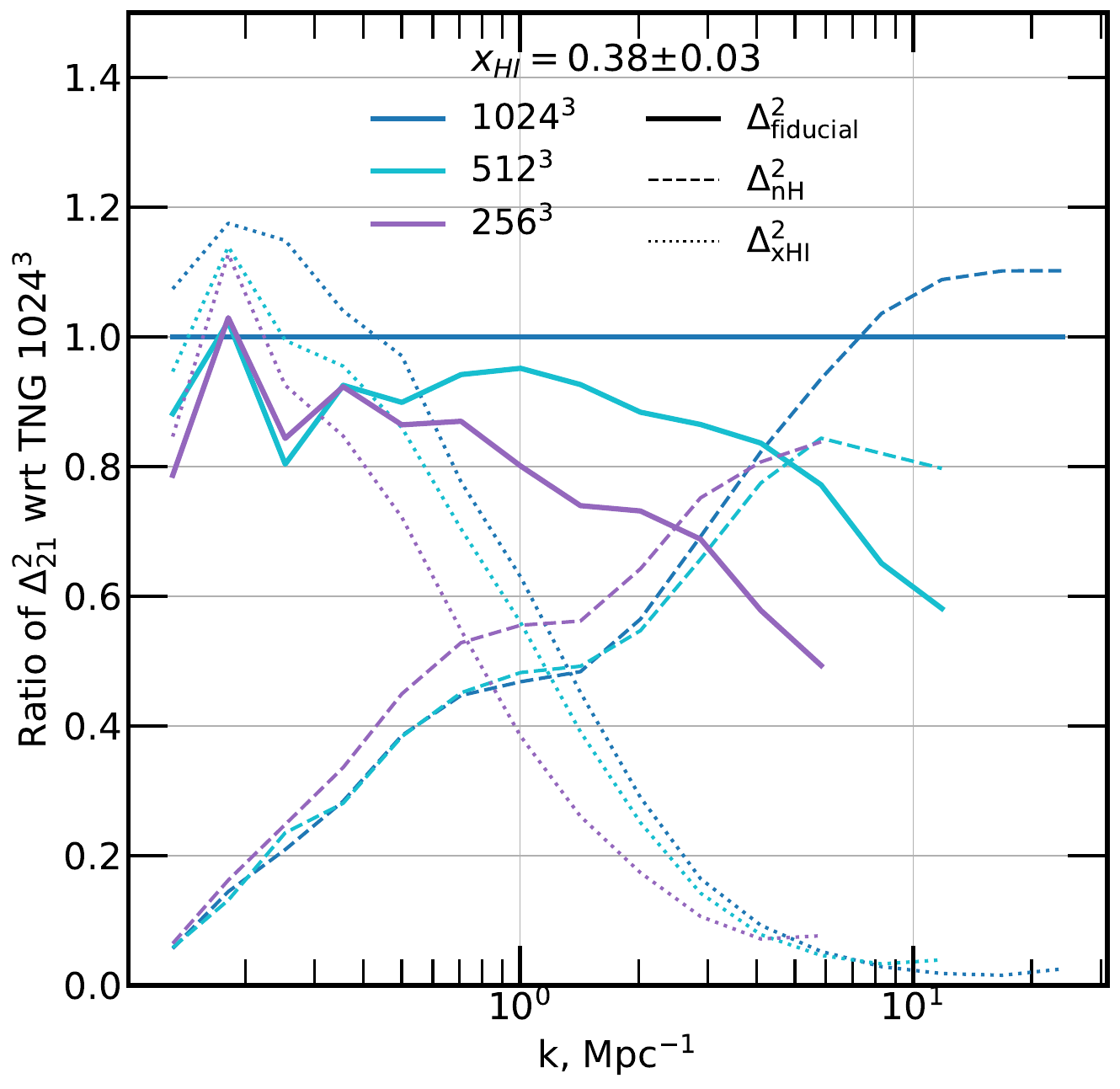}
    \includegraphics[width=0.32\textwidth]{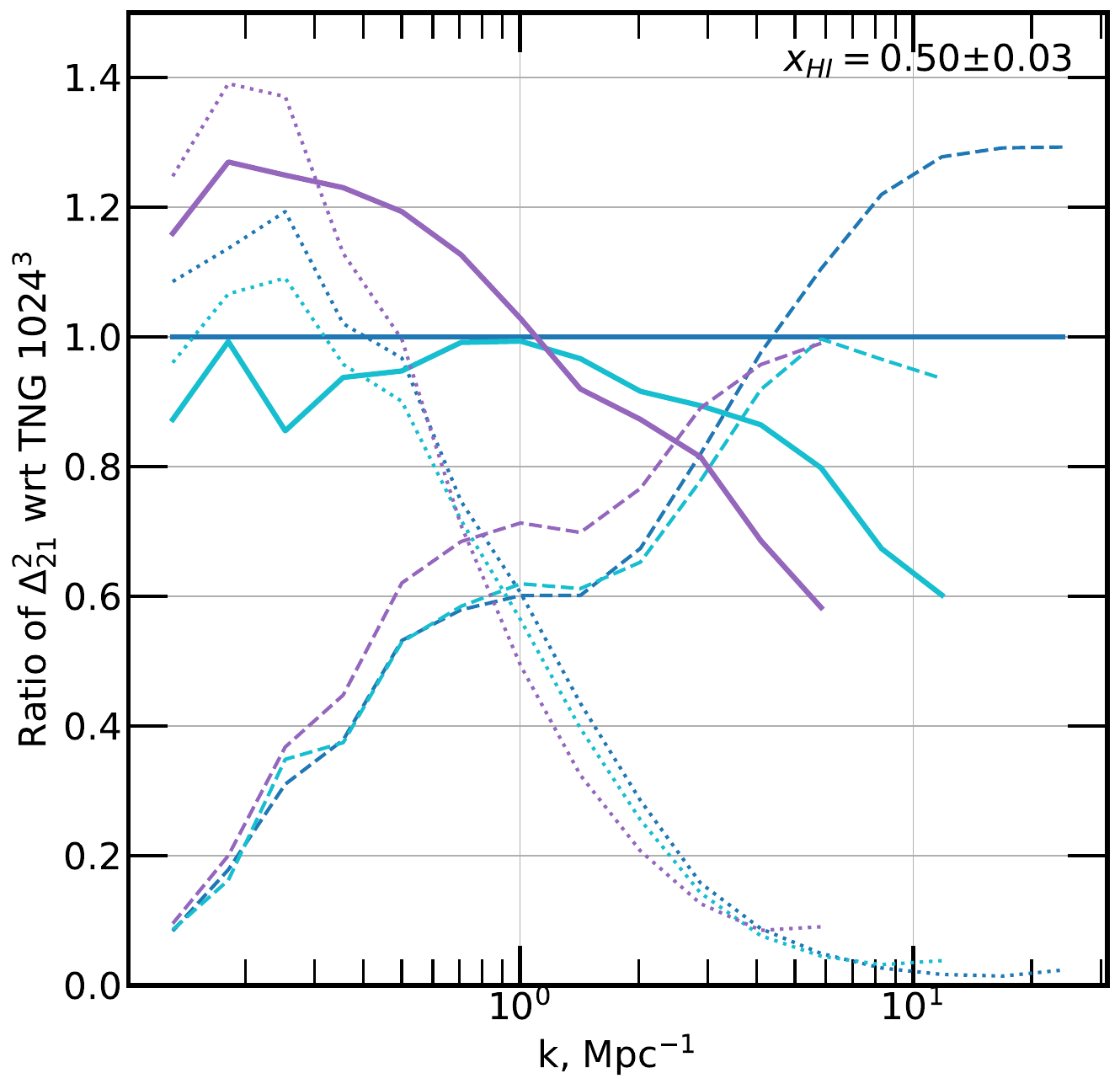}
    \includegraphics[width=0.32\textwidth]{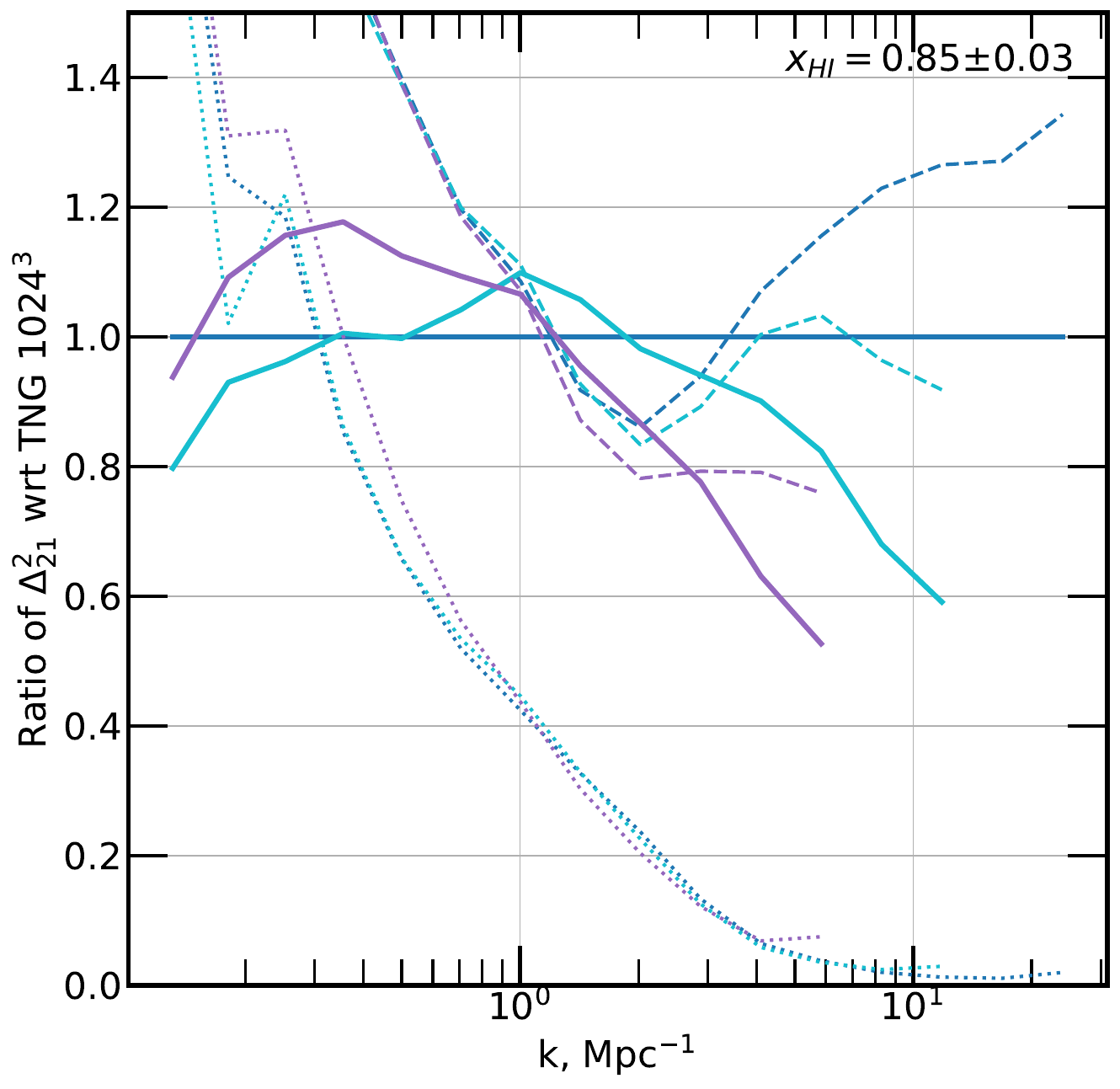}
    
    \caption{Comparison between 21cm brightness temperature, density, and ionization power spectra from the TNG100 simulation. In each panel, power is computed at redshifts where the average ionized fraction is similar. using three different underlying Eulerian grids of a different resolutions. The results are given in fractions of the $1024^3$ results.}
    \label{fig:res_conv}
\end{figure*}

For the halo mass model (or \feschalo{}; dotted lines), we find moderate to large differences near $k \leq 1 \, \rm Mpc^{-1}$, with either a slight re-normalisation or a gentle decrease towards smaller scales. This is reminiscent of the results from the ``TNG populations'' (\figref{fig:tng_pop}), and ``Illustris sources'' (\figref{fig:ill_src_ps}). Indeed, by introducing the right halo mass dependence to the escape of ionizing photons, we can mimic the source field of a different simulation that has a different stellar mass to halo mass relation.

For the evolving model (or \fescevol{}; dashed lines), and for \xhi{}$ \, > 0.30$, we find little difference at $k = 1 \, \rm Mpc^{-1}$, with all simulations showing differences at larger and smaller scales. Whereas when \xhi{}$ \, = 0.30$, the evolving model is close to the constant model in TNG and Illustris (barring some re-normalisation). The scale dependent changes seen in the evolving model with respect to the constant model may seem surprising, as at fixed redshift, we do not alter the escaping luminosities amongst galaxies, but apply a common escape fraction to all sources. However, the driving sources of reioinzation evolve over time. Thus, by introducing a time evolution in \fesc{} we can favour the time integrated contribution of specific halo populations, affecting the sizes and distributions of ionized regions. By \xhi{}$ \, = 0.30$, these inherited differences wash out, explaining the shift with respect to the constant model.

\subsection{Brightness temperature model variants}
\label{app:model_var}

Throughout this work, we consider several model variants, in addition to our fiducial model:

\begin{itemize}
    \item[--] model A: Simplest model, assumes a null line of sight velocity gradient, and $ T_{\rm CMB}\approx T_{\rm s}$.
    \item[--] model B: Model A, but $ T_{\rm CMB}\neq T_{\rm s}$, and $ T_{\rm s} \approx T_{\rm gas}$.
    \item[--] fiducial: As described in \ref{sec:mths}. Model B, but accounting for the LoS velocity gradient from the gas (Note that for TNG100-NR we assume $ T_{\rm CMB}\approx T_{\rm s}$).
\end{itemize}

Figure \ref{fig:ps_dtb_model_variants} shows the ratio of \dtbPS{} for the different models in TNG,TNG100-NR, Illustris, and Eagle with respect to the fiducial model of each simulation. Each panel shows the results from the simulations when average \xhi{} was similar. The model A and model B results, that do not include any velocity effects (redshift space distortion along the LoS), are very similar to each other, but substantially different to the fiducial model that does account for these effects.

Accounting for velocity effects incurs up to a 20\% reduction in power (near $k$\SI{\approx 10}{\per\mega\parsec}) for all simulations, with the scale at which the maximum difference occurs being slightly different between simulations. At large scales ($k$\SI{\leq 1}{\per\mega\parsec}) and when average \xhi{}$<0.6$, velocity effects increase power in Eagle. As reionization progresses, velocities have a much greater large-scale effect in Eagle, increasing \dtbPS{} by up to $ \approx 35 \%$ at the largest scales.

Since the model A and B curves are indistinguishable, there are no significant effects from the temperature of the gas. At large scales, this is understandable as we purposefully place ourselves at times when the IGM has already been heated (and the temperature term in Eq.~\ref{eq:dTb2} is $\approx 0$) However, at small scales this could appear surprising. Indeed, different stellar feedback prescriptions alter the density and velocities of the gas at such scales, therefore one could have reasonably expected to see differences arising from the hot gas expelled by supernova and AGN feedback. In fact, for there to be an appreciable difference, one would require the differences in hot winds to significantly alter the term $ 1 - T_{\rm gas}/T_{\rm CMB}$ (i.e. for the unheated gas to be cool or close to the temperature of the CMB), and this may only occur at very small scales which are smoothed over in our predictions.

\subsection{Resolution and convergence}
\label{app:conv}

Figure \ref{fig:res_conv} shows the ratio of \dtbPS{} at $\mean{x_{\rm HI}}=0.85,0.50,0.38$ and at two different resolutions ($256^3$ and $512^3$) with respect to the full resolution of $1024^3$ cells (full curves). For each resolution, we perform the entire post-processing pipeline as described in Sec. \ref{sec:mths}. For all average \xhi{}, the two highest resolution results agree to within 10\% for $  k \lesssim 5.0 \, \rm Mpc^{-1}$ (accounting for noise and aliasing effects at the largest scales). 
At smaller scales (larger k), where the \dtbPS{} is dominated by the the gas density terms, the $512^3$ and $1024^3$ are different by up to 40\% (at the smallest resolved scales). This makes intuitive sense, as lower resolution smooths the small scales and suppresses the variance in the gas density terms of \dtbPS{}, leading to lower power. Thus, were the grids higher resolution, one could expect higher small scale \dtbPS{} predictions. Seemingly, larger more resolved grids would be required (e.g. at least $ 4096^3$) for our predictions to be converged at \SI{\geq 10}{\per\mega\parsec}. This is confirmed by examining the dashed curves that show the power contribution of the hydrogen density. Indeed from $  k \lesssim 5.0 \, \rm Mpc^{-1}$, there are $>20\%$ differences in the hydrogen density power, explaining all the small scale resolution effects. 
The large scale differences in power arise because the $256^3$ \code{CIFOG} output is at such a different redshift, and a different TNG100 snapshot was used to compute the power spectra. The remaining large scale discrepancies can be explained by differing ionization power, which is sensitive to changes in the average ionization of the various \code{CIFOG} outputs used for this comparison.

\label{lastpage}
\end{document}